\documentclass[MNA]{mna} %

\usepackage{framed}  

\SHORTTITLE{%
    Neural Field Models: A mathematical overview and unifying framework
}

\TITLE{%
    Neural Field Models: A mathematical overview and unifying framework
}   

\AUTHORS{%
    Blake~J.~Cook
    \footnote{%
        Centre for Systems Modelling \& Quantitative Biomedicine, Institute for Metabolism and Systems
        Research, University of Birmingham, B15 2TT, UK
    }
    \footnote{Equal contribution as first author}%
    \and 
    Andre~D.~H.~Peterson
    \footnote{%
        Department of Medicine, University of Melbourne, Melbourne, Australia
    }
    \footnotemark[2]
    \and
    Wessel~Woldman
    \footnotemark[1]
    \footnote{%
        Equal contribution as last author
    }
    \and
    John~R.~Terry
    \footnotemark[1]
    \footnotemark[4]
    \footnote{%
        \EMAIL{J.R.Terry@bham.ac.uk}
    }
}

\SHORTAUTHOR{Blake J. Cook, Andre D. H. Peterson, Wessel Woldman, John R. Terry}


\AMSSUBJ{35A08; 35A24; 35R09; 37G35; 92-08; 92-10; 92B20; 92C20} 

\SUBMITTED{March 22, 2021} 
\ACCEPTED{February 4, 2022} 

\ARXIVID{2103.10554v2} 

\VOLUME{2}
\YEAR{2022}
\PAPERNUM{2}
\DOI{10.46298/mna.7284}

\ABSTRACT{%
    Mathematical modelling of the macroscopic electrical activity of the brain is highly
    non-trivial and requires a detailed understanding of not only the associated mathematical
    techniques, but also the underlying physiology and anatomy. Neural field theory is a
    population-level approach to modelling the non-linear dynamics of large populations of neurons,
    while maintaining a degree of mathematical tractability. This class of models provides a solid
    theoretical perspective on fundamental processes of neural tissue such as state transitions
    between different brain activities as observed during epilepsy or sleep. Various anatomical,
    physiological, and mathematical assumptions are essential for deriving a minimal set of
    equations that strike a balance between biophysical realism and mathematical tractability.
    However, these assumptions are not always made explicit throughout the literature. Even though
    neural field models (NFMs) first appeared in the literature in the early 1970's, the
    relationships between them have not been systematically addressed. This may partially be
    explained by the fact that the inter-dependencies between these models are often implicit and
    non-trivial. Herein we provide a review of key stages of the history and development of neural
    field theory and contemporary uses of this branch of mathematical neuroscience. First, the
    principles of the theory are summarised throughout a discussion of the pioneering models by
    Wilson and Cowan, Amari and Nunez. Upon thorough review of these models, we then present a
    unified mathematical framework in which all neural field models can be derived by applying
    different assumptions. We then use this framework to i) derive contemporary models by Robinson,
    Jansen and Rit, Wendling, Liley, and Steyn-Ross, and ii) make explicit the many significant
    inherited assumptions that exist in the current literature. This framework ensures that the
    validity of these assumptions will be carefully considered when interpreting results derived
    from NFMs and also in light of further developments in mathematical modelling, neurophysiology
    and neuroanatomy that enable these assumptions to be updated. We then provide an overview of
    key application areas of neural field theory and critically examine the limitations inherent in
    this class of models. Finally, we will discuss several avenues that may be able to address
    these limitations, including novel methods from stochastic processes, random neural networks,
    networked dynamical systems, and ensemble methods that connect the dynamics from the microscale
    to the macroscale.
}

\newcommand{\Z}{\mathbb{Z}}              
\newcommand{\lap}{\nabla^2}              
\newcommand{\di}{\,\mathrm{d}}            

\newcommand{\br}[1]{\left \lbrace #1 \right \rbrace}     
\newcommand{\abs}[1]{\left \vert #1 \right \vert}        
\newcommand{\tx}[1]{\textnormal{#1}}

\newcommand{\ddx}[3][]{  
  \frac{\mathrm{d}^{#1} {#2}}     
    {\mathrm{d}{#3}^{#1}}
}

\newcommand{\ppx}[3][]{      
  \frac{\partial^{#1} {#2}}  
    {\partial{#3}^{#1}}
}

\begin{document}

\section{Overview} 
\label{S:intro}

The use of mathematics to understand fundamental principles in neuroscience can be traced back to
the early 20$\mbox{th}$ century. In 1907, Lapicque
\cite{lapicqueQuantitativeInvestigationsElectrical2007} introduced what could be termed the first
mathematical model of the neuron. Using a set of ODEs for the RC circuit, this approach is an early
precursor of many subsequent mathematical studies. Building on advances in experimental methods for
isolating neurons \textit{in vitro} and recording electrophysiological activity, the seminal work
of Hodgkin and Huxley  \cite{hodgkinQuantitativeDescriptionMembrane1952} introduced a
non-linear dynamical systems model to describe the membrane dynamics of a single neuron.
Soon thereafter, Beurle  \cite{beurlePropertiesMassCells1956} formulated the first
population model of neural tissue. Building from here and inspired by statistical distributions 
of
interacting particles (Reichl \cite{reichlModernCourseStatistical1999}), the neural field
approximation permits a focus on averaged properties of the neuronal ensemble, rather than the
effect of each individual neuron on all of the others (Deco et 
al.~\cite{decoDynamicBrainSpiking2008}). Effectively, these so-called \textit{neural field 
models}
relate the average pre-synaptic firing rates to the average post-synaptic membrane potentials of
neural populations. Over the past several decades, large-scale brain modelling  has continued to
mature and proliferate,  providing theoretical frameworks to study open questions in neurobiology.

One of the primary purposes of this review is to provide a unifying mathematical framework for
deriving and explaining neural field models. To achieve this, we first summarise important concepts
of neurobiology and give a brief summary of single-neuron models that underpin a ``bottom-up''
approach to modelling brain dynamics. We then introduce ensemble approaches to modelling brain
activity, and the underlying concepts of neural field models. Alongside, an historical summary of
the field, we explicitly derive the Wilson-Cowan model, the Amari model, and the Jansen-Rit model,
focussing  on the assumptions and formalisms used in these ``first-order'' models. We then consider
``second-order'' models which stratify the state variables, enabling a more detailed description of
the aggregate interaction of multiple neural populations. From a general integral equation
description, we use Green's functions to decouple the spatial and temporal components and derive
the cortical model described by Robinson et 
al.~\cite{robinsonPropagationStabilityWaves1997}. Using
state-dependent input coefficients on the state variables (i.e. conductance-based synapses), the
cortical model of Liley et al.~\cite{lileyAlphaRhythmEmerges1999} is derived. Finally 
replacing 
the
spatial kernel with a point source: reducing the neural field to a neural mass, we demonstrate how
the Jansen and Rit  \cite{jansenElectroencephalogramVisualEvoked1995} and Wendling et 
al.~\cite{wendlingEpilepticFastActivity2002} models can be derived. In all cases, significant
assumptions are made by these authors which we highlight and discuss their justifications in
detail. The assumptions made typically take the form of  mathematical approximations and
physiological and anatomical simplifications.

Following these historical reviews and unifying derivations,  we turn our focus to summarise the
background of brain activity measurements using electroencephalography (EEG) and how this is used
to inform neural field models. Contemporary applications of neural field theory are then discussed,
including studies on  the emergence of rhythms in brain activity,  sleep cycles, epilepsy and
recurrent seizure activity, the action of anaesthesia on brain rhythms, and the progression of
Alzheimer's disease. Upon review of these important applications,  we examine the issues with
neural field models, specifically the myriad of assumptions on connectivity, the phenomenological
nature of these models, and the experimental validity of these theories. We end our review by
highlighting areas in which improvements in neural field theory can be made and focus on
unaddressed issues between neuroscience and mathematics and the role of neural field theory.
Throughout the review, it is assumed that the reader is familiar with  differential equations,
integral equations,  Green's functions, integral transforms,  and dynamical systems.

\section{Mathematical approaches to modelling emergent brain activity} 
\label{S:macroscale-models}

Mathematical modelling provides an essential method of quantifying the behaviour of complex
biological systems, which can be utilised when developing or refining our understanding of the
anatomy and physiology of neural systems. This is achieved by proposing and verifying new
hypotheses, identifying important mechanisms and parameters, and making predictions based on
experimental data. Advancements in accuracy and resolution of experimental measurements of
microscale neural activity have led to quantitative mathematical models, such as the Hodgkin-Huxley
model, that characterise the signalling processes between neurons from a dynamical systems
perspective. Macroscale neural activity however, provides many significant challenges in both
measuring and modelling. Small sections of brain tissue typically have a high density of neurons
--- which themselves are highly complex units, that are coupled together in irregular and
non-uniform structures. These challenges require innovative methods for quantifying emergent
dynamics of neural populations that reduce the complexity of the system to key parameters, which in
turn can be measured using techniques from experimental neuroscience.

From a dynamical systems perspective, there are two main approaches to constructing a mathematical
model of the brain: the first is a `bottom-up approach' where evolution equations are based upon
averaged physiological properties of the individual neurons. The second is a `top-down approach'
where evolution equations are defined based upon direct observations of the emergent behaviour of
neuronal populations. The sheer number of neurons involved in the generation of 
macroscopic
electrical activity along with the anisotropic and heterogeneous non-local connectivity of neurons
makes a `bottom-up' approach highly non-trivial and largely impractical for modelling brain
dynamics from single neuron dynamics. To this end, a `top-down' approach may be better suited for
understanding which key population-level physiological parameters influence macroscopic 
dynamics.
For both these approaches a critical challenge is to relate and bridge the descriptions of
neural activity at different spatiotemporal scales. Multiscale neural activity is often temporally
separated by several orders of magnitude i.e., milliseconds versus hour timescales). Similarly, for
different spatial scales the connection between electro-chemical activities at the micro-metre
scale is vastly different to activity at the centimetre scale. The relationship between neural
activities at different spatiotemporal scales remains an open topic of research within both
theoretical and experimental neuroscience.

In this section we first provide the necessary neurophysiological background for constructing
neural field models from microscopic neural activity. A summary of the microscopic dynamics of
individual neurons is given along with a discussion of macroscale dynamics of the brain. The issues
inherent in a ``bottom-up'' perspective, which seeks to construct macroscale models by coupling
together microscale models of individual neurons is then discussed. We then provide an overview of
the alternate ``top-down'' modelling approach utilised in neural field modelling. This top-down
approach is the primary focus of this review, for which we provide a detailed discussion of the key
mathematical machinery involved. In this review we explicitly specify the assumptions made
in these models. As with most mathematical models, many assumptions are inherited, and therefore
implicit, over many iterations and applications of the model. We have categorised these assumptions
into three main types:
\begin{itemize}
    \item \textbf{Mathematical assumption:} Where a substitution of an equation or expression has
        been made that is believed to be a reasonable qualitative and/or quantitative approximation
        and/or description of the physical system being described.
    \item \textbf{Physiological assumption:} A simplification of physiological function, usually of
        neurons and synapses. There are many aspects of the structure-function relationship of
        brain constituents and neurons that are either unknown or not quantitatively measurable.
    \item \textbf{Anatomical assumption:} A simplification of the anatomical structure of the
        brain, usually with respect to connectivity. There are many properties of neural
        connectivity that are not covered by quantitative neuroanatomy, and these are often
        simplified to a qualitative conjecture or estimate.
\end{itemize}

\subsection{Physiology of the brain} 
\label{S:physiology}

At the microscale, the fundamental cells of the brain are neurons and glia. Both are
specialised cells that transmit signals using electrochemical mechanisms. Although glia play a role
in neural information processing (Ransom and Sontheimer 
\cite{ransomNeurophysiologyGlialCells1992}, Araque and Navarrete 
\cite{araqueGlialCellsNeuronal2010}), their role is typically less defined. Consequently, the
models discussed in this review are primarily centred around describing the activity of neurons and
not glia.

\begin{figure}[htbp]
    \centering
    \includegraphics[width=0.98\textwidth]{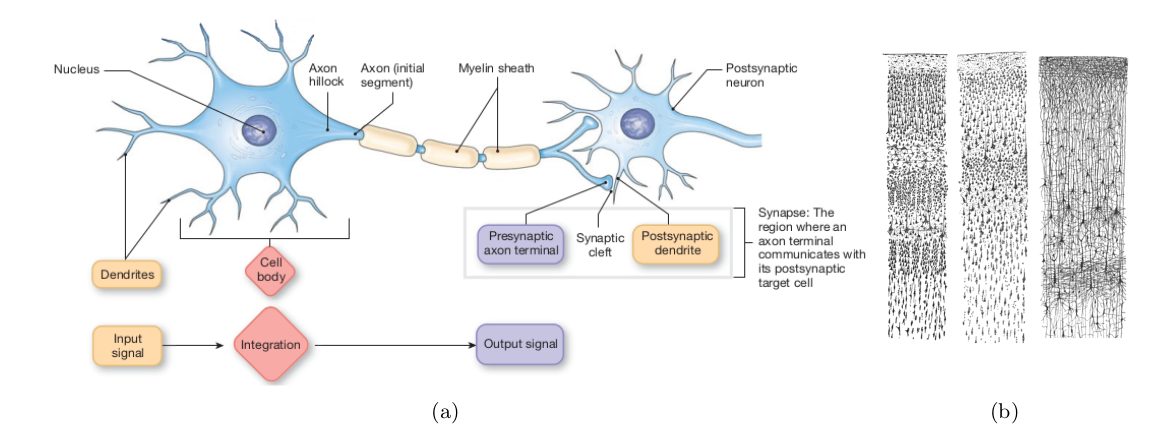}
    \caption{%
        \textbf{Neuron and synapse anatomy}. (a) Simplified illustration of neuron anatomy
        (modified from Silverthorn et al.~\cite{silverthornHumanPhysiologyIntegrated2013}). The
        cellular membrane of neurons maintains a steady state of electrochemical equilibrium at a
        potential difference around -70 mV. Incoming signals from other neurons perturb the
        membrane potential at the cell body, until a threshold potential is reached. At this point,
        the neuron sends a unidirectional self-propagating signal (an action potential) along its
        axon to the synaptic terminals. These terminals interface with the dendrites and cell
        bodies of other neurons at junctions called synapses. The anatomy of the synapse is
        intricate and relies on a number of sub-cellular interactions of neurotransmitters.
        Neurotransmitters are stored in vesicles at the terminals, which are released into the
        synaptic junction upon receiving an action potential. The neurotransmitters trigger the
        movement of ions on the next neuron, depending on which ion channel it activates. Different
        types of neurotransmitters activate different receptors, which can either excite or inhibit
        the connecting neuron by increasing or decreasing the membrane potential respectively. (b)
        Sketches of the lateral arrangement of neuron in the cortex (Ram\'on y Cajal
        \cite{ramonycajalHistologieSystemeNerveux1911}), which form functional units of cortex
        known as `cortical columns'. The activity of these units are described by neural field
        models, and are thought to be a primary source of the EEG signal due to location and
        geometric arrangement.
    }
    \label{F:physiology}
\end{figure}

Neurons generally receive signals at extensions from their cell body called dendrites and
send signals along a filament called the axon (see Figure \ref{F:physiology} (a)). Neurons
communicate by transmitting signals along axons, referred to as action potentials. These are neural
signals where the membrane potential at subsequent points along the axon rapidly rises and falls in
a matter of milliseconds. This propagating signal terminates at the axon terminals, where it causes
the release of chemical neurotransmitters at the interface with the dendrite, known as the synapse.
Additionally, neural communication can also occur at gap-junctions directly between cells via
direct electrical signalling, as well as by chemical communication. While all neurons share similar
features, the shape and size of neurons can vary significantly. For instance, the number of
dendrites and axon terminals can differ considerably, and the length of the axon is generally much
longer in peripheral neurons outside of the brain. Moreover, some neurons (such as the neuron
depicted in Figure \ref{F:physiology} (a)) have insulated axons called ``myelination'', which
enables faster axonal signal transmission. The myelin is formed by glial cells called
oligodendrocytes wrapped around the axon, which insulate and decrease the capacitance across the
axonal membrane. Consequently, this increases the conduction velocity and is essential for normal
brain function. Myelinated neural tissue is referred to as ``white matter'' whereas unmyelinated
neural tissue is called ``grey matter''.

Neurotransmitters activate receptors, which in turn have excitatory or inhibitory effects on the
postsynaptic neuron. For example, the key receptors for glutamate all have excitatory effects,
whereas for GABA the main receptors result in inhibition. There are many other neurotransmitters,
the receptors for which can be both excitatory and inhibitory (e.g. acetylcholine), meaning care
should be taken when describing a neurotransmitter as ``excitatory'' or ``inhibitory''. These
chemical changes result in a transient change to the postsynaptic membrane potential, termed
excitatory postsynaptic potentials (EPSPs), or inhibitory postsynaptic potentials (IPSPs). These
PSPs propagate along the dendrites to the cell body where they are summed and the resulting
alteration to the membrane potential of the postsynaptic neuron can lead to the generation of a new
action potential and so the process continues. The description above of neurotransmitter kinetics
has been vastly simplified, as our definition of microscopic addresses the dynamics of cellular to
multi-cellular spatiotemporal scales. Whereas neurotransmitters and all of their numerous
associated processes such as extra-cellular dynamics and molecular biochemical pathways constitute
entire fields of study within neuroscience and other disciplines such as biophysics.

The vast majority of the volume of the brain, the \textit{cerebrum}, is divided into two
hemispheres which can be organised into layers and sections with different physiological structure
or class of responsibilities in information processing. Populations of densely
interconnected neurons give rise to complex and rich emergent dynamics at the macroscale. 
Rhythmic
oscillations in brain activity are correlated with a variety of functions, which change and learn
over time (Zeki \cite{zekiVisionBrain1993}). Electroencephalography (EEG) is used to
measure the aggregated electrical activity in the brain, and as such provides a useful reference
point for models of neural activity. It has been demonstrated that the main contributors
to the EEG signal are primarily the post synaptic potentials occurring throughout the neuronal
population, rather than the action potentials (Buzs\'aki
\cite{buzsakiOriginExtracellularFields2012}). This is due to the geometry of the axons and
dendrites as the apical dendrites form branched structures that orientate vertically to generate an
equivalent current dipole that is normal to an EEG electrode (Nunez and Srinivasan
\cite{nunezElectricFieldsBrain2006}). Beyond that however, from an 
electromagnetism
perspective, there is a fundamental gap of several orders of magnitude between the local field
potentials measure on the microscale and the putative generators of the EEG signal on the
macroscale. Because of the lack of data over several orders of magnitude, the complex
geometry of the brain and the fact that the microscopic generators are not point sources like
electrons but ionic dynamics, it is difficult to apply Maxwell's equations (Gratiy et 
al.~\cite{gratiyMaxwellEquationsTheory2017}).

More than half of the surface area of the brain lies within folds and valleys, known as
``gyri'' and ``sulci''. On average, the surface area amounts to 2300 cm$^{2}$ with a volume of
1200 cm$^{3}$ in adult humans (Elias and Schwartz \cite{eliasSurfaceAreasCerebral1969}). 
The
outer 2--3 mm cerebral tissue is called the ``cortex'', which can be further subdivided
into four sections called ``lobes''. Cortical tissue is composed of layers of grey matter
surrounding layers of white matter. Neurons of the cortex are arranged in specific layers in a
somewhat regular column pattern,  called ``cortical columns'', which are illustrate in Figure
\ref{F:physiology} (b). The cortex can be regarded as the part of brain responsible
for cognition, and can be broadly partitioned into regions (or ``lobes'')  according to the types
of activities that are processed. The frontal lobe is responsible for higher thought processes,
such as behaviour, self-awareness, and problem solving, whereas the occipital lobe at the back of
the brain is responsible for visual processing. Between these regions lies the parietal lobe which
is primarily involved in sensory processing and motor function. The temporal lobes on the sides of
a cortex are mainly utilised in memory processing, and language comprehension.

The base of the brain is where the cerebellum and brainstem are located. The cerebullum has a
tightly packed grey matter morphology and has crucial roles in fine motor function and balance. The
brainstem connects the brain to the spinal cord, and also contains circuits that are crucial for
basic survival functions such as breathing regulation. Deep structures towards the centre of the
cerebrum involve structures such as the cigulate cortex, thalamus, hypothalamus, pituary gland.
These structures are essential for relaying information throughout the brain, modulating
neurohormones and homeostasis, survival instincts, and processing short-term and long-term memory.

Connectivity in the brain is an ongoing topic of research, most notably with the human connectome
project (Van Essen \cite{vanessenWUMinnHumanConnectome2013}). While the brain is a 
largely
heterogeneous structure, some key trends in connectivity on the macroscale have been observed.
White matter fibres diffuse out from the thalamus to the cortex, where interestingly there are one
hundred times more cortico-thalamic connections than thalamo-cortical connections. A well known
principle in neuroscience, often used to simplify models in mathematical neuroscience, is Dale's
law (Eccles et al.~\cite{ecclesElectricalChemicalTransmission1976}). This asserts that a 
neuron
releases the same neurotransmitters at each axon terminal, regardless of the identity of the target
cell. Although there are exceptions in reality, this simplification allows neurons to be grouped
together into populations by the characteristics of afferent neurotransmission, in it's simplest
form allows partitioning of neurons into ``excitatory'' and ``inhibitory'' populations. In the
cortex, 80\% of cortical neurons are excitatory whereas the remaining 20\% are inhibitory and there
is a somewhat regular column pattern in the way neurons are arranged radially. It should be noted
that 97\% of connections in cortical tissue are to other regions of cortex (Braitenberg and 
Sch\"uz
\cite{braitenbergCortexStatisticsGeometry2013}).

Heterogeneous connectivity of neurons in brain tissue remains one of the significant hurdles in
experimental and theoretical neuroscience. Arrangements and connections of neurons are unclear from
microscopic examination of slices of neural tissue. Neurons in the central nervous system, known as
interneurons, feature many more dendrites and axon terminals connecting to many other interneurons
in highly non-trivial ways. Often, identifying excitatory and inhibitory neurons is also a
difficult task. MRI technologies can be used to investigate bulk connectivities of neurons in
neural tissue non-invasively by the diffusion of water along axons, but this is an imperfect
imaging process that constitutes its own research sub-domain
(Fink and Fink \cite{finkPrinciplesModernNeuroimaging2018}).

\subsection{Microscopic descriptions of emergent brain dynamics} 
\label{S:microscale-modelling}

The complexity of the brain and its multiple spatial and temporal scales of description makes
mathematical modelling an attractive tool to aid understanding. At the microscale, the
Hodgkin-Huxley model provides a simplified description of the cellular dynamics of a single neuron
and describes how a neuron produces an action potential (Hodgkin and Huxley
\cite{hodgkinQuantitativeDescriptionMembrane1952}). This model was arguably the first great
achievement of mathematical neuroscience, recognised through the 1963 Nobel prize in physiology or
medicine. It describes the release of an action potential in terms of ion flow across a cellular
membrane via four nonlinear ordinary differential equations. Hodgkin and Huxley formulated the 
mechanisms responsible for the initiation and propagation of an action potential in terms of ionic 
currents, conductances and membrane potentials. These were based on Kirchoff's laws and R-C circuits 
using measurements from a giant axon of the squid.

Whilst the Hodgkin-Huxley equations accurately capture the voltage dynamics, they contain a level
of detail that is not always necessary, and may in fact hinder progress by convoluting primary
dynamical mechanisms. This is particularly the case, when we consider coupling together interacting
neurons to form networks. One approach to address this complexity is to reduce the dimensionality
of the system by identifying the crucial steady states and stabilities and deriving a simpler
dynamical system that has similar dynamics.

An example of such an abstracted approach is seen in the Morris-Lecar model
\cite{morrisVoltageOscillationsBarnacle1981}, which combines ion gating variables into one 
state
variable and so simplifies the Hodgkin-Huxley equations. Another abstraction of the Hodgkin-Huxley
model is the FitzHugh-Nagumo model  \cite{fitzhughImpulsesPhysiologicalStates1961},
which reduces the key dynamics to a two-variable system by generalising the van der Pol oscillator
model.

Often the full description of the voltage curve during an action potential is not necessary to
describe the fundamentals of neural dynamics, as it remains reasonably identical in shape and it is
the timings of the action potentials that determine the dynamics (Burkitt
\cite{burkittReviewIntegrateandfireNeuron2006, 
burkittReviewIntegrateandfireNeuron2006a}).
Hence, further simplifications can be made of which the most simple is the
integrate-and-fire (IF) neuron (Dayan and Abbott
\cite{dayanTheoreticalNeuroscienceComputational2001}, Burkitt
\cite{burkittReviewIntegrateandfireNeuron2006}). This model describes sub-threshold
membrane potential dynamics and implements a reset condition at the threshold potential which
resets the potential to resting state and represents an action potential spike with a Dirac-delta
distribution. Although, there are variations of IF models that represent the action potential,
particularly for fast spiking (Izhikevich \cite{izhikevichDynamicalSystemsNeuroscience2007}).

From these simplified microscale models, we can straightforwardly create large-scale networks. For
example, many spiking models of epilepsy utilise computer simulations of large networks of
multi-compartmental Hodgkin-Huxley style neurons. An advantage of this is that they can be informed
by \textit{in vitro} studies and so parameters have a direct correspondence to measurable
electrophysiological properties. This correspondence can help us understand physiological
and pathophysiological brain dynamics by allowing more detailed descriptions of ion
channelopathies, such as mutations, channel up/down regulations, glial cell interactions, synaptic
changes, and network connectivity. An early pioneer of such an approach was Roger Traub 
 \cite{traubNeocorticalPyramidalCells1979,traubCellularMechanismNeuronal1982}
whose detailed models of both the cortex and the thalamus were amongst the first to reproduce both
healthy and abnormal brain dynamics. Critically, these findings were similar to that found in
\textit{in vitro} studies
(Traub et al.~\cite{traubModelOriginRhythmic1989, traubModelCA3Hippocampal1991, 
traubCombinedExperimentalSimulation2005}).
Building on these early approaches, there are modern-day projects that utilise this
simulation-focused ``bottom-up'' description of the brain: the most highly publicised being the
Human Brain Project (Markram \cite{markramHumanBrainProject2012}).

However, there are disadvantages to pursuing this bottom-up approach. For example, there quickly
become issues with model complexity, data fitting, and interpretability of results. With 10$^{12}$
neurons in the brain, even an abstracted and simplified spiking model (Giannakopoulos et 
al.~\cite{giannakopoulosEpileptiformActivityNeocortical2001}, Larter et 
al.~\cite{larterCoupledOrdinaryDifferential1999}, Ursino and La Cara 
\cite{ursinoTravellingWavesEEG2006}),
will still effectively have the same level of complexity as the brain itself. This requires large
scale simulations on a super-computer to be effective. It also challenges the traditional notion of
a model: namely that it simplifies the order of complexity of the system being modelled. For a more
detailed discussion of the implications of modelling macroscopic brain tissue with ensembles of
microscale models, we refer the reader to several review articles including
de Garis et al.~\cite{degarisWorldSurveyArtificial2010} and Eliasmith and Trujillo 
\cite{eliasmithUseAbuseLargescale2014}.

\subsection{Toward macroscopic descriptions of emergent brain dynamics} 
\label{S:macroscale-modelling}

An alternative approach is to consider the system from a more ``top-down'' perspective. A useful
metaphor in this regard is that of temperature in a room. We know that temperature arises as the
net kinetic energy of atoms within a room. However, it is impractical to calculate temperature by
considering interactions between every particle. Instead, we model statistical quantities of
particles in the room, leading to much simpler models than can easily be solved.

\begin{figure}[htbp]
    \centering
    \includegraphics[width=0.98\textwidth]{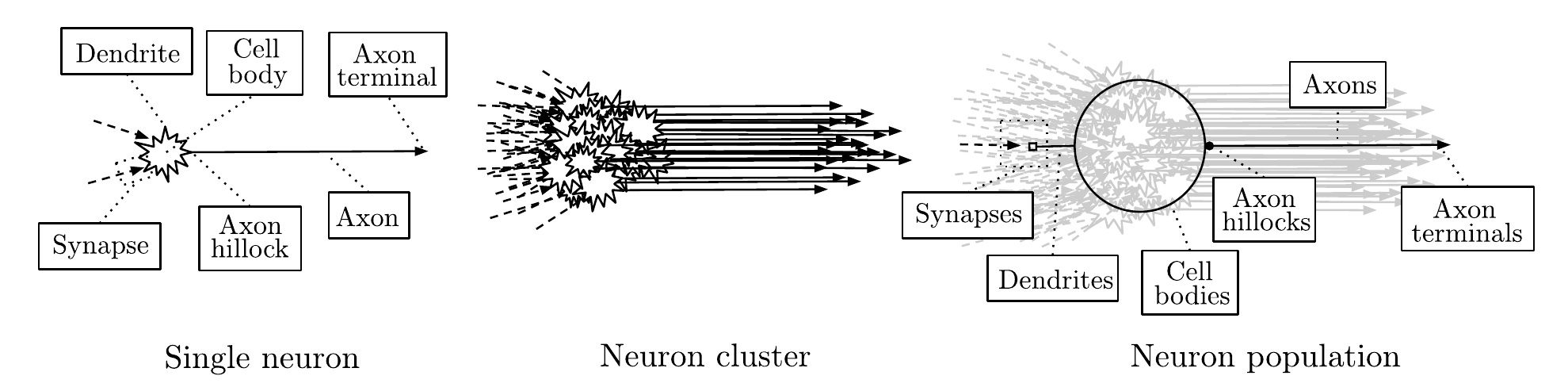}
    \caption{%
        \textbf{Diagram of neural mass approximation.} A single neuron description starts on the
        left, where models such as Hodgkin-Huxley are appropriate to describe quantities such as
        \textit{membrane potential} and \textit{spike frequency}. Increasing in scale, abstracted
        neuronal models such as leaky integrate-and-fire are used to model signal transmission
        amongst interconnected neurons. At larger scales on the right, individually addressing
        neurons becomes mathematically intractable, so a continuum approximation is typically made.
        Models at this scale include neural field models and neural mass models, which describe the
        evolution of aggregate quantities across different populations of neurons. Neurons are
        grouped together by the type and time-scale of post-synaptic signals they send. Aggregate
        quantities used in ensemble descriptions of neural dynamics include \textit{mean membrane
        potential} and \textit{mean spike rate} of sub-populations, which are closely related by a
        sigmoidal wave-to-pulse transfer function.
    }
    \label{F:neural-population-approx}
\end{figure}

In a similar manner, rather than considering the behaviour of each individual neuron, neural field
models describe the bulk behaviour of an area of cortical tissue \emph{en masse}. When constructing
neural field models, neurons are typically segregated into populations with shared statistical
factors. As illustrated in Figure \ref{F:neural-population-approx}, the state variables of the
system are population-level statistics of large collections of neurons. Following Dale's Law
(introduced earlier), a common characteristic used to partition neurons is the type of synapses
they connect to. For most models it is typical to partition neurons into ``excitatory'' and
``inhibitory'' populations.

As discussed earlier, these models typically don't consider non-neuronal populations, such as glial
cells, even those these are estimated to outnumber neurons around ten-fold (von Bartheld et 
al.
\cite{bartheldSearchTrueNumbers2016}). This is a significant assumption, since many non-neuronal
cells support and actively facilitate connections between neurons. However, including this layer of
detail significantly increases the complexity of any neural model, regardless of spatiotemporal
scale. Furthermore, the neuronal constituents of these continuum populations are also
non-compartmentalised, so we do not consider the full detail of diffusion of neurotransmitters and
inputs relative to the neuron cell body  using the dynamics of the cable equation.

Focusing on the statistical quantities of the emergent behaviour results in a simpler theoretical
framework. One where the models remain deterministic, and in principle tractable, even if the
dynamics of the collective units do not. This provides an approach to define a straightforward
neural field model, where the emergent behaviour is defined as the mean of its density.
Effectively, this describes the dynamics of the averaged activity across a region of space. Since
many techniques used to analyse the brain, such as local field potentials (LFPs),
electroencephalography (EEG), or magnetoencephalography (MEG), are observations of population level
activity, such models are well suited to understand these data recordings. A crucial advantage of
the neural field approach is that the resulting model is considerably less complex than an
equivalent microscopic network model. Therefore they are amenable to both forward modelling
approaches, such as exact solutions using Green's functions (Coombes et 
al.~\cite{coombesWavesBumpsNeuronal2003}), bifurcation theory (Kuznetsov
\cite{kuznetsovElementsAppliedBifurcation1997}), model inversion techniques such as DCM 
(Kiebel et al.~\cite{kiebelDynamicCausalModelling2008}), Kalman filtering (Freestone et 
al.~\cite{freestonePatientspecificNeuralMass2013}, Balson et 
al.~\cite{balsonSeizureDynamicsComputational2014}), as well as machine learning 
approaches 
(Ferrat et al.~\cite{ferratClassifyingDynamicTransitions2018}).

\begin{figure}[htbp]
    \centering
    \includegraphics[width=0.98\textwidth]{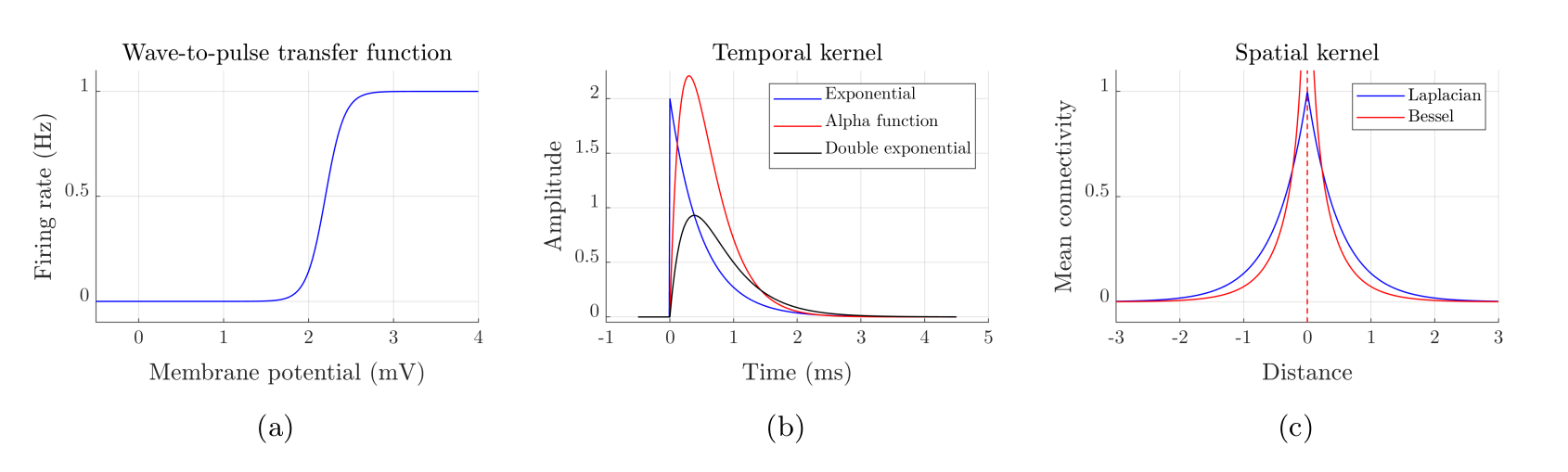}
    \caption{%
        \textbf{Transfer functions and integral kernels of neural field models}. (a) The sigmoidal
        wave-to-pulse transfer function is responsible for relating the mean membrane potential to
        the mean firing rate of a neural population. (b) The temporal integral kernel represents
        how incoming action potentials propagate and decay over time across the neural membrane. An
        incoming action potential has maximal influence on the neurons membrane potential shortly
        after it arrives, when neurotransmitters are released at the synaptic junction and interact
        with the neuron membrane, which perturbs the membrane potential by net ion movement across
        the neuron membrane. The strength of this potential perturbation depends on the proximity
        of the synapse to cell body, and decays over time. (c) The spatial kernel assigns weight to
        neural signals between populations depending on the population types and their spatial
        proximity, which takes into account the density of connections between populations. The
        connectivity between populations decays as the spatial distance between two points
        increases, hence there is a spatial limit on the maximal mean firing rate from one
        population to another.
    }
    \label{F:kernel-gallery}
\end{figure}

\begin{figure}[htbp]
    \centering
    \includegraphics[width=0.98\textwidth]{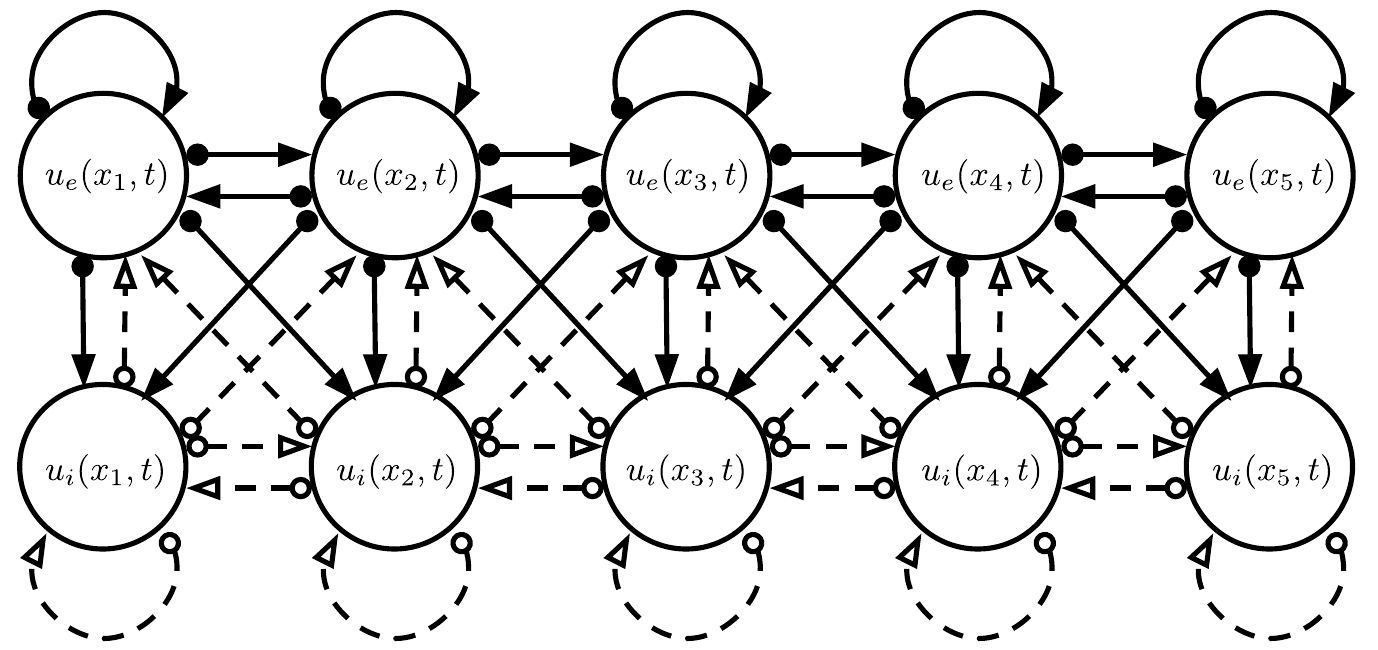}
    \caption{%
        \textbf{Diagram of basic two population neural field model} (adapted from
        Wilson and Cowan  \cite{wilsonMathematicalTheoryFunctional1973}). This illustration 
        gives a magnified
        representation of two interconnected neural continuums across one spatial dimension (with
        only nearest connections shown for brevity). Here we consider an `excitatory' population
        and an `inhibitory' population, where excitatory connections are represented by solid
        arrows whereas inhibitory connections are represented by dashed arrows. Neural field models
        in practice are indexed by a continuous spatial variable, as opposed to the discrete
        objects represented here. Although long-range connections are not shown for brevity, the
        strength of connections between neural populations tend to decrease exponentially with
        distance.
    }
    \label{F:basic-two-population-nfm}
\end{figure}

\section{A potted history of neural field models} 
\label{S:first-wave-nfms}

In this section, we will review key aspects of early neural field models. We first define the
overall principles and elements of a neural field model, using the Amari field equations as a
scaffold for addressing the main assumptions necessary for the construction of all neural
population models. We then provide a brief overview of the history of this class of models from the
1950s to the 1970s before discussing the canonical neural field model by Wilson and Cowan. We then
compare this with concurrent formulations of neural field theory by Amari and Nunez and discuss the
differences and similarities in mathematical structure. This section sets the scene for a
discussion of more recent advances in neural field modelling considered in Section
\ref{S:second-wave-nfms}.

Before we begin, it is important to note that as neural field modelling approaches have evolved, a
variety of names have appeared throughout the literature  including `neural mass models', `mean
field models', `neural population models'. Although these terms are all closely related,
they are conceptually different and have been used interchangeably and ambiguously in the
literature. The origin of `neural mass models' can be traced back to the work of Walter 
Freeman
and the concept of \textit{mass action}: the observed population-level behaviour of a mass of
neurons (Freeman \cite{freemanMassActionNervous1975}). In his seminal work Freeman 
defined a
second-order differential equation to describe the evolution of the population-level response to an
external stimulus within the olfactory bulb. Over time, `neural mass models' came to refer
to the subset of neural field models constrained to a single point in space that have no spatial
dependence.  will continue this discussion in Section \ref{S:neural-mass-models}. Since 
neural field models describe the average behaviour of neural populations, they're
also commonly referred to as `mean field models'. However it must be highlighted that
this is not the same as utilising the `mean-field approximation' from statistical mechanics (see
for example Sompolinsky et al.~\cite{sompolinskyChaosRandomNeural1988}), where a 
network of
interacting single neurons is replaced by a Gaussian process describing the average behaviour of a
single unit. For more details, we refer the interested reader to Sompolinsky et 
al.~\cite{sompolinskyChaosRandomNeural1988}, Vollhardt 
\cite{vollhardtDynamicalMeanfieldTheory2012}, Mart\'inez-Cancino and Sotero Diaz 
\cite{martinez-cancinoDynamicalMeanField2011}, and the many references therein. The
connection between this class of models and neural field models will be examined in Section
\ref{S:random-neural-networks}. In this Section our focus is on exploring the origins and
principles of the more general class of neural field models.

\subsection{Principles of neural field models} 
\label{S:nfm-priciples}

Whilst the origin of the modern formulations of neural field models is found in the seminal works
of Wilson and Cowan  
\cite{wilsonExcitatoryInhibitoryInteractions1972,wilsonMathematicalTheoryFunctional1973}, 
the
neural field model proposed by Amari  \cite{amariDynamicsPatternFormation1977} provides a 
clearer
framework for our definitions and the essential assumptions common to all neural field models by
construction.

Within the neural field framework, neural interactions occur in a two-dimensional ``cortical
sheet''. This can be thought of as unfolding the sulci and gyri of the cortex into a flat sheet
that is, on average, about 2.6 m$^{2}$ (Haberly and Shepherd
\cite{haberlySynapticOrganizationBrain1998}). 
To mathematically do this, we must consider two limits, the first countable and the 
second uncountable. First we take the thermodynamic limit, where the number of neurons 
approaches infinity whilst keeping the neuron density constant). 
Then we apply the continuum limit, where the neurons are treated as true mathematical points and the lattice spacing between the neurons shrinks to zero resulting in a continuum across the sheet.
\begin{framed}
  \textbf{Mathematical assumption:}
  In neural field models, the cortex is treated spatiotemporally as a continuous two-dimensional
    sheet, in the sense that radial redundancy exists in the cortical columns and that the cortex
    is functionally two-dimensional. This is a significant approximation as neurons are actually
    discrete multi-compartmented entities embedded in a densely connected three-dimensional
    laminated volume. This assumption is made so that neural tissue can be treated as an averaged
    bulk entity (Freeman \cite{freemanMassActionNervous1975}) with approximated spatial and
    temporal distributions.
\end{framed}
Neurons are grouped together in populations based on shared characteristics such as the polarity of
signals, timescales of operation, the depth of cortex. The state variables are aggregate quantities
of these populations, such as \textit{mean membrane potential} $u_a(x,t)$ where $a$ is a population
index, $x$ is a point in the neural field, and $t$ is a time. Overall, the mean membrane
potential is dictated by the \textit{mean firing rate} of afferent populations. The magnitude of an
incoming signal at a neuron depends on which population it arrived from, how far away the afferent
neuron is, and how long ago the signal arrived. In turn, the mean membrane potential of a
population dictates the mean firing rate of the population, which in turn influences the mean
membrane potential of connected populations.

The strength of an incoming signal at a neuron decays soon after the signal is received, which can
be represented as a simple temporal integral convolution. We can therefore describe the evolution
of the mean membrane potential as
\begin{equation}
  u_{a}(x,t) =
    \int_{-\infty}^{t}
      \psi(t - T)\, I_{a}(x,T)
    \di T,
    \label{eq:amari-temp-conv}
\end{equation}
where the kernel $\psi$ represents the temporal decay on incoming signals, and $I_{a}(x,t)$
represents the incoming signals received by population $a$. Contemporary neural field models
typically represent these as temporal convolutions to allow for non-linear temporal operators such
as those found in conductance-based synapses (Peterson et 
al.~\cite{petersonHomotopicMappingCurrentbased2018}).

The input term can be represented as a delta spike train  corresponding to incoming action
potentials with a spatial integral transformation that includes the spatial weighting of signals
received by population $a$,
\begin{equation}
    I_{a}(x,t) =
        \sum_b
        \int_{\Omega}
            w_{a,b}(x,X) \sum_{m\in\Z} \delta(t - T^{(m)}_{b}(X))
        \di X + q_a(x,t),
    \label{eq:amari-spatial-kernels}
\end{equation}
where $\Omega$ denotes the spatial neural medium, the kernel $w_{a,b}(x,X)$ represents the spatial
weighting of signals from population $b$ at position $X$ to population $a$ at position $x$,
$T^{(m)}_{b}(x)$ is the time of the $m$'th action potential spike from population $b$ at the
position $x$, $\delta(t)$ is the Dirac delta distribution, and $q_a(x,t)$ is an external firing
rate. The value of $w_{a,b}(x,X)$ can be interpreted as the probability of connection
between neurons. In other words, the kernel represents the mean synaptic connectivity from
population $b$ at point $X$ to population $a$ at point $x$.
In the vast majority of neural field models, the spatial kernel is assumed to be homogeneous so
that $w_{a,b}(x,X) = w_{a,b}(x-X)$ and the corresponding integral transform is a spatial
convolution.
\begin{framed}
    \textbf{Anatomical assumption:}
    For every point in the neuronal medium, each neuron is assumed to be randomly connected,
    following some spatially homogeneous distribution. Collections of axons connect to nearby
    neurons, and can connect to neurons of the same population. This is a significant assumption,
    as a multitude of emergent dynamics of neural tissue arises from the non-trivial and non-local
    connectivity between constituents, which appears to follow fundamentally non-random trends.
\end{framed}

The delta spike train can be replaced with a phenomenologically derived \textit{wave-to-pulse
transfer function} (also known as a \textit{firing rate function}). Effectively this converts the
mean membrane potential into a mean firing rate (Freeman
\cite{freemanTutorialNeurobiologySingle1992}) if the synaptic response is on a slower timescale
than that of the interspike interval and the variance is small (Coombes et 
al.~\cite{coombesWavesBumpsNeuronal2003}, Bressloff and Coombes 
\cite{bressloffDynamicsStronglyCoupled2000}).
Axonal conduction velocities can be assumed to be infinitely fast at this point in the derivation,
however many contemporary models include a finite conduction velocity and a time-dependent
component in their definition of $w_{a,b}$. Hence, we can express incoming signals from population
$b$ to population $a$ as
\begin{equation}
    I_{a}(x,t) =
        \sum_b
        \int_{\Omega}
            w_{a,b}(x,X)\, f_b[u_b(X,t - \abs{x-X}/v)]
        \di X + q_a(x,t),
    \label{eq:amari-input-delta}
\end{equation}
where $v$ is the axonal conduction velocity, and $f_b$ is the wave-to-pulse transfer function.
Traditionally, this is a static nonlinearity such as a sigmoidal function that is continuously
differentiable, monotonically increasing, and saturating to a constant for small and large input
values. This represents populations of neurons saturating to a maximal firing rate as the mean
membrane potential increases. The most common form of this transfer function is the single logistic
sigmoid, which is given as
\begin{equation}
    f_a(u_a) = \frac{F_a}{1+\exp[-\lambda_a \, (u_a - \theta_a)]},
    \label{eq:wc-sigmoid}
\end{equation}
where $u_a$ is the membrane potential of population $a$, $F_a$ is the maximum firing rate of
population $a$, and $\lambda_a/\theta_a$ are the shape/location parameters.
\begin{framed}
    \textbf{Mathematical assumption:} We interpret the sigmoidal function, or in this case the
    logistic equation, as a close approximation of the cumulative function of a normal distribution
    (Marreiros et al.~\cite{marreirosPopulationDynamicsVariance2008}), instead of the usual 
    error
    function. In this way it has an associated mean and standard deviation of thresholds, that are
    averaged over a neuronal population. However, this assumes that the thresholding function is
    that of a McCulloch-Pitts neuron 
    \cite{mccullochLogicalCalculusIdeas1943},
    which is artificial and over-simplified compared to an actual spiking neuron model.
    Essentially, the sigmoidal function is a nonlinear function that is continuously
    differentiable, increasing, and smoothly bounded at both ends. The error function and the
    logistic equation are examples of sigmoid functions and are found in many natural systems
    (DeLean et al.~\cite{deleanSimultaneousAnalysisFamilies1978}). For further information on 
    its
    derivation, see the paper by Freeman  \cite{freemanNonlinearGainMediating1979} which 
    was
    one of the first papers to try and derive the sigmoid function from experimental data, although
    it did appear in relation to neural models earlier than this (Freeman
    \cite{freemanLinearAnalysisDynamics1972}, Wilson and Cowan
    \cite{wilsonExcitatoryInhibitoryInteractions1972}, Lopes da Silva et 
    al.~\cite{lopesdasilvaModelBrainRhythmic1974}, Nunez 
    \cite{nunezBrainWaveEquation1974}).
\end{framed}

In Figure \ref{F:kernel-gallery} we present illustrations of examples of kernels $\psi(t)$
and $w_{a,b}(x)$, as well as a plot of a wave-to-pulse transfer function. In Figure
\ref{F:kernel-gallery}(b), three different temporal kernels are shown. An exponential kernel
assumes an instantaneous rise time, for example, where there is no synaptic decay and a decay
defined by the membrane time-constant. An alpha function has a finite rise time and a synaptic
time-constant that is the same as the membrane time-constant. A double exponential kernel has
separate synaptic and membrane time-constants so that the rise and decay can be independent. In
Figure \ref{F:kernel-gallery}(c), two different spatial kernels are shown where the Bessel function
has a singularity at zero, whereas the Laplacian is finite at zero. Given that the region
containing the singularity is immeasurable (zero distance) and largely irrelevant (only the tail of
the distribution can be measured), the differences between these two functions are effectively
negligible.

Combining the above equations \eqref{eq:amari-temp-conv} and \eqref{eq:amari-input-delta} and
assuming an infinitely fast conduction velocity ($v \to \infty$) gives us the integral form of a
basic neural field model:
\begin{equation}
  u_a(x,t) =
    \int_{-\infty}^t
      \psi(t-T)
      \left\lbrace
        \sum_b
        \int_\Omega
          w_{a,b}(x-X) \, f_b[u_b(X,T)]
        \di X + q_a(x,T)
      \right\rbrace
    \di T . \label{eq:amari-field}
\end{equation}
This model represents the classical definition of the Amari equations.
\begin{framed}
    \textbf{Physiological assumption:} The spatiotemporal delay resulting from the finite
    conduction velocity in Equation \eqref{eq:amari-input-delta} can be omitted if we assume the
    conduction velocity is infinitely fast. This can be a suitable assumption for modelling smaller
    sections of neural tissue. For brevity, we will make this assumption here so as to eliminate
    the $\abs{x-X}/v$ term in the temporal argument of $u(x,t)$ in the integrand. However, not all
    neural field models make this assumption.
\end{framed}

From this foundation, we are in a better position to understand how neural field models evolved in
the following decades, all of which are derived from these same fundamental principles. Referring
to this structure also gives a clearer view of concurrent modelling approaches of this era, helping
us to understand how this type of modelling was initially conceived between works of key research
groups. Before we embark on the discussion of modern neural field modelling, we will revisit some
of the early attempts to model neural populations and further highlight the structures and
assumptions made. Many of these assumptions continue to be utilised in contemporary neural field
models today.

\subsection{Origins of neural field theory}  
\label{S:wc-derivation}

The method of studying the average behaviour of a neuronal population  can be traced back to Beurle
 \cite{beurlePropertiesMassCells1956}. He explored the emergent properties of a 
 continuum
approximation to a mass of densely connected, uniformly distributed excitable cells, observing
plane waves, spherical waves and vortex effects. Subsequently, in the early 1960s Griffith expanded
these ideas, proposing a ``nonlinear field theory'' of neural networks 
\cite{griffithFieldTheoryNeural1963}. Here the concept of a connectivity kernel that related 
the
level of excitation to distance was introduced and the stability of emergent waves was subsequently
studied  \cite{griffithFieldTheoryNeural1965}. Alongside these developments,  it
was also shown that inhibitory neural action was critical  for wave stability 
\cite{griffithStabilityBrainLikeStructures1963}.

What we might now consider to be the beginnings of modern neural field theory, was established
through the canonical models of Freeman  \cite{freemanWavesPulsesTheory1972}, Wilson 
\cite{wilsonExcitatoryInhibitoryInteractions1972}, Wilson and Cowan 
\cite{wilsonMathematicalTheoryFunctional1973}, Nunez  \cite{nunezBrainWaveEquation1974},
Lopes da Silva et al.~\cite{lopesdasilvaModelBrainRhythmic1974} and Amari 
\cite{amariHomogeneousNetsNeuronlike1975,  amariDynamicsPatternFormation1977}. 
Previous
approaches only considered a single neuron population, whereas Wilson and Cowan derived the first
canonical neural field model in terms of spatially extended masses of excitatory and inhibitory
neurons.

Wilson and Cowan start with a description of  mean neuronal activities across excitatory and
inhibitory populations, and propose expressions for aggregated neural quantities based on concepts
from neuronal dynamics. This approach effectively extends the work of Beurle from a single
population to multiple populations. A sketch of the physiology considered in this model is given in
Figure \ref{F:basic-two-population-nfm}. The state variables used by Wilson and Cowan are
\textit{mean neuronal activity} per unit time, where $\hat{A}_e(x,t)$ refers to the activity of the
excitatory population and $\hat{A}_i(x,t)$ refers to the activity of the inhibitory population.
That is, activity is considered as the proportion of excitatory or inhibitory neurons at $(x,t)$
that are firing action potentials. These variables are related to the average firing rate: when
this is one, the neurons are all firing action potentials at their maximum rate whereas an average
firing rate of zero represents typical background activity. Small negative values are permitted and
represent a dampening of background activity. Wilson and Cowan state their model in terms of update
equations, given by
\begin{equation}
    \hat{A}_e(x,t+\tau) =
        R_e(x,t) \; S_e[N_e(x,t)],
    \qquad
    \hat{A}_i(x,t+\tau) =
        R_i(x,t) \; S_i[N_i(x,t)],
    \label{eq:wc-update-eqs}
\end{equation}
where $\tau$ represents the synaptic operating delay (time taken for a neuron to fire an action
potential after its membrane potential surpasses threshold), $R_a(x,t)$ is the fraction of
population $a$ that is primed to fire an action potential (outside the refractory phase),
and $S_a(u) = f_a(u) - f_a(0)$ (using $f_a$ from Equation \eqref{eq:wc-sigmoid} with $F_a =
1$) represents a centred wave-to-pulse transfer function. The expression $S_a(N_a(x,t))$ 
relates
mean population input, $N_a(x,t)$, to the fraction of population surpassing threshold excitation
and firing action potentials. Wilson and Cowan originally use a centred activation function
that passes through the origin and approaches a negative asymptote as the argument decreases below
zero, but this is likely due to the choice in definition of the state variables of the system.
\begin{framed}
    \textbf{Mathematical assumption:} This formulation assumes that the primed fraction of the
    population, $R_a(x,t)$, and the firing fraction of the population $S_a(N_a(x,t))$ are
    independent. However, in general this is not true (Cowan et 
    al.~\cite{cowanWilsonCowanEquations2016}, Smith and Davidson 
    \cite{smithMaintainedActivityNeural1962}).
    As highlighted by Wilson and Cowan, for finite populations of neurons there will be a
    correlation between the recency of a neuron firing and its likelihood of firing in the future.
    However, Wilson and Cowan argue that if the number of neurons in these population is large
    enough, this correlation will become small, and so independence is considered a reasonable
    approximation.
\end{framed}
The synaptic operating delay $\tau$ refers to the period between a neuron's membrane potential
surpassing the threshold potential and the neuron becoming active and initiating an action
potential. This quantity is not well described in 
\cite{wilsonExcitatoryInhibitoryInteractions1972} and 
\cite{wilsonMathematicalTheoryFunctional1973}, but it is effectively a computational
timestep for the neural system assumed to have an order of 0.5 ms. The model posed by Beurle
 \cite{beurlePropertiesMassCells1956} utilised the same terminology, and the work by Wilson
and Cowan sought to improve upon these early findings. It should be highlighted here that neural
population models at this time were largely conceptual and the data on neuronal activity and
synaptic interactions was not well-developed. Significant assumptions about the values of these
timescales had to be made in order to begin the study of these particular dynamical systems, which
could then be evolved and improved over time.

The proportion of sensitive neurons in the population is complementary to the proportion of neurons that are in the refractory period,
\begin{equation}
    R_e(x,t) =
        1 -
        \int_{t-r_e}^t
            \hat{A}_e(x,T)
        \di T
    \qquad
    R_i(x,t) =
        1 -
        \int_{t-r_i}^t
            \hat{A}_i(x,T)
        \di T
\end{equation}
where $r_e/r_i$ are the refractory periods of excitatory/inhibitory neurons. Each spiking input to
the population has a magnitude that depends on the connectivity or synaptic strength to the
upstream population. Additionally, it is assumed that the populations of neurons sum their net mean
inputs and the effect of stimulation decays over time. This can be thought of as the influence of
spiking inputs to individual neurons on neuronal activity diminishing as time from the input signal
increases. It can be represented as a temporal integral convolution with a kernel denoted by
$\alpha(t)$, which is generally taken as a simple exponential.
\begin{framed}
    \textbf{Physiological assumption} It is assumed in the Wilson and Cowan derivations that
    $\alpha(t) = \hat{\alpha} \exp(-t/\mu) \Theta(t)$ where $\hat{\alpha}$ is the maximum
    post-synaptic potential amplitude, $\mu$ is a time constant of decay and $\Theta(t)$ is the
    Heaviside step function. It is assumed that the effect of stimulation decays with the same rate
    across excitatory and inhibitory neurons, so that $\mu$ is a population-independent time
    constant. Future models incorporate this detail into their choice of spatial kernels and
    setting $\hat{\alpha} = 1$. This alternate method tends to be more common since the magnitude
    (as well as polarity of signals) tend to be dependent on the afferent population.
\end{framed}
In addition to the temporal decay of afferent signals, the initial magnitude of those signals will
depend on where they originated from, as we conceptually described in Equation
\eqref{eq:amari-spatial-kernels}. Using homogeneous spatial kernels, Wilson and Cowan express the
mean population inputs by
\begin{equation}
\begin{split}
    N_e(x,t) &=
        \int_{-\infty}^t
            \alpha(t-T)
            \left\lbrace
                [\beta_{e,e} \otimes_x \hat{A}_e] (x,T)
                - [\beta_{e,i} \otimes_x \hat{A}_i] (x,T)
                + P_e(x,T)
            \right\rbrace
        \di T, \\
    N_i(x,t) &=
        \int_{-\infty}^t
            \alpha(t-T)
            \left\lbrace
                [\beta_{i,e} \otimes_x \hat{A}_e](x,T)
                - [\beta_{i,i} \otimes_x \hat{A}_i](x,T)
                + P_i(x,T)
            \right\rbrace
        \di T,
\end{split}
\end{equation}
where $[\beta_{a,b} \otimes_x f](x) = \int_{\Omega} \beta_{a,b}(x-X) f(X) \di X$ represents a
spatial convolution, and $P_e$/ $P_i$ represent external inputs to the excitatory/inhibitory
populations.
\begin{framed}
    \textbf{Physiological assumption:} Connectivity is assumed to be of the form $\beta_{a,b}(x) =
    \bar{\beta}_{a,b} \exp(-\abs{x}/\sigma_{a,b})$ where $\sigma_{a,b}$ denotes the space constant
    and $\bar{\beta}_{a,b}$ denotes the mean synaptic weight (Uttley and Matthews
    \cite{uttleyProbabilityNeuralConnexions1955}, Sholl
    \cite{shollMeasurableParametersCerebral1956}).
\end{framed}

A simpler expression of this system is possible by interchanging the order of temporal and spatial
integrals and using the time coarse grained variables
\begin{equation}
    A_a(x,t) =
        \frac{1}{\mu}
        \int_{-\infty}^t
            \exp\left( -\frac{t-T}{\mu} \right) \,
            \hat{A}_a(x,T)
        \di T.
\end{equation}
These new variables provide a first-order simplification of  the integro-differential equations.
\begin{framed}
    \textbf{Physiological assumption:} It is assumed here that the membrane time constant is an
    order of magnitude more significant than the synaptic dynamics. This effectively makes the
    synapses instantaneous interfaces, which is not the case in reality. Additionally, it is also
    assumed in the final Wilson and Cowan model that signal propagation along axons is also
    instantaneous. This may be a suitable simplification on smaller spatial regions,  but may be
    inappropriate for modelling large sections of cortex such as a lobe of the brain.
\end{framed}

This change of variables is not without consequence. Wilson and Cowan highlight that such a
transformation inevitably removes high frequency variations, but they argue that such variations
are not relevant to the implications results. Since the kernel of this integral transformation is
$\frac{1}{\mu}\exp(-t/\mu) \Theta(t)$, the Green's function of the differential operator $\left(1 +
\mu \ppx{}{t}\right)$. This gives us the relation
\begin{equation}
    \hat{A}_a(x,t) = \mu \ppx{}{t} A_a(x,t) + A_a(x,t). \label{eq:tcgid}
\end{equation}
Using the fact that the refractory period of cortical neurons $r_a$ is an order of magnitude lower
than the membrane time constant $\mu$ (Oshima and Jasper 
\cite{oshimaBasicMechanismsEpilepsies1969}), we
have $A_e(x,t) \approx A_e(x,t-r)$. We can then integrate both sides of the above relation and
apply this approximation to obtain
\begin{equation}
\begin{split}
    \int_{t-r_a}^{t} \hat{A}_a(x,T) \di T &=
        \mu\left[
          A_a(x,t) - A_a(x,t-r)
        \right]
        + \int_{t-r_a}^{t} A_a(x,T) \di T, \\
    &\approx
        r_a A_a(x,t),
        \quad\implies\quad
        R_a(x,t) \approx [1 - r_a A_a(x,t)].
\end{split}
\end{equation}
The derivation by Wilson and Cowan  \cite{wilsonMathematicalTheoryFunctional1973} initially 
asserts
that axonal velocity is finite, but they relax that assumption towards the end 
of their
derivation. In this derivation we will assume that the conduction velocity is infinitely fast so
that we may keep our guided derivation simpler. Expanding the left hand side of the Wilson and
Cowan Equations (\ref{eq:wc-update-eqs}) in a Taylor series gives:
\begin{equation}
    \hat{A}_a(x,t+\tau) \approx \hat{A}_a(x,t) + \tau \ppx{\hat{A}_a}{t}(x,t),
\end{equation}
where we have omitted the terms of higher order $\tau^2$, since $\tau \ll 1$. Taking the
derivative of $\hat{A}_a$ with respect to $t$ in Equation \eqref{eq:tcgid} and substituting the
result into this expression gives
\begin{equation}
    \hat{A}_a(x,t+\tau) \approx A_a(x,t) + (\mu + \tau) \ppx{A_a}{t}(x,t)
        + \tau \mu \ppx[2]{A_a}{t}(x,t)
        = \left( 1 + \tau \ppx{}{t} \right)\, \left( 1 + \mu \ppx{}{t} \right) A_a(x,t).
\end{equation}
Here it is assumed that the synaptic operating delay is at least an order of magnitude
smaller than the membrane decay constant, enabling a further approximation:
\begin{equation}
    \hat{A}_a(x,t+\tau) \approx A_a(x,t) + \mu \ppx{A_a}{t}(x,t) .
\end{equation}
We can then substitute this result back into Equation \eqref{eq:wc-update-eqs} along with 
the above
results to obtain the system
\begin{equation}
    \begin{split}
      \mu \ppx{}{t}&A_e(x,t) =
        -A_e(x,t) \\
        &+ [1 - r_e A_e(x,t)] \,
        S_e\left\lbrace
            [w_{e,e} \otimes_x A_e](x,t)
            - [w_{e,i} \otimes_x A_i](x,t)
            + Q_e(x,t)
        \right\rbrace,
        \label{eq:wc-final-1}
    \end{split}
\end{equation}
\begin{equation}
    \begin{split}
      \mu \ppx{}{t}&A_i(x,t) =
        -A_i(x,t) \\
        &+ [1 - r_i A_i(x,t)] \;
        S_i\left\lbrace
            [w_{i,e} \otimes_x A_e](x,t)
            - [w_{i,i} \otimes_x A_i](x,t)
            + Q_i(x,t)
        \right\rbrace,
        \label{eq:wc-final-2}
    \end{split}
\end{equation}
where $w_{a,b}(x) = \hat{\alpha} \mu \beta_{a,b}(x)$ is a scaled connectivity kernel, and $Q_a(x,t)
= \int_{-\infty}^{t}\alpha(t-T) \,\hat{Q}_a(x, T) \di T$ represents a filtered external input. It
should be noted that most neural field models disregard the refractory period at this scale of
description, and thus set $r = 0$ implicitly or explicitly as an additional approximation.
Disregarding the refractory period, we note this system can also be represented in the more general
form:
\begin{equation}
  \mu \ppx{}{t} A_{a}(x,t) =
    - A_a(x,t) +
    S_a
    \left\lbrace
        \sum_{b} w_{a,b}(x) \otimes_x A_b(x,t) + Q_a(x,t)
    \right\rbrace
    ,\label{eq:wc-final}
\end{equation}
where $a,b$ are generic population indices, $\mu$ is the membrane time constant of population $a$,
and $w_{a,b}(x)$ is the connectivity kernel summarising the signed magnitude of post-synaptic
potentials.

The Wilson and Cowan model represented a major step forward for the field of macroscale neural
modelling, significantly extending the previous models of Beurle and Griffith. Critically, the
model proposed by Wilson and Cowan recognised and incorporated the influence on emergent neural
dynamics of inhibitory neurons. In contrast,  most early neural field models 
only considered  a
single excitatory population. This permanently changed the course of development of models in this
field,  and directly led to the modern formulations of neural field theory. The dynamical system
analyses  of resulting fixed points and limit cycles  that led to these conclusions and consequent
developments are summarised by Zetterberg et 
al.~\cite{zetterbergPerformanceModelLocal1978},
Ermentrout and Terman  \cite{ermentroutMathematicalFoundationsNeuroscience2010}, and 
Neves and Monteiro 
\cite{nevesLinearAnalysisCoupled2016}. A brief summary of the applications and extensions of the
Wilson and Cowan model can be found in the reviews by Kilpatrick 
\cite{kilpatrickWilsonCowanModel2013} and Chow and Karimipanah  
\cite{chowWilsonCowanEquations2020}.

Following the work of Wilson and Cowan, the neural field models were further developed by other
authors. Of note is a publication by Amari who investigated spatiotemporal patterns and information
processing in homogeneous neural networks  
\cite{amariHomogeneousNetsNeuronlike1975}.
Amari further developed these approaches in a subsequent publication 
\cite{amariDynamicsPatternFormation1977}, that considered an extension from discrete 
neural
networks to a neural field continuum. In contrast to Wilson and Cowan who segregated populations
according to whether neurons were excitatory or inhibitory, the neural field model put forth by
Amari segregated neurons into populations according to cortical depth. Here, each layer contained
both excitatory and inhibitory neurons, whose signals were encoded within the spatial kernels
$w_{a,b}$ representing close inhibitory connections and long-range excitatory connections 
(Amari \cite{amariHomogeneousNetsNeuronlike1975}).
\begin{framed}
    \textbf{Physiological assumption:} Amari assumes that each layer of cortical neurons  have
    non-identical time constants. As with the Wilson and Cowan model, it is still assumed that
    incoming postsynaptic potentials  have infinitely fast rise,  followed by an exponential decay.
    This is in fact a generalisation of the work of Wilson and Cowan, which only allowed one time
    constant for all populations of neurons.
\end{framed}
The derivation of the Amari field equations closely follows the principles we discussed in Section
\ref{S:nfm-priciples}, but instead using population-dependent temporal kernels $\psi_a$:
\begin{equation}
  u_a(x,t) =
    \int_{-\infty}^{t}
        \psi_{a}(t - T)
        \left\lbrace
            \sum_b
            \int_{\Omega}
                w_{a,b}(x - X) \,  f_b[u_b(X, T)]
            \di X
        + q_a(x,T)
        \right\rbrace
    \di T,
  \label{eq:amari-basic}
\end{equation}
where $u_a(x,t)$ is the mean membrane potential of neurons in population $a$ at position $x$ and
time $t$, $\psi_{a}(t)$ is the temporal kernel of population $a$, $w_{a,b}(x)$ is the spatially
homogeneous spatial kernel  that assigns signed weights to incoming signals  from population $b$
according to distance, and $q_a(x,t)$ is the external input to population $a$. Note if we use a
linear description of temporal decay of signals,  we may utilise Green's functions to invert the
convolution and  express the Amari model in differential form. As with many other models in this
period,  it is assumed that spikes are instantaneous  and exponentially decay. Hence, Amari
utilises the temporal kernel given by $\psi_{a}(t) = \frac{1}{\mu_a} \exp(-t/\mu_a) \, \Theta(t)$,
where $\mu_a$ is the membrane time constant of population $a$ and $\Theta(t)$ is the Heaviside step
function, which allows us to express the Amari equations in differential form,
\begin{equation}
  \mu_a\ppx{}{t} u_a(x,t) =
    -u_a(x,t) +
    \sum_b
    \int_{\Omega}
      w_{a,b}(x-X) f_b[u_b(X,t)]
    \di X
    + q_a(x,t).
  \label{eq:amari-eq}
\end{equation}

\begin{figure}[htbp]
    \centering
    \includegraphics[width=0.94\textwidth]{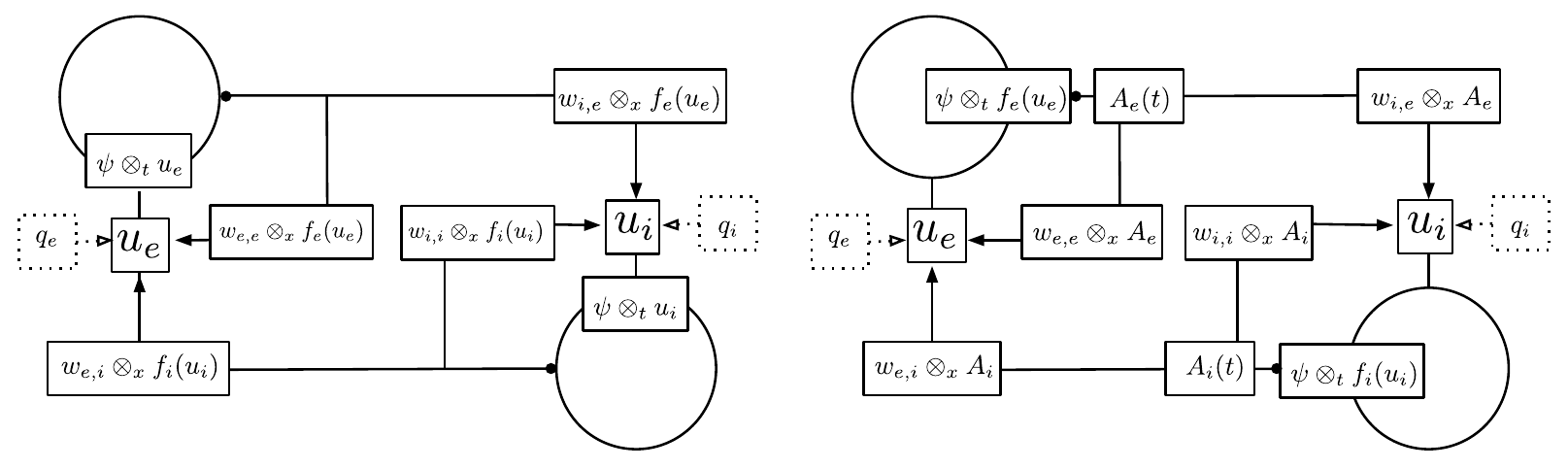}
    \caption{%
        \textbf{Comparison of first order neural field by Amari (left) and Wilson and Cowan
        (right).} The relation between the Wilson-Cowan model and the Amari model is illustrated
        with annotations of the corresponding state variables $A_e, A_i$ and $u_e, u_i$. Although
        the mathematical structure of these models appears different, they describe similar
        dynamical systems from different perspectives. The differential operator in these models
        can be represented as a temporal convolution with a kernel $\psi$ using standard methods of
        integral transforms. Temporal convolution is represented here with $\otimes_t$ and spatial
        convolution is represented with $\otimes_x$. Crucially, the Amari formulation considers a
        sum of sigmoids of inputs whereas the Wilson and Cowan formulation considers a sigmoid of a
        sum of inputs. These operations do not commute with the temporal kernel $\psi$ unless the
        kernel is population-independent, hence these formulations are only equivalent when the
        temporal kernel is population independent.
    }
  \label{F:wc-model}
\end{figure}

This form of the Amari equations appears to resemble the field equations derived by Wilson and
Cowan, which we summarised in Equation \eqref{eq:wc-final} with some key mathematical differences.
The major difference appears to be the ordering of the sigmoidal wave-to-pulse transform and the
sum over afferent signals from different populations. Specifically, the Wilson and Cowan model
takes the sigmoid of a sum of inputs whereas the Amari model takes the sum of sigmoids of inputs.
Another minor difference can be identified in the definition of the wave-to-pulse
transformations in either formulation. The Wilson and Cowan model uses a centred sigmoid $S_a(u)$,
whereas the Amari equation uses a non-centred sigmoid $f_a(u)$, but this difference can be
rectified by a translation mapping given by $S_a(u) = f_a(u) - f_a(0)$. Despite this difference 
in
formulation, both equations give rise to similar dynamics. This is unsurprising since they describe
the same physical system from two different perspectives. The Wilson and Cowan model state
variables start with a description of ``proportion of active neurons at a point'', which can be
interpreted as analogous to a mean firing rate of a population. This can be easily visualised by a
comparison of the diagrammatic representation of these systems in Figure \ref{F:wc-model}. From a
purely mathematical perspective, the Wilson and Cowan model is equivalent to an Amari model with a
single membrane time constant $\mu_a = \mu$ subject to the following transformations:
\begin{equation}
  u_a(x,t) =
    \sum_b A_b(x,t) \otimes w_{a,b}(x) + Q_a(x,t)
    \label{eq:amari-wc}
    ,
\end{equation}
\begin{equation}
  \mu \ppx{}{t} A_a(x,t) + A_a(x,t) = f_a[u_a(x,t)]
    \label{eq:wc-amari}
    .
\end{equation}
This substitution and calculation were performed by Miller and Fumarola 
\cite{millerMathematicalEquivalenceTwo2011}, but is extended here for spatially-dependent neural
fields. By taking the first time derivative of Equation \eqref{eq:amari-wc} we have
\begin{equation}
\begin{split}
  \mu \ppx{}{t} \left( u_a(x,t) - Q_a(x,t) \right) &=
    \sum_b
    \left(
      \mu \ppx{}{t} A_b(x,t)
    \right)
    \otimes w_{a,b}(x)
    , \nonumber\\
  &=
    - \sum_b A_b(x,t) \otimes w_{a,b}(x)
    + \sum_b S_b[u_b(x,t)]
      \otimes w_{a,b}(x)
    , \qquad \textrm{\eqref{eq:wc-final}}
    ,\nonumber\\
  &=
    - (u_a(x,t) - Q_a(x,t))
    + \sum_b S_b[u_b(x,t)]
      \otimes w_{a,b}(x)
    , \qquad \textrm{\eqref{eq:amari-wc}}
    ,\nonumber\\
  \iff\mu \ppx{}{t} u_a(x,t) &=
    - u_a(x,t)
    + \sum_b f_b[u_b(x,t)] \otimes w_{a,b}(x)
    + q_a(x,t)
    ,
\end{split}
\end{equation}
where $q_a(x,t) = \left( \mu \ppx{}{t} + 1 \right)Q_a(x,t) + \sum_b f_b(0) \int_\Omega w_{a,b}(x-X)
\di X$. This transformation only works for an Amari model with one membrane time constant instead of
multiple population-specific time constants (or if there is only one neural population).

While these models are closely related, the operations follow a different order when
applied to their respective state variables. In the Wilson and Cowan model the temporal filter
occurs after the sigmoid and before the sum, whereas in the Amari model the temporal filter occurs
after the sum and before the sigmoid. We highlight this in Figure \ref{F:wc-model}. Equivalence of
these models occurs only if the temporal kernel is population-independent. Under this assumption
the operations commute and so are interchangeable. This is due to the nature of the approximations
taken by Wilson and Cowan and the original (without time coarse graining) state variables used.
However, one can still apply this transformation to obtain an extended Wilson and Cowan model with
multiple timescales,  but the resulting ``activity'' state variables will not be identifiable with
the original state variables proposed by Wilson and Cowan. Fundamentally, the separation in time
scales across populations is a different model that expands beyond the scope of the original Wilson
and Cowan equations.

Mathematically, the Wilson and Cowan formulation is arguably more attractive, since linearising a
dynamical system over a single sigmoid is less complex than linearising over a sum of sigmoids.
However, this formulation is delicate and requires a large amount of assumptions and conceptual
leaps to derive and simplify from first principles as we have seen throughout this Section.
Nonetheless, the ideas behind this transformation can be used to transform any neural field model
into the single sigmoid form, and we will formalise this within a general mathematical framework in
Section \ref{S:modern-nfms-derivation}.

These early neural field models were an important milestone  for the development of mathematical
descriptions of large-scale neural activity. Analysing dynamical systems became a common goal
amongst a variety of research groups as this methodology became more established throughout the
1970's. However, there were significant limitations of the utility of these models, and some of
these limitations have been inherited by modern approaches of neural field theory. In particular,
the quantities modelled and parameters used in these constructions were difficult to estimate and
fit from methods of data acquisition in experimental neuroscience at the time. This is still a
problem that exists today due to limitations in experimental neuroscience, but there are ongoing
efforts to improve how we quantify connectivity in neural tissue that can be utilised in new
iterations of neural field theory. Synaptic dynamics in particular have  more complexity and
variability than this, the frameworks put forth by Wilson, Cowan, and Amari can be regarded as
\textit{first order neural fields}, as they neglect the variability in the synaptic dynamics and
focus only on the overall mean membrane potentials. However, the models proposed by 
Nunez 
\cite{nunezBrainWaveEquation1974} (as well as Lopes da Silva et 
al.~\cite{lopesdasilvaModelBrainRhythmic1974} and Freeman  
\cite{freemanMassActionNervous1975})
address this shortcoming in synaptic interactions. Before we discuss modern neural field models, we
will review the development of this additional principle through a general overview of the
framework by Nunez.

\subsection{Synaptic population dynamics} 
\label{S:nunez-model}

Around the same time as the work of Wilson and Cowan  
\cite{wilsonExcitatoryInhibitoryInteractions1972}
and Freeman  \cite{freemanWavesPulsesTheory1972}, another distinct approach to modelling
neural population dynamics emerged in the work of Nunez  
\cite{nunezBrainWaveEquation1974}.
The mathematical framework presented by Nunez uses different state variables and finds another
balance between biological plausibility and mathematical tractability. The equations appear to be
distinct from the Wilson and Cowan equations, as well as the Amari field equations, but as
highlighted by Jirsa and Haken  \cite{jirsaDerivationMacroscopicField1997}, the Nunez brain 
wave
equation is grounded in the same mathematical principles. Instead of connecting the Nunez model to
the frameworks of Amari, Wilson, and Cowan, we will instead focus on what made this model distinct
in its formulation. Although based in the same principles, this model was constructed on the basis
of synaptic interactions, similarly to the models of Freeman 
\cite{freemanModelMutualExcitation1974} and Lopes da Silva et 
al.~\cite{lopesdasilvaModelBrainRhythmic1974}, all of which foreshadow key developments 
in modern
neural field theory.

Instead of modelling multiple populations, the Nunez brain wave equation effectively models one
population with both excitatory and inhibitory synapses. This model aims to maintain a close
relationship with established biophysics and begins with a decomposition of electroencephalography
(EEG) signals in terms of contributions from excitatory/inhibitory post-synaptic potentials and
action potentials. We will return to our discussion of EEG in Section \ref{S:applications}, as
there are significant assumptions necessary for this connection to EEG data, but here we will
simply consider the conceptual electrical potential of the neural field. Consider a one-dimensional
neural field $\Omega$ consisting of a single population of cortical neurons. The instantaneous
gross potential of the field, $\Phi$, is related to the decomposition
\begin{equation}
    \Phi(x,t) \sim \phi\, \rho\, A(x,t) + \phi_e\, \rho_e(x,t) - \phi_i\, \rho_i(x,t),
\end{equation}
where $\rho$ is the density of cortical cells, $\phi$ is the local surface potential of a single
active cortical cell, $A$ is the fraction of active cortical cells (similar to the state variables
in the Wilson and Cowan equations), $\phi_e/\phi_i$ is the surface potential of a single
excitatory/inhibitory post-synaptic potential, and $\rho_e(x,t)/\rho_i(x,t)$ is the density of the
excitatory/inhibitory post-synaptic potentials. The underling relationship between $A$ and
$\rho_e/\rho_i$ is the principle driver of cortical dynamics.

The excitatory/inhibitory post-synaptic potentials $\rho_e/\rho_i$ depend on the aggregated
fraction of active cortical cells $A$, weighted by the mean connectivity. The connectivity in the
neural field is described by a spatial kernel within an integral transform, as we have seen for the
Amari equations and the Wilson and Cowan equations. The model by Nunez is explicitly intended to
model large-scale neural activity, so axonal conduction velocities $c$ are considered. The duration
of signal transmission between two points in neural tissue is highly variable, depending on the
complex geometry and topology of the relevant neural networks. It is impossible to include this
level of detail when modelling at this scale, but the integral kernel can be expanded to
incorporate the broad impact of variable conduction velocities. Let $w_a(c, x, X)$ represent the
mean connectivity from $X$ to synapse group $a\in\lbrace e, i \rbrace$ at the point $x$ than have
an average axonal velocity $c$. Then the mean excitatory/inhibitory post-synaptic potentials
$\rho_e/\rho_i$ can be represented in terms of the fraction of active cortical cells $A$ with
\begin{equation}
    \rho_a(x,t) =
        Q_a(x,t) +
        \int_0^\infty
            \int_\Omega
                w_a(c, x, X) \,
                A\left(
                    X, t - \frac{\abs{x - X}}{c}
                \right)
            \di X
        \di c,
    \label{eq:nunez-psp}
\end{equation}
where $a\in\lbrace e, i \rbrace$ is a synapse index, $Q_a(x,t)$ is an external/sub-cortical
synaptic input. With similar justifications as found in the Wilson and Cowan model (Nunez
\cite{nunezBrainWaveEquation1974, nunezQuantitativeDescriptionLargescale2000}, Nunez 
and Srinivasan 
\cite{nunezElectricFieldsBrain2006}), a typical choice of spatial kernel is given by
\begin{equation}
    w_a(c,x,X) =
        \sum_{n=1}^N
        \lambda^{(a)}_n f_n^{(a)}(c)
        \exp(-\lambda^{(a)}_n \abs{x - X}).
\end{equation}

In turn, the aggregated fraction of active neurons $A$ also responds to changes in the
excitatory/inhibitory post-synaptic densities $\rho_e/\rho_i$. The mean membrane potential $u(x,t)$
can be assumed to be represented as a linear sum $u(x,t) = \nu_e \rho_e(x,t) + \nu_i \rho_i(x,t)$,
where $\nu_e/\nu_i$ is the magnitude of excitatory/inhibitory post-synaptic potentials. Using the
principle of mass action (Freeman \cite{freemanWavesPulsesTheory1972}), the mean 
membrane potential is
related to the firing rate of the population by a sigmoidal wave-to-pulse transfer function $f(u)$
(Equation \eqref{eq:wc-sigmoid}). Hence, we have the relation $A(x,t) = f(\nu_e \rho_e(x,t) + \nu_i
\rho_i(x,t))$ (Jirsa and Haken \cite{jirsaDerivationMacroscopicField1997}). In the simplest 
case, this can be
expanded about a fixed point of $A$ and this non-linearity can be replaced with a linear
relationship, given by $A(x,t) = A_0 + \nu_i \rho_e - \nu_i \rho_i$ (Nunez
\cite{nunezBrainWaveEquation1974, nunezGenerationHumanEEG1989}, Jirsa and Haken
\cite{jirsaDerivationMacroscopicField1997}).

To transform these integral equations into partial differential equations, a spatial-temporal
Fourier transform can be used. Let $\hat{\rho}_a(x,\omega) = \mathcal{F}\lbrace{\rho_a : t \to
\omega\rbrace}$ be the temporal Fourier transform of $\rho_a$. Then from Equation
\eqref{eq:nunez-psp}, we have
\begin{equation}
    \hat{\rho}_a(x,\omega) =
    \hat{Q}_a(x,\omega) +
    \int_0^\infty
        \int_\Omega
            w_a(c, x, X) \, e^{-i \omega\, \tau(c,x-X)} \, \hat{A}(X,\omega)
        \di X
    \di c,
\end{equation}
where $\tau(c,x-X) = \abs{x-X}/c$ is the spatiotemporal delay. Assuming the connectivity is
homogeneous, we can write $w_a(c,x,X) = w_a(c, x-X)$. For brevity, we set $W_a(c, x-X, \omega) =
w_a(c, x-X) \, e^{-i\omega\,\tau(c, x-X)}$. Let $\tilde{\rho}_a(k,\omega) =
\mathcal{F}\lbrace{\hat{\rho_a} : x \to k \rbrace}$ be the spatial Fourier transform. Combining the
above results, we have
\begin{equation}
    \tilde{\rho}_a(k,\omega) =
    \tilde{Q}_a(k,\omega) +
    \tilde{A}_a(k,\omega)
    \int_0^\infty
        \tilde{W}_a(c, k)
    \di c.
\end{equation}

A wave equation can be obtained from this relation by evaluating the transformed connectivity
kernel, rearranging this relation, expressing $A$ in terms of $\rho_e/\rho_i$ (with linear
approximation or non-linear sigmoidal relation), and inverting the Fourier transforms. Details of
further steps in this general derivation, as well as simplifications and approximations, can be
found in the works of Nunez  \cite{nunezGenerationHumanEEG1989, 
nunezQuantitativeDescriptionLargescale2000}, Nunez and Cutillo 
\cite{nunezNeocorticalDynamicsHuman1995}, Nunez and Srinivasan 
\cite{nunezElectricFieldsBrain2006,nunezNeocorticalDynamicsDue2014}.

Although the Nunez brain equation only models one cortical population, it resembles a
two-population model due to the way it factors the intertwined excitatory and inhibitory synaptic
interactions rather than using distinct neural populations. This alternative focus on the synapses
rather than the cells was also seen in the models of Freeman 
\cite{freemanModelMutualExcitation1974}, Lopes da Silva et 
al.~\cite{lopesdasilvaModelBrainRhythmic1974}, van Rotterdam et 
al.~\cite{vanrotterdamModelSpatialtemporalCharacteristics1982}, and became a dominant 
concept in the
development of modern neural field modelling. Nonetheless, the wave equation derived by Nunez was a
significant theoretical development in neural field modelling, and continues to be a source of
discussion throughout the literature
(Ross et al.~\cite{rossBrainwaveEquationIncorporating2020}). We will see in practice 
throughout the next Section
how essential this synaptic viewpoint is to the construction of modern neural field models.

\section{Modern formulations of neural field modelling} 
\label{S:second-wave-nfms}

Following the canonical works of Wilson and Cowan, Freeman, Amari, and Nunez in the 1970's, neural
field theory continued to evolve as did the mathematical conceptualisation of a neural mass.
Neurons are complex multi-compartmental units that transmit an excitatory or inhibitory signal to
other neurons if and only if a weighted sum of temporally decaying afferent stimuli  exceeds a
threshold potential. Each of these constituent units of neural systems exhibit non-local and
non-trivial connectivity, and are also susceptible to microscopic stochastic fluctuations. The
accumulation of these complexities leads to patterns of complex oscillatory activities - even in
relatively small systems -  that appear to be crucial for information processing in neural systems.
Key results from these early neural field models demonstrated that  a statistical physics approach
may be viable,  perhaps essential,  to explain fundamental modes of function and signal propagation
in neural systems. Developments on this front are critical  to advancing neuroscience,
understanding information processing,  and developing novel solutions and therapies to common
neurological diseases  that exhibit pathological oscillation in brain activity. As this branch of
mathematical neuroscience progressed, necessary developments were made in the level of detail of
these models, particularly as more neural populations were integrated in this mathematical
framework.

This increased granularity and level of detail occurred throughout the resurgence of neural field
models in the 1990's (Freeman \cite{freemanTutorialNeurobiologySingle1992}, Wright et 
al.~\cite{wrightComputerSimulationElectrocortical1994}, Wright and Liley
\cite{wrightSimulationElectrocorticalWaves1995}, Nunez and Cutillo
\cite{nunezNeocorticalDynamicsHuman1995}, Jansen and Rit
\cite{jansenElectroencephalogramVisualEvoked1995}), especially when methods to transform the
integro-differential equations into partial differential equations were employed (Jirsa and 
Haken
\cite{jirsaFieldTheoryElectromagnetic1996,  jirsaDerivationMacroscopicField1997},
Robinson et al.~\cite{robinsonPropagationStabilityWaves1997}). At this point there was a 
significant
increase in the literature as researchers realised that neural field models were a viable way to
describe brain activity as measured by the EEG, for
example, brain rhythms (Liley et 
al.~\cite{lileyAlphaRhythmEmerges1999,lileySpatiallyContinuousMean2002}, Nunez and 
Cutillo 
\cite{nunezNeocorticalDynamicsHuman1995},
Robinson et al.~\cite{robinsonPredictionElectroencephalographicSpectra2001,
robinsonNeurophysicalModelingBrain2003}, Rennie et 
al.~\cite{rennieUnifiedNeurophysicalModel2002}), the effects of anaesthesia (Steyn-Ross
\cite{steyn-rossTheoreticalElectroencephalogramStationary1999}, Bojak and Liley
\cite{bojakModelingEffectsAnesthesia2005}), the onset of dementia (Bhattacharya
\cite{bhattacharyaThalamoCorticoThalamic2011}, Pons et 
al.~\cite{ponsRelatingStructuralFunctional2010}, M\"uller 
\cite{mullerUnifiedNeuralField2017}), 
and
epilepsy (Lopes da Silva et al.~\cite{lopesdasilvaEpilepsiesDynamicalDiseases2003}, 
Breakspear et al.~\cite{breakspearUnifyingExplanationPrimary2006}, Blenkinsop et 
al.~\cite{blenkinsopDynamicEvolutionFocalonset2012}, Kramer et 
al.~\cite{kramerHumanSeizuresSelfterminate2012}, Martinet et 
al.~\cite{martinetHumanSeizuresCouple2017}).

In this section,  we will look at central concepts prevalent in all modern neural field (and neural
mass) models, where the model output can be interpreted as an integral equation typical of a signal
processing problem. The physiological basis for these modern neural field approaches are addressed,
and we discuss how this gives rise to differences in mathematical structure between neural field
models that describe the same phenomena. Following this, we will give an example derivation of a
general multi-population neural field model, utilising the assumptions found in the work of Amari
 \cite{amariDynamicsPatternFormation1977}, Robinson et 
 al.~\cite{robinsonPropagationStabilityWaves1997}, Jirsa and Haken 
\cite{jirsaDerivationMacroscopicField1997}, and Coombes et 
al.~\cite{coombesWavesBumpsNeuronal2003}. We show how the integral form of these 
equations are
reformulated as coupled partial differential equations. The original reduction of the field
equation and subsequent derivation of the partial differential wave equation was performed by Jirsa and Haken
 \cite{jirsaDerivationMacroscopicField1997} and Robinson et 
 al.~\cite{robinsonPropagationStabilityWaves1997}. This was recapitulated by Coombes et 
 al.~\cite{coombesWavesBumpsNeuronal2003} and is summarised and annotated below for 
 completeness sake,
and for future reference. Special attention is paid to the underlying assumptions that lead to the
model equations. We will then briefly discuss non-linear and stochastic extensions of the modern
neural field modelling framework and conclude with a discussion on the limitations of neural field
theory before we show how this framework can be simplified further by considering connectivity at a
point only. This eliminates spatial dependence, and provides the common structure to neural mass
models.

\subsection{A unifying framework for neural fields} 
\label{S:unifying-nfm-framwork}

Historically, the level of description of neural field models  has been at the level of mean
membrane potentials (or equivalently the mean firing rate) across neuronal populations. This
characterisation using these state variables seems intuitive at first, but it lacks essential
detail on  the inputs to the mean membrane potential: synaptic transmission. Modelling mean
membrane potentials directly often does not allow for the inclusion for different temporal dynamics
corresponding to each type of post synaptic potential. Instead, modern neural field models use the
\textit{mean post-synaptic potentials} as their state variables to better capture this detail and
provide a more realistic model of neuronal dynamics. Neural field models following the
publications of Wilson and Cowan addressed the temporal dynamics of afferent signals in more
detail, since the magnitude, sign, and timescale of a received signal is dependent on the
particular neurotransmitter used (Zetterberg
\cite{zetterbergStochasticActivityPopulation1973}, Lopes da Silva et 
al.~\cite{lopesdasilvaModelBrainRhythmic1974}, van Rotterdam et 
al.~\cite{vanrotterdamModelSpatialtemporalCharacteristics1982}). As mentioned in the 
physiological
background in Section \ref{S:physiology}, synapses are remarkably complex interfaces subject to
variability in neurotransmitters. This can result in drastically different time-scales and signed
magnitudes. Differences can arise from the location of the synapse relative to the cell body, and
the kinetics of the neurotransmitter and the ion channels involved, as well as extracellular
dynamics.

The representation of neural field models   can differ significantly between publications from
different authors, sometimes even between publications from the same authors. Throughout the
literature, neural field models can be conceptualised as either a temporal differential operator
acting on state variables or as a type-II Volterra integral equation (Stone and Goldbart
\cite{stoneMathematicsPhysicsGuided2009}). Both forms are equivalent, but have different advantages
in practice. For instance, differential operator forms are more useful for the numerical studies of
these systems. Theoretically, it is simpler to work with the integral equation formulation, as the
temporal integral convolution operator  clearly depicts the pulse-to-wave transfer operation, which
converts the mean firing rate to a mean post-synaptic potential. All linear differential
formulations can be expressed in integral form by using Fourier transforms and Green's functions.
This was demonstrated in the previous section, where we began with an integral description of the
Amari field equations in \eqref{eq:amari-basic}. and used elementary techniques to translate to a
partial differential equation description of the Amari model in \eqref{eq:amari-eq},
\begin{equation}
\begin{split}
u_a(x,t) &=
    \int_{-\infty}^t
      \psi_a(t-T)
      \left\lbrace
        \sum_b
        \int_\Omega
          w_{a,b}(x-X) \, f_b[u_b(X,T)]
        \di X
        + q_a(x, T)
      \right\rbrace
    \di T , \nonumber\\
    \iff &\mu_a \ppx{}{t} u_a(x,t) =
    -u_a(x,t)
    + \sum_b
    \int_\Omega
      w_{a,b}(x-X) \, f_b[u_b(X,t)]
    \di X
    + q_a(x, t),
\end{split}
\end{equation}
where $\psi_a(t) = \frac{1}{\mu_a} e^{-\frac{t}{\mu_a}} \Theta(t)$  is referred to as the 
temporal
kernel (and $\Theta(t)$ is the Heaviside step function). This formulation allows for a clearer
interpretation of neural fields as a signal processing problem, where the temporal kernel
represents post-synaptic processing. The temporal kernel can be any suitable function that is zero
for all negative $t$, has a sharp rise to maximum, and decays to zero at $t$ increases.

The Amari field equations were a crucial extension of  the model by Wilson and Cowan,  allowing for
population-specific timescales. This permitted the inclusion of  more realistic population
interactions,  as different membrane timescales can account for  different trends in synapse
locations and/or different neurotransmitter dynamics across different populations. Neural field
models following the work of Amari  expanded on this concept,  and further generalised the input to
the field equations  to allow for the inclusion of more detailed interactions at the synapse
junctions between populations. We can illustrate this critical expansion by extending the Amari
field equations directly.

Consider a neural field across a one dimensional spatial domain $\Omega$ composed of $N$ generic
neural populations and let $u_a(x,t)$ denote the mean membrane potential of population $a$ at the
point $x\in\Omega$ at time $t$. The mean membrane potential is determined by the mean post-synaptic
potential $V_{a,b}(x,t)$ where $b$ is a population that connects to population $a$. The
current mean membrane potentials will influence the conductance, hence we can summarise the
magnitude and polarity of the post-synaptic potentials with state-dependent constants
$\nu_{a,b}(u_b(x,t))$. This weighted sum of post synaptic potentials (as well as any external
input) will be temporally filtered, since the strength of a signal decays rapidly. We can
mathematically express this relation between $u_a(x,t)$ and $V_{a,b}(x,t)$ as
\begin{equation}
  u_a(x,t) =
    \int_{-\infty}^t
      \eta_a(t-T) \,
      \left\lbrace
        \sum_b \nu_{a,b}(u_b(x,T)) \, V_{a,b}(x,T) + u_a^0(x,T)
      \right\rbrace
    \di T, \label{eq:rob}
\end{equation}
where $\eta_a(t)$ represents the synaptic kernel of population $a$. We note here that some models
(such as the cortical model from Liley et al.~\cite{lileyAlphaRhythmEmerges1999}) include an
external input term within the braces in the integral,  but the formulation of external inputs
differs between publications  so we will omit these terms for now. More elaborate neural field
models,  such as those seen in the work of Liley et al.~\cite{lileyAlphaRhythmEmerges1999} 
and
Steyn-Ross et al.~\cite{steyn-rossTheoreticalElectroencephalogramStationary1999}, include 
more
realistic conductance-based synaptic dynamics (Peterson et 
al.~\cite{petersonHomotopicMappingCurrentbased2018}) at the expense of introducing 
further
non-linearity to the model equations. However, most neural field models in practice tend to use the
simplest mapping  from mean post-synaptic potentials to mean membrane potentials  by assuming
synapses act instantaneously  and that synaptic weights are state-independent. That is, most
authors set $\eta_a(t) = \delta(t)$ where $\delta$ denotes the Dirac delta distribution, and set
$\nu_{a,b}(u_b)$ to be a constant value $\nu_{a,b}$  for each $a$ and $b$. This gives the mapping
\begin{equation}
  u_a(x,t) =
    \sum_b \nu_{a,b} \, V_{a,b}(x,t) + u_a^0(x,t)
    .
    \label{eq:mean-mp-ext}
\end{equation}
Each of these post-synaptic potentials can be  expressed as a spiking input that is temporally
weighted, which is represented by a spatial convolution with a connectivity kernel,
\begin{equation}
  V_{a,b}(x,t) =
    \int_{-\infty}^{t}
      \psi_{a,b} (t-T) \, \phi_{a,b}(x,T)
    \di T
    .
    \label{eq:mean-psp-noext}
\end{equation}
where $\phi_{a,b}(x,t)$ denotes the mean pulse density at the synapses of population $a$
from the population $b$. For most neural field models,  the external input to the system is added
on the post-synaptic potential level,  rather than the mean membrane potential level. This can be
achieved by splitting the membrane external input  $u_a^0(x,t) = \sum_b \nu_{a,b} V_{a,b}^0(x,t)$
and rewriting equations \eqref{eq:mean-mp-ext} and \eqref{eq:mean-psp-noext} as
\begin{equation}
  u_a(x,t) =
    \sum_b \nu_{a,b} \, V_{a,b}(x,t)
    ,
    \label{eq:mean-mp-noext}
\end{equation}
\begin{equation}
  V_{a,b}(x,t) =
    \int_{-\infty}^{t}
      \psi_{a,b} (t-T) \, \phi_{a,b}(x,T)
    \di T
    + V_{a,b}^0(x,t)
    .
    \label{eq:mean-psp-ext}
\end{equation}

The temporal decay and magnitude of post-synaptic potentials will depend on which population they
originate from. Using the same principles as for the Amari equation, this can be expressed in the
form of a temporal convolution describing the rise and fall of an incoming synaptic transmission.
\begin{equation}
  \phi_{a,b}(x,t) =
    \int_{\Omega}
      w_{a,b}(x - X) \, f_b[u_b(X,t)]
    \di X.
    \label{eq:psp-input-term}
\end{equation}
The average firing rate is related to the average membrane potential using the same non-linear
sigmoidal transfer function $f_b$ employed in early neural field models,  as we saw in the previous
section. The logistic function is a common choice of sigmoid across this period of neural field
models as well. A parameterisation of this used by
Robinson et al.~\cite{robinsonPredictionElectroencephalographicSpectra2001} is
\begin{equation}
  f_a[u_a(x,t)] =
    \frac{F_{a}}{1+\exp[-(u_a(x,t)-\theta)/\sigma']},
\end{equation}
where $F_{a}$ is the maximum firing rate of neurons in population $a$, and $\sigma' = \sigma
\sqrt{3}/\pi$ so that $\theta$ and $\sigma$ are the mean and standard deviation of a distribution
of threshold values over the neuronal population. The normalisation of $\sigma$ tends to be done so
that the derivative of the sigmoid approximates a Gaussian (Rennie et 
al.~\cite{rennieEffectsLocalFeedback1999}). The voltage scale used here is with respect to 
the resting
potential of the population, which is approximately the same as the reset voltage after firing in a
single neuron (Koch \cite{kochBiophysicsComputationInformation2004}). In a single neuron,
this is approximately -70 mV, and the threshold would be approximately -55mV (Destexhe
\cite{destexheConductanceBasedIntegrateandFireModels1997}, Meffin et 
al.~\cite{meffinAnalyticalModelLarge2004}). Therefore the threshold relative to resting 
potential is a
difference of 15mV, of which Robinson et 
al.~\cite{robinsonEstimationMultiscaleNeurophysiologic2004} uses a value of $\theta$ = 13.3 
mV. Common
parameter values seen in the work of Robinson et al. include $F_a = 340$ Hz and $\sigma = 3.8$ mV.

Combining equations \eqref{eq:mean-mp-noext}, \eqref{eq:mean-psp-ext}, and
\eqref{eq:psp-input-term} results in the expression
\begin{multline} 
  u_a(x,t) = \sum_b \nu_{a,b}
    \left\lbrace
          \int_{-\infty}^t
            \psi_{a,b}(t-T)
        \left[
      \int_\Omega
        w_{a,b}(x - X)
            \, f_b
            \left(
              \sum_c \nu_{b,c} \, V_{b,c}(X,T)
            \right)
          \di X
        \right]
      \di T\right.\\
\left. \vphantom{\int_{-\infty}^t
	\psi_{a,b}(t-T)
	\left[
	\int_\Omega
	w_{a,b}(x - X)
	\, f_b
	\left(
	\sum_c \nu_{b,c} \, V_{b,c}(X,T)
	\right)
	\di X
	\right]
	\di T}     + V_{a,b}^0(x,t)
    \right\rbrace
    .
\end{multline}

\begin{figure}[htbp]
    \centering
    \includegraphics[width=0.96\textwidth]{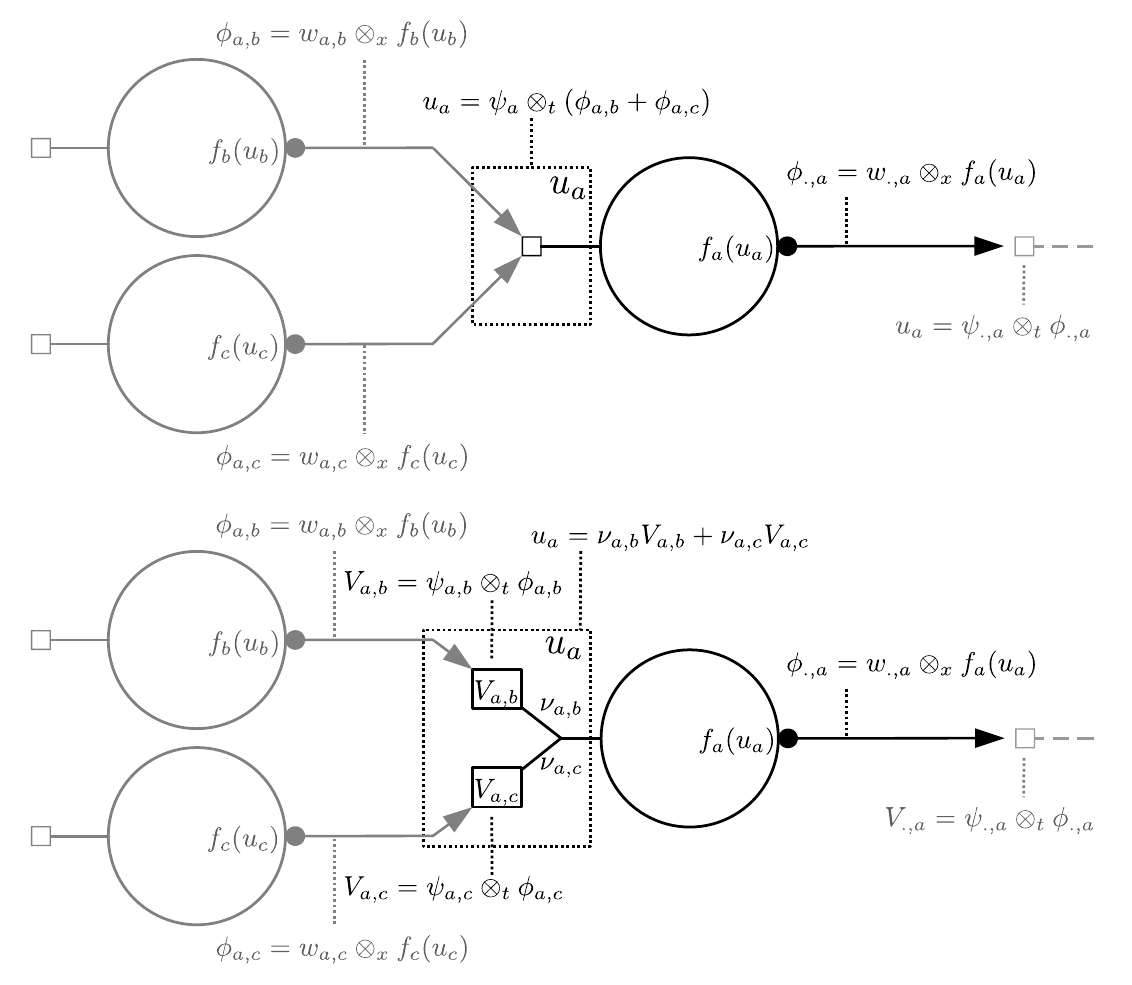}
    \caption{%
        \textbf{Comparison of 1st order NFM (top) to 2nd order NFM (bottom).} We illustrate this
        difference using a generic junction in a neural field model with three populations where
        population $b$ and $c$ connect to population $a$. A key development in modern Neural Field
        Models is the stratification of mean membrane potential state variables $u_a$ into a set of
        mean post-synaptic potential state variables $V_{a,b}, V_{a,c}$ with weights $\nu_{a,b},
        \nu_{a,c}$. This allows for the inclusion of specific hypotheses and assumptions about the
        differences in temporal dynamics of synaptic transmission.
    }
    \label{F:nfm-order-comparison}
\end{figure}

Using the mean membrane potential in analysis of these models can be cumbersome in this more
detailed framework, so typically the dynamics of the post-synaptic potentials are analysed  before
linearly combining the results  to give the mean membrane potential. Upon interchanging the order
of  spatial and temporal convolutions,  a general framework for modern second-order neural field
models is given by
\begin{equation}
  u_a(x,t) = \sum_b \nu_{a,b}
    \int_{-\infty}^t
      \psi_{a,b}(t-T)
      \left[
        \int_\Omega
          w_{a,b}(x - X) \, f_b
          \left(
            u_b(X,T)
          \right)
        \di X
      \right]
    \di T
    + u_a^0(x,t)
    ,
    \label{eq:gen-nfm-mean-potential}
\end{equation}
\begin{equation}
  V_{a,b}(x,t) =
    \int_{-\infty}^t
      \psi_{a,b}(t-T)
      \left[
        \int_\Omega
          w_{a,b}(x - X) \, f_b
          \left(
            \sum_c \nu_{b,c} \, V_{b,c}(X,T)
          \right)
        \di X
      \right]
    \di T
    + V_{a,b}^0(x,t),
    \label{eq:gen-nfm-psps}
\end{equation}
where $\int_{-\infty}^t \psi_{a,b}(t - T) \, q_{a,b}(x,T) \di T = V_{a,b}^0(x,t)$ and $u_a^0(x,t) =
\sum_b \nu_{a, b} V_{a,b}^0(x,t)$ are external inputs. The generalised parameters derived
throughout this Section are summarised in Table \ref{T:unified-nfm-params}. An illustration of the
advantages that come with additional complexity of these NFM models is given in Figure
\ref{F:nfm-order-comparison}

\begin{table}
    \centering
    \begin{tabular}{ | p{1.5cm} | p{11.0cm} | }
        \hline
        \textbf{Quantity}&
          \textbf{Physical interpretation}\\
        \hline

        \multicolumn{2}{|l|}{%
            \rule{0pt}{24pt}1st order:
            $\begin{array}{l}
                u_a(x,t) = \int_{-\infty}^t
                    \psi_a(t - T)\, \lbrace\sum_b \phi_{a,b}(x, T) + q_a(x,T)\rbrace
                \di T,
                \\[4pt]
                \phi_{a,b}(x,t) = \int_{\Omega}  w_{a,b}(x-X) f_b[u_b(X,t)]  \di X,
            \end{array}$
        }\\[14pt]
        \hline

        $u_a(x,t)$&
          Mean membrane potential of neuronal population $a$ at position $x$ and time $t$ \\
        $\psi_a(t)$ &
          Membrane Temporal kernel of population $a$ \\
        $w_{a,b}(x)$ &
          Spatial kernel of connections from population $b$ to population $a$ at position $x$ \\
        $\phi_{a,b}(x,t)$ &
          Pulse density of action potentials from population $b$ to population $a$ at position $x$ and time $t$ \\
        $q_a(x,t)$ &
          External input to population $a$ \\
       $T_a^{(m)}(x)$ &
          Time of $m$'th action potential from population $a$ at position $x$ \\
        $f_a(u)$ &
          Wave-to-pulse transfer function for population $a$ \\
        \hline

        \multicolumn{2}{|l|}{%
            \rule{0pt}{24pt}2nd order:
            $\begin{array}{l}
                u_a(x,t) = \sum_b \nu_{a,b}\,V_{a,b}(x,t),
                \\[4pt]
                V_{a,b}(x,t) = \int_{-\infty}^t
                    \psi_{a,b}(t - T)\, \lbrace\phi_{a,b}(x,T) + q_{a,b}(x,T)\rbrace
                \di T,
                \\[4pt]
                \phi_{a,b}(x,t) = \int_{\Omega} w_{a,b}(x-X) f_b[u_b(X,t)]  \di X,
            \end{array}$
        }\\[14pt]
        \hline
        $V_{a,b}(x,t)$ &
          Mean post-synaptic potential at population $a$ from population $b$ \\
        $\nu_{a,b}$ &
          Magnitude and polarity of post-synaptic potentials at population $a$ from population $b$ \\
        $\eta_{a}(t)$ &
          Synaptic temporal kernel of population $a$ \\
        $\psi_{a,b}(t)$ &
           Membrane temporal kernel of transmissions from population $b$ to population $a$ \\
        $q_{a,b}(x,t)$ &
          External input to population $a$ from population $b$ \\
        \hline

    \end{tabular}
    \caption{
      \textbf{Table of quantities used in general neural field models.} Physical interpretations of
      quantities are provided for the notation used throughout discussions and derivations of the
      first order neural field models in Section 3 and the second order neural field models
      discussed in Section 4.
    }
    \label{T:unified-nfm-params}
\end{table}

\subsection{Derivation of modern neural field models} 
\label{S:modern-nfms-derivation}

Although they are rarely explicitly derived, most neural field models in the literature can
be manipulated and rearranged to resemble Equations \eqref{eq:gen-nfm-mean-potential} and
\eqref{eq:gen-nfm-psps}, revealing common underlying structures that differ only in their choices
of spatiotemporal kernels, populations and connectivity structures, and assumptions made. 
This
common mathematical framework allows for a more direct comparison of neural field models and
explicit signposting and discussion of their underpinning assumptions. To illustrate this,
we consider a common neural field model of cortical tissue by Robinson et 
al.~\cite{robinsonPropagationStabilityWaves1997}. This models the evolution of the mean 
membrane
potential via the interactions of mean post-synaptic potentials. Here we will derive the cortical
model and show how the integral equation formulation is transformed into a set of partial
differential equations, which can then be expressed as a system of ordinary differential equations
(or stochastic differential equations, depending on the nature of external inputs).

The cortical neural field model from Robinson et 
al.~\cite{robinsonPropagationStabilityWaves1997}
contains two neuronal populations, an excitatory $(e)$ population and an inhibitory $(i)$
population, in similar fashion to the Wilson and Cowan model. This can be interpreted as modelling
the most prevalent types of neurons in the cortex, which secrete the excitatory neurotransmitter
AMPA and the inhibitory neurotransmitter GABA-A. A key simplification is made to the general neural
field model in Equation \eqref{eq:gen-nfm-mean-potential} where the kinetics and electrodynamic
response of these neurotransmitters, and hence the synaptic kernel of both populations, is assumed
to be the same.

\begin{framed}
    \textbf{Physiological assumption:} We assume that the shape of the synaptic filtering is the
    same for both excitatory and inhibitory inputs i.e. both AMPA and GABA-A receptors. Their
    responses are in fact very close with GABA-A having only a slightly higher rise time than AMPA,
    which is itself very small (Dayan and Abbott 
    \cite{dayanTheoreticalNeuroscienceComputational2001}). This has
    the consequence that the synaptic $\tau$ and membrane time constants $\mu$ are the same for
    both inhibitory and excitatory populations.
\end{framed}

This gives us a single temporal kernel in our integral equation $\psi_{a,b}(t) = \psi(t)$ for all
$a,b$ in $e,i$. This allows us to push the sum over afferent populations into the temporal
integral, so we have
\begin{multline}
  u_a(x,t) =
    \int_{-\infty}^t
      \psi(t-T)
      \left[
        \sum_b \nu_{a,b}
        \int_\Omega
          w_{b}(x - X) \, f
          \left(
            u_b
            \left(
              X,T - \frac{\abs{x-X}}{v}
            \right)
          \right)\right.
        \di X\\
        \left.\vphantom{ \sum_b \nu_{a,b}
        	\int_\Omega
        	w_{b}(x - X) \, f
        	\left(
        	u_b
        	\left(
        	X,T - \frac{\abs{x-X}}{v}
        	\right)
        	\right)
        	\di X}
        + q_a(x,T)
      \right]
    \di T,
\end{multline}
where we have combined $\displaystyle \sum_b \nu_{a,b}\, q_{a,b}(x,t) = q_a(x,t)$ for  brevity. It
should be noted here that we only consider a neural field with one spatial dimension throughout
this derivation, and can be readily extended to two dimensions.

The external input is conceptualised as  an ``external'' population $(s)$, but for our purposes
this is notational. The significance of this term will not largely be discussed here, as the form
tends to depend on the model and application. The notation used throughout the literature is
$q_a(x,t) = \nu_{a,s} \phi_s(x,t)$, where $\nu_{a,s}$ is a scalar synaptic weighting and $\phi_s$
is the firing rate of the external population.

The spatial connectivity,  in addition to assuming random connectivity in the thermodynamic limit,
is also formulated such that neurons of either population indiscriminately synapse onto other
populations. That is, the spatial weighting on the incoming firing rate from surrounding
populations only depends on the spatial density of afferent neurons. This allows us to make an
additional notational simplification on the synaptic kernel $w_{a,b}(x) = w_{b}(x)$. In this sense,
the spatial kernels dictate the spatial density of neurons rather than the spatial density of
synapses. In addition to this, the same sigmoidal potential-to-rate transfer function is used for
both populations, making the notational simplification $f_b(u_b) = f(u_b)$. When considering finite
conduction velocities, spatiotemporal delays are incorporated into the model as
\begin{multline}
  u_a(x,t) =
    \int_{-\infty}^t
      \psi(t-T)
      \left[
        \sum_b \nu_{a,b}
        \int_\Omega
          w_{b}(x - X) \, f
          \left(
            u_b
            \left(
              X,T - \frac{\abs{x-X}}{v}
            \right)
          \right)
        \di X\right.\\
\left.\vphantom{  \sum_b \nu_{a,b}
	\int_\Omega
	w_{b}(x - X) \, f
	\left(
	u_b
	\left(
	X,T - \frac{\abs{x-X}}{v}
	\right)
	\right)
	\di X}        + \nu_{a,s} \phi_s(x,T)
      \right]
    \di T.
\end{multline}
where we assume a constant axonal propagation velocity $v$ (Coombes et 
al.~\cite{coombesWavesBumpsNeuronal2003}).

\begin{framed}
    \textbf{Physiological assumptions:} The axonal conduction velocity is assumed to be a constant
    and identical for all populations on the same spatial scale. Axonal velocities are different
    for different spatial scales depending on whether the axons are myelinated or not, and are
    proportional to the thickness or the diameter of the axon (Lohmann and R\"orig
    \cite{lohmannLongrangeHorizontalConnections1994}). Experimental values also show that there is
    a distribution of axonal conduction velocities between populations on the same scale 
    (Bojak and Liley \cite{bojakAxonalVelocityDistributions2010}) and that these act as a linear 
    filter
    (Roberts and Robinson \cite{robertsModelingDistributedAxonal2008}). This distribution is 
    neglected in
    this model for the sake of simplicity and the conduction velocity will be constant unless
    otherwise explicitly stated.
\end{framed}

This equation now serves as the starting point of the derivation of the wave equation (Jirsa 
and Haken \cite{jirsaDerivationMacroscopicField1997}). The neural field (in terms of the firing 
rate)
is a convolution of the spatial kernel with the spatiotemporally delayed sigmoid
\begin{equation}
  \phi_{b}(x, t) =
    \int_\Omega
      w_{b}(x-X) \,
      f\left[u_b\left(X,t-\frac{\abs{x-X}}{v}\right)\right]
    \di X,
\end{equation}
and the membrane potential is a convolution of the synaptic (temporal) kernel with the neural field
\begin{equation}
  u_a(x,t) =
    \int_{-\infty}^t
      \psi(t-T)
      \sum_b \nu_{a,b} \phi_{b}(x, T)
    \di T.
  \label{eq:neuralfield_convolution}
\end{equation}
Green's functions and Fourier transforms are now used to reduce the canonical neural field equation
\eqref{eq:gen-nfm-mean-potential} into coupled partial differential equations (Jirsa and 
Haken
\cite{jirsaDerivationMacroscopicField1997}, Coombes et 
al.~\cite{coombesWavesBumpsNeuronal2003}) by
decoupling the spatial and temporal components.

According to the theory of Green's functions, we have
\begin{equation}
  \mathbf{D}_t\br{u_a(x,t)} = \sum_b \nu_{a,b} \phi_{b}(x, t),
    \qquad
    \mathbf{D}_t \br{\psi(t)} = \delta(t)
    ,
\end{equation}
\begin{equation}
  \mathbf{D}_x\br{\phi_{b}(x,t)} =
    f\left[u_b\left(X,t-\frac{\abs{x-X}}{v}\right)\right],
    \qquad
    \mathbf{D}_x \br{w_{b}(x)} = \delta(x),
\end{equation}
where the differential operators over time $\mathbf{D}_t$ and space $\mathbf{D}_x$ are dependent on
the exact forms of $\psi(t)$ and $w_{b}(x)$, respectively. In order to deal with the spatiotemporal
delay, the neural field is redefined so that the delay is absorbed into the spatial kernel with a
delta distribution where
\begin{equation}
  \phi_{b}(x, t) =
    \int_\Omega
      \int_{-\infty}^{t}
        G_{b}(x-X, t-T) \,
        f\left[u_b\left(X,T\right)\right]
      \di T
    \di X,
\end{equation}
where $G_{b}(x,t) = \delta(t - \abs{x}/v) w_{b}(x)$ so that
\begin{equation}
  \mathbf{D}_t\br{u_a(x,t)} = \sum_b \nu_{a,b} \phi_{b}(x, t),
    \qquad
    \mathbf{D}_t \br{\psi(t)} = \delta(t)
    ,\label{eq:robinson-pde-a}
\end{equation}
\begin{equation}
  \mathbf{D}_x\br{\phi_{b}(x,t)} =
    f\left[u_b\left(x,t\right)\right],
    \qquad
    \mathbf{D}_x \br{\delta(t - \abs{x}/v) w_{b}(x)} = \delta(x).
    \label{eq:robinson-pde-b}
\end{equation}
To express this neural field in terms of coupled partial differential equations (PDEs), we need to
derive the operators $\mathbf{D}_x$ and $\mathbf{D}_t$ that correspond to the choice of temporal
kernel $\psi(t)$ and spatial kernels $w_{b}(x)$ we make. In what follows, we will derive the
operator $\mathbf{D}_x$ using Fourier transforms to deal with the spatial time delay. After
deriving the spatial PDE, we will then derive the temporal differential operator $\mathbf{D}_t$
from experimental principles and show how this is equivalent to using a double-exponential temporal
kernel.

Finding the partial differential equation associated with the spatial differential operator
$\mathbf{D}_x$ is a matter of determining a realistic connectivity kernel $w_b(x)$ for each
population and then taking spatiotemporal Fourier transform. Most neural field models tend to use
the Laplacian distribution, $w_b(x) = \exp(-|x|/r_{b})/2 r_{b}$ where $r_{b}$ is the axonal range
of population $b$. The Robinson neural field model uses a different kernel instead, given by
$w_{b}(x) = K_0(x/r_{b})/2\pi r_{b}^2$ where $K_0$ is a Bessel function of the second kind. This
has a similar decay rate to the Laplacian distribution, but this function is singular at the
origin, implying infinite connectivity between neuronal populations at a point. From a modelling
perspective, the difference in kernels, whether they have a singularity or not, does not
significantly change the model. This is because the region containing the singularity is
immeasurable and largely irrelevant to the millimetric scale at which a neural field model
describes. After choosing the Bessel spatial kernel,  equation \eqref{eq:robinson-pde-b} further
simplifies to the resulting PDE,
\begin{equation}
  \left[
    \left(
      1 + \frac{1}{\gamma_{b}} \ppx{}{t}
    \right)^2
    - r_{b}^2 \lap
  \right]
  \phi_{b}(x,t) = f(u_b(x,t)),
  \label{eq:rob-wave}
\end{equation}
where $\gamma_{b} = v/r_{b}$ as a temporal dampening term.

To obtain the corresponding temporal PDE,  we can derive the temporal differential operator
$\mathbf{D}_t$ (and the analogous integral kernel) from biophysical principles. We start from
canonical R-C circuit equations, inspired by simplified spiking models such as integrate-and-fire
neurons with current based synapses (Burkitt
\cite{burkittReviewIntegrateandfireNeuron2006a}). In this case there is no spiking mechanism and
the state variables are treated as population averages with rate based inputs. The membrane is
described as a leaky capacitor with $C_a u_a(x,t) = q_a(x,t)$ where $C_a$ is the capacitance ---
assumed to be constant for a given population --- and $q_a(x,t)$ is the charge at population $a$ at
position $x$ at time $t$. Taking the derivative of this gives us the current which can be separated
into the leak and synaptic currents
\begin{equation}
\begin{split}
    C_a\ppx{}{t} u_a(x,t) &=
        I_{a}^{\mathrm{leak}}(x,t) + I_{a}^{\mathrm{syn}}(x,t), \\
   \implies \ppx{}{t} u_a(x,t) &=
       -\frac{1}{\mu} u_a(x,t) + \frac{1}{C_a} \sum_{b} N_{a,b} I_{a,b}(x,t),
    \label{eq:current1}
\end{split}
\end{equation}
where the membrane potential leaks over the membrane time constant $\mu$ (which we had
previously assumed is the same for all populations) and the synaptic current is the linear sum of
all the incoming currents $I_{a,b}(x,t)$, with coefficients $N_{a,b}$ that quantify the number of
connections from population $b$ to population $a$.

\begin{framed}
    \textbf{Physiological assumption:} We approximate the synaptic and dendritic filtering of the
    post-synaptic current $I_{a,b}(t)$ as exponentially decaying from a maximal current, over a
    synaptic time constant. This is a reasonably good fit to experimental data as can be seen in
    previous modelling work by Destexhe et al.~\cite{destexheSynthesisModelsExcitable1994}.
\end{framed}

Continuing the derivation,  the form of the decaying exponential for the post-synaptic current 
is defined as
\begin{equation}
  \ppx{}{t} I_{a,b}(x,t) =
    -\frac{1}{\tau} I_{a,b}(x,t) + A_{b}\phi_{b}(x,t).
  \label{eq:current2}
\end{equation}
where $A_b$ is the maximal current amplitude and $\tau$ is the synaptic time-constant.
Differentiating Eq.(\ref{eq:current1}) with respect to time and substituting Eq.(\ref{eq:current2})
gives
\begin{equation}
  \ppx[2]{u_a}{t} =
    -\frac{1}{\mu} \ppx{u_a}{t} -
    \frac{1}{C_a\tau} \sum_{b} N_{a,b} I_{a,b}(x,t) +
    \frac{1}{C_a} \sum_{b} N_{a,b} A_{b} \phi_{b}(x,t).
\end{equation}
Then back-substituting Eq.(\ref{eq:current1}) into the second term on the right hand side
of the above equation gives
\begin{equation}
  \ppx[2]{u_a}{t} =
    -\frac{1}{\mu} \ppx{u_a}{t} -
    \frac{1}{\tau}
    \left[
      \ppx{u_a}{t} + \frac{u_a(x,t)}{\mu}
    \right] +
    \frac{1}{C_a} \sum_{b} N_{a,b} A_{b} \phi_{b}(x,t),
\end{equation}
which, after collecting terms and multiplying through by $\tau \mu$, becomes
\begin{equation}
  \left[
    \tau \mu \ppx[2]{}{t} + (\tau + \mu)\ppx{}{t} + 1
  \right] u_a(x,t)
    = \sum_{b} \nu_{a,b} \phi_{b}(x,t),
    \qquad \nu_{a,b} = N_{a,b} s_{a,b},
    \quad s_{a,b} = \frac{A_{b} \tau \mu}{C_a}.
\end{equation}
A more through discussion of this derivation for single neuron models can be found in works by
Gerstner and Kistler  \cite{gerstnerSpikingNeuronModels2002}, Ermentrout and Terman 
\cite{ermentroutMathematicalFoundationsNeuroscience2010}. More complex descriptions of synaptic
dynamics, such as non-linear conductance-based synapses, can be utilised instead which is seen in
the work of Liley et al.~\cite{lileyContinuumModelMammalian1997}, Pinotsis et 
al.~\cite{pinotsisConductancebasedNeuralField2013}, and Peterson et 
al.~\cite{petersonHomotopicMappingCurrentbased2018}. We will comment on these 
non-linear extensions
later. For this derivation,  we will consider only linear synaptic kernels  so that we may derive
the differential operator  using Green's functions, which gives rise to the following partial
differential equation
\begin{equation}
  \left(
    \tau \ppx{}{t} + 1
  \right)
  \left(
    \mu \ppx{}{t} +1
  \right) u_a(x,t)
  = \sum_{b} \nu_{a,b} \phi_{b}(x,t),
  \label{eq:rb-input}
\end{equation}
which corresponds to the double exponential synaptic kernel,
\begin{equation}
  \psi(t) =
    \frac{1}{\tau - \mu}[e^{-t/\tau} - e^{-t/\mu}] \, \Theta(t),
\end{equation}
where $\mu$ and $\tau$ are the membrane and synaptic time constants, respectively.

\begin{figure}[htbp]
    \centering
    \includegraphics[width=0.90\textwidth]{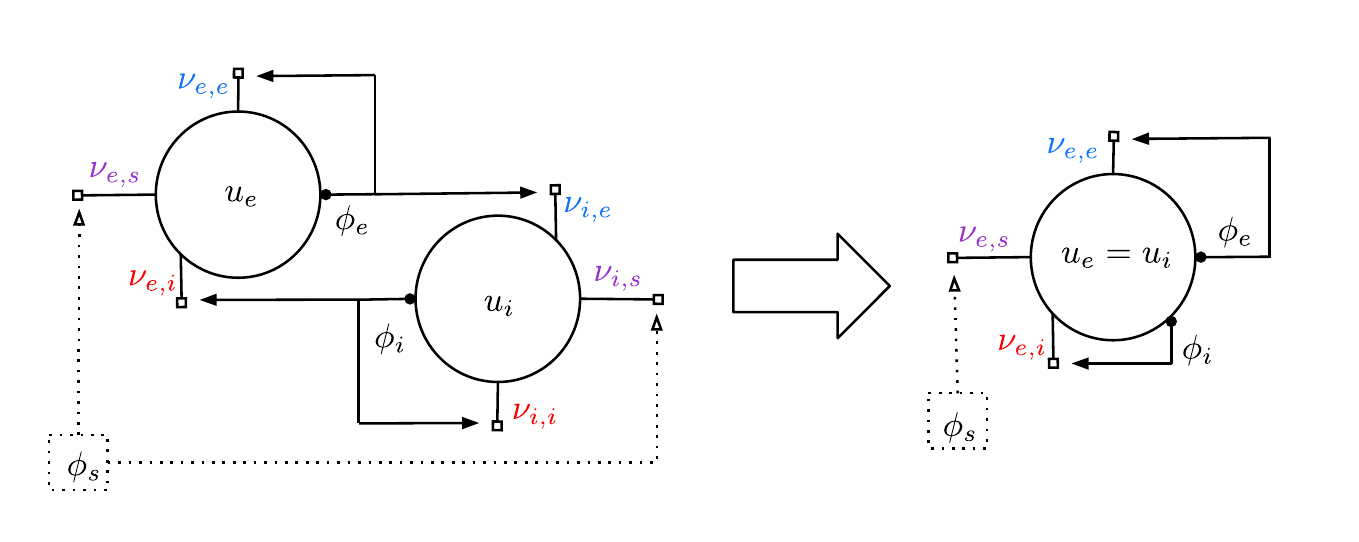}
    \caption{%
        \textbf{Diagram of simplification in the Robinson cortical NFM.} This diagram shows the
        simplification made by equating the synaptic coefficients $\nu_{e,b} = \nu_{i,b}$
        (highlighted by matching colours), which effectively combines the cortical excitatory and
        inhibitory populations into one cortical population with two self-connections. External
        inputs are represented by dotted arrows.
    }
    \label{F:robinson-simplification}
\end{figure}

The basic form of the Robinson model  utilises three populations: excitatory cortical neurons
$(e)$, inhibitory cortical neurons $(i)$, and external non-specific neurons $(s)$. At this point,
the Robinson model begins to deviate from other neural field models by making a major anatomical
assumption that effectively combines the cortical populations. A diagrammatic comparison
contrasting the Robinson model with typical neural field models can be found in Figure
\ref{F:robinson-simplification}.

\begin{framed}
    \textbf{Anatomical assumption:} The local connectivity approximation
    Rennie et al.~\cite{rennieEffectsLocalFeedback1999} is used i.e. that the intra-cortical 
    connectivities are
    proportional to the density of neurons, as well as the average number of synapses involved.
    That is, there is a symmetry between the type of post-synaptic neurons that a pre-synaptic
    neuron will synapse onto. For example, if $N_{a,b}$ is the connectivity from a population of
    neurons b, to a population a, then $N_{e,b} = N_{i,b}$. That is, there are approximately the
    same proportion of connections from population b, to both the excitatory $e$, and inhibitory
    $i$, populations. This approximation is also made with regards to the synaptic response
    $s_{a,b}$ where $s_{e,b} = s_{i,b}$. Since the synaptic constant for each population interface
    is given by the product $\nu_{a,b} = N_{a,b} s_{a,b}$, we have $\nu_{e,b} = \nu_{i,b}$ for all
    populations $b$. Please refer to Rennie et al.~\cite{rennieEffectsLocalFeedback1999} for 
    further details of
    this approximation. We also assume that the sigmoidal functions used for both populations are
    identical.
\end{framed}

The main consequence of this approximation is that the membrane potential for both excitatory
neurons and inhibitory neurons are equivalent. Upon expansion of the input equation Eq.
\eqref{eq:rb-input},
to both excitatory and inhibitory populations $a\in\lbrace e,i \rbrace$ with an external input
$\phi_s$, and $b \in \lbrace e,i,s \rbrace$, we have
\begin{equation}
  \left(
    \tau \ppx{}{t} + 1
  \right)
  \left(
    \mu \ppx{}{t} +1
  \right) u_e(x,t)
    = \nu_{e,e} \phi_{e}(x,t)
    +  \nu_{e,i} \phi_{i}(x,t)
    +  \nu_{e,s} \phi_{s}(x,t),
    \label{eq:rb-e-potential}
\end{equation}
\begin{equation}
  \left(
    \tau \ppx{}{t} + 1
  \right)
  \left(
    \mu \ppx{}{t} +1
  \right) u_i(x,t)
    = \nu_{i,e} \phi_{e}(x,t)
    +  \nu_{i,i} \phi_{i}(x,t)
    +  \nu_{i,s} \phi_{s}(x,t).
    \label{eq:rb-i-potential}
\end{equation}
When we apply the local connectivity approximation to these equations where $\nu_{e,b} =
\nu_{i,b}$, it is clear that Eq. \eqref{eq:rb-i-potential} becomes redundant as it is identical to
Eq. \eqref{eq:rb-e-potential} . For simplicity, we set $u_i = u_e$ and $\nu_{a,b}$ becomes
$\nu_{b}$, as there is only one afferent neural population.

\begin{framed}
    \textbf{Physiological assumption:} As a consequence of both populations having the same time
    constants and the local connectivity approximation, inhibitory neurons are assumed to have
    exactly the same characteristics as excitatory neurons, which is obviously not true for
    biological neurons.
\end{framed}

Even though there is only effectively one type of neuron, we can still distinguish between
populations as well as an external input. This is due to the different synaptic products $\nu_b =
N_b s_b$ of the different populations and the different axonal dampening coefficients $\gamma_b$.
Although it is possible to treat the populations individually, this effectively doubles the number
of parameters and dimensionality of state space of the model. Analytical and computational studies
with spiking models have shown that when the populations are treated differently, the results of a
stability analysis are not substantially different (Brunel
\cite{brunelDynamicsSparselyConnected2000}, Meffin et 
al.~\cite{meffinAnalyticalModelLarge2004}).

\begin{framed}
    \textbf{Mathematical assumption:} As is commonly done in the literature, we will only examine
    the spatially invariant case of the dampened wave equation i.e. the global spatial modes.
\end{framed}

To investigate the spatially uniform activity,  the spatially inhomogeneous term in the wave
equation Eq. \eqref{eq:rob-wave} is set to zero $r_{b}^2 \lap \phi_b(x,t) = 0$. To make this a
neural mass model, we would set the axonal range to zero $r_{b} = 0$. However, setting $\lap
\phi_b(x,t) = 0$ allows us to solve for the spatially uniform solutions of the neural field model.
This removes the spatial variation  while still preserving the axonal range, conduction velocity
and intra-cortical connectivities. The damped wave equation now becomes
\begin{equation}
  \left(
    \frac{1}{\gamma_{a}} \ppx{}{t} + 1
  \right)^2 \phi_{a}(x,t) = f(u_a(x,t)),
\end{equation}
where we have removed the spatially dependent term. Since solutions to these equations are
spatially homogeneous, we may drop the $x-$dependence and treat this as though it were a neural
mass model. Even though the model is assumed to be spatially isotropic and homogeneous, there is
still some spatial structure contained in the temporal dampening term $\gamma_{b} = v/r_{b}$, which
contains information about the spatiotemporal delay. This makes the category of the model somewhat
ambiguous. Technically it is not a neural field model as the field is spatially invariant, but it
layers on top of one by including only the spatially uniform solutions. Yet it could also be
described as an atypical neural mass model with spatiotemporal delays, so it is somewhere in
between both categories. In the literature it is described as a neural field model with only global
spatial modes.

Previously, this cortical neural field model has been used to  describe the whole cortex 
(Robinson et al.~\cite{robinsonPropagationStabilityWaves1997}, Rennie et 
al.~\cite{rennieEffectsLocalFeedback1999}) and extended to a thalamo-cortical model 
(Rennie et 
al.~\cite{rennieUnifiedNeurophysicalModel2002}, Robinson et 
al.~\cite{robinsonEstimationMultiscaleNeurophysiologic2004}, Breakspear et 
al.~\cite{breakspearUnifyingExplanationPrimary2006}). In both instances, the axonal range 
$r_a$ is
different for both the excitatory, $r_e$, and the inhibitory, $r_i$, neurons. By taking the ``local
inhibition approximation'' (Rennie et al.~\cite{rennieEffectsLocalFeedback1999}) it is 
assumed 
that
all long range connections are purely excitatory, hence we can use $r_i \ll r_e \implies \gamma_i
\gg \gamma_e$. So as $r_i \to 0$, $\gamma_i \to \infty$, which means that if separate wave
equations are written for both the excitatory $\gamma_e, \phi_e$, and inhibitory $\gamma_i, \phi_i$
components, the inhibitory wave equation
\begin{equation}
  \left(
    \frac{1}{\gamma_i} \ppx{}{t} + 1
  \right)^2 \phi_i(t)
  = f(u_i(t)) = f(u_e(t))
\end{equation}
becomes $\phi_i(t) = f(u_e(t))$. Therefore the inhibitory firing rate is governed by the sigmoidal
function of the membrane potential. Another consequence of this assumption is that the long range
excitatory to inhibitory connections are not included.
\begin{framed}
    \textbf{Anatomical assumption:} In this model, since there are not distinct excitatory and
    inhibitory cortical populations, there are no local inhibitory-inhibitory connections. These
    are hypothesised in other neural models to play a significant role in brain dynamics such as
    alpha rhythmogenesis (Liley et al.~\cite{lileyAlphaRhythmEmerges1999}), and epilepsy 
    (Wendling et al.~\cite{wendlingEpilepticFastActivity2002}). Further, long-range 
    excitatory-inhibitory
    connections have also not been included i.e. afferent connections from pyramidal neurons
    synapsing on inhibitory interneurons.  These would also significantly affect overall cortical
    dynamics.
\end{framed}
This gives us the full cortical neural field model proposed by
Robinson et al.~\cite{robinsonPropagationStabilityWaves1997},
\begin{equation}
\begin{split}
  \left(
    \frac{1}{\gamma_e} \ppx{}{t} + 1
  \right)^2 \phi_e(t)
    &= f(u_e(t)), \\
  \left(
    \mu \ppx{}{t} + 1
  \right)
  \left(
    \tau \ppx{}{t} + 1
  \right) u_e(t)
    &= \nu_{e,e} \phi_e(t) + \nu_{e,i} f(u_e(t)) + \nu_{e,s}\phi_s(t),
\end{split}
\end{equation}
which, when rearranged and put into a coupled ordinary differential equation (ODE) form, are
usually expressed in the literature as
\begin{equation}
\begin{split}
  \ddot{\phi}_e(t) &=
    \gamma_e^2[f(u_e(t)) - \phi_e(t)] - 2 \gamma_e \dot{\phi}_e(t), \\
  \ddot{u}_e(t) &=
    \alpha \beta [
      \nu_{e,e}\phi_e(t)
      + \nu_{e,i} f(u_e(t))
      + \nu_{e,s}\phi_s(t) - u_e(t)
    ]
    - (\alpha + \beta) \, \dot{u}_e(t),
\end{split}
\end{equation}
where $\alpha = \mu^{-1}$ and $\beta = \tau^{-1}$.

\begin{figure}[htbp]
    \centering
    \includegraphics[width=0.47\textwidth]{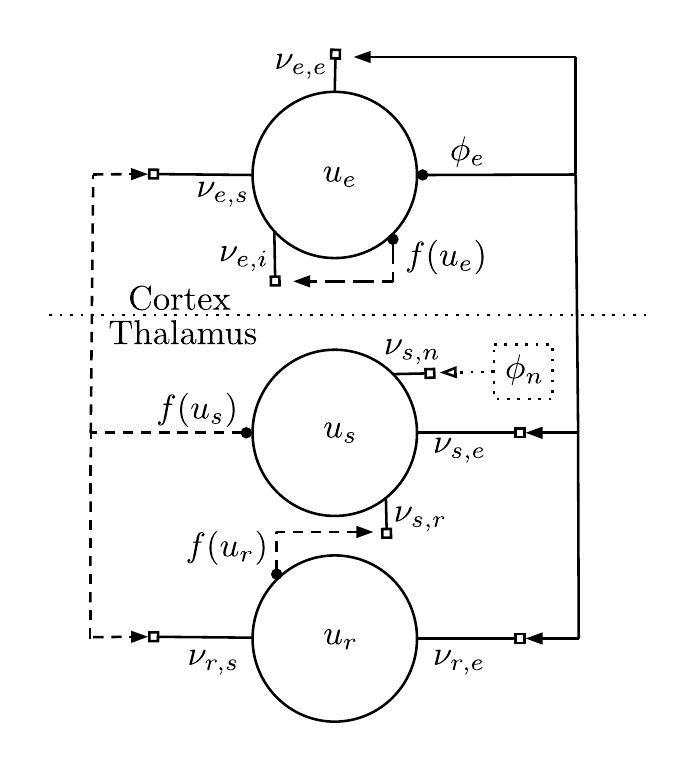}
    \caption{%
        \textbf{Illustration of Robinson cortico-thalamic NFM.} External inputs are represented by
        dotted arrows and connections that utilise the limiting axonal range assumption $r_a \to 0$ are
        shown in dashed lines. Connections that cross the cortex-thalamus boundary incur a time delay
        of $t_0/2$.
    }
    \label{F:robinson-model}
\end{figure}

This cortical model by Robinson et al. was extended to include thalamic populations. Recent
publications generally analyse the full thalamo-cortical model, which includes a time delay between
cortical and thalamic populations, but overall uses the same principles as discussed in this
section so far. In a sense, we can interpret this thalamic extension as an expansion of the
non-specific external population term $\nu_{e,s}\phi_s(t)$ in the cortical model. In addition to
the cortical neurons, composed of excitatory $(e)$ and inhibitory $(i)$ sub-populations, Robinson
et al. also consider populations from the specific thalamic nucleus $(s)$, the reticular thalamic
nucleus $(r)$, and a new input term arising from non-specific subcortical neurons $(n)$. The
equations for the thalamo-cortical neural field model are given by
\begin{equation}
\begin{split}
  \left(
    \frac{1}{\gamma} \ppx{}{t} + 1
  \right)^2 \phi_e(t)
    &= f(u_e(t)), \\
  \left(\mu \ppx{}{t} + 1\right)
  \left(\tau \ppx{}{t} + 1\right) u_e(t)
    &= \nu_{e,e}\phi_e(t)
    + \nu_{e,i} f(u_e(t))
    + \nu_{e,s} f \left( u_s\left( t - \frac{t_0}{2} \right) \right), \\
  \left(\mu \ppx{}{t} + 1\right)
  \left(\tau \ppx{}{t} + 1\right) u_s(t)
    &= \nu_{s,e}\phi_e\left(t-\frac{t_0}{2}\right)
    + \nu_{s,r} f(u_r(t))
    + \nu_{s,n} \phi_n(t), \\
  \left(\mu \ppx{}{t} + 1\right)
  \left(\tau \ppx{}{t} + 1\right) u_r(t)
    &= \nu_{r,e}\phi_e\left(t-\frac{t_0}{2}\right)
    + \nu_{r,s} f(u_s(t)),
\end{split}
\end{equation}
which, when simplified and expressed in ODEs,
\begin{equation}
\begin{split}
  \ddot{\phi}_e(t) &= \gamma^2[f(u(t)) - \phi(t)] - 2 \gamma \dot{\phi}(t), \\
  \ddot{u}_e(t) &=
    \alpha \beta
    \left[
      \nu_{e,e}\phi_e(t)
      + \nu_{e,i} f(u_e(t))
      + \nu_{e,s}
      f \left(
        u_s\left( t - \frac{t_0}{2} \right)
      \right)
      - u_e(t)
    \right]
    - (\alpha + \beta) \, \dot{u}_e(t), \\
  \ddot{u}_s(t) &=
    \alpha \beta
    \left[
      \nu_{s,e}\phi_e\left(t-\frac{t_0}{2}\right)
      + \nu_{s,r} f(u_r(t))
      + \nu_{s,n} \phi_n(t)
      - u_s(t)
    \right]
    - (\alpha + \beta) \, \dot{u}_s(t), \\
  \ddot{u}_r(t) &=
    \alpha \beta
    \left[
      \nu_{r,e}\phi_e\left(t-\frac{t_0}{2}\right)
      + \nu_{r,s} f(u_s(t))
      - u_r(t)
    \right]
    - (\alpha + \beta) \, \dot{u}_r(t),
\end{split}
\end{equation}
where $\alpha = \mu^{-1}$ and $\beta = \tau^{-1}$, and $t_0$ is the time delay of the full
thalamo-cortical loop (hence $t_0/2$ represents the unidirectional delay from a thalamic population
to a cortical population or vice versa). An illustration of the full thalamo-cortical neural field
model is given in Figure \ref{F:robinson-model}.

Building off of the prior derivation and assumptions, the cortical model by Robinson has
lower dimensionality in state space and parameter space in comparison to other neural field models.
This allows extensions of the model to include more sophisticated anatomical and physiological
factors that are generally not possible in other neural field frameworks. The neural field model
proposed by Robinson and colleagues has undergone a variety of modifications since its first
appearance in the literature in 1997, and has inspired other authors to make their own
simplifications and extensions Marten et al.~\cite{martenOnsetPolyspikeComplexes2009}. 
Another
canonical neural field that is contemporaneous of the Robinson model is by Liley et 
al.~\cite{lileyAlphaRhythmEmerges1999}. This NFM was the first model to extend neural field 
models to
include more physiologically realistic conductance-based synaptic dynamics as we discussed in
Equation \eqref{eq:rob},
\begin{equation}
    u_a(x,t) =
        \int_{-\infty}^{t}
            \eta_a(t-T)
            \left\lbrace
              \sum_{b} \nu_{a,b}(u_b(x,T)) \; V_{a,b}(x,T)
              + u_a^0(x,T)
            \right\rbrace
        \di T.
\end{equation}
That is, the nonlinear mapping from mean postsynaptic activities $V_{a,b}$  to mean membrane
potential $u_a$ is temporally filtered (Ermentrout and Terman
\cite{ermentroutMathematicalFoundationsNeuroscience2010}) and the synaptic weights are state
dependent
\begin{equation}
    \nu_{a,b}(u_b) = \frac{
        u_{a}^{\tx{rev}} - u_{b}
    }{
        \abs{u_{a}^{\tx{rev}} - u_{a}^{\tx{rest}}}
    },
\end{equation}
where $u_{a}^{\tx{rev}}$ is the reversal potential  and $u_{a}^{\tx{rest}}$ is the resting potential.
The synaptic kernel considered is a simple exponential, which gives us
\begin{equation}
\begin{split}
  \eta_a(t) =
    \frac{1}{\tau_a}\exp(-t/\tau_a) \, \Theta(t), \\
  \implies \left( \frac{1}{\tau_a}\ppx{}{t} + 1 \right)u_a(x,t) =
    \sum_b
    \frac{
        u_{b}^{\tx{rev}} - u_{a}
    }{
        \abs{u_{b}^{\tx{rev}} - u_{a}^{\tx{rest}}}
    } \, V_{a,b}(x,t)
    + u_a^0(x,t)
    . \label{eq:liley}
\end{split}
\end{equation}
From Eq.(\ref{eq:liley}), we can see that because conductance-based synapses are linear in both
$u_a$ and $V_{a,b}$, they are bilinear, which introduces another nonlinearity into the equations.
Eq.(\ref{eq:liley}) cannot be written as a linear operator because the sum is nonlinear. This makes
this model fundamentally different to the other models described in this paper. The Robinson model
has been derived with conductance-based synapses and its dynamics directly compared to linear
current-based synapses (Peterson et 
al.~\cite{petersonHomotopicMappingCurrentbased2018}). Although,
this is more biologically realistic, it introduces another layer of complexity and makes the
analysis more difficult. Unlike the model by Robinson et 
al.~\cite{robinsonPropagationStabilityWaves1997}, the Liley model also includes long-range 
excitatory
to inhibitory connections and different PSP's for excitatory and inhibitory populations.
\begin{framed}
    \textbf{Physiological assumption:} In this relatively simple formulation of conductance-based
    synapses, it is assumed that the reversal potentials are static parameters.  More accurate
    models also include additional  extra-cellular dynamics where the reversal potentials are state
    variables computed from extra- and intra-cellular ionic concentrations (Cressman et 
    al.~\cite{cressmanInfluenceSodiumPotassium2009}, Krishnan and Bazhenov
    \cite{krishnanIonicDynamicsMediate2011}).  Ionic dynamics are known to affect overall brain
    dynamics, particularly seizures (Ziburkus et 
    al.~\cite{ziburkusSeizuresImbalancedStates2012}).
\end{framed}

The Liley model was in turn extended by Steyn-Ross et 
al.~\cite{steyn-rossTheoreticalElectroencephalogramStationary1999} to a system of 
stochastic
differential equations that includes gap junctions. Gap junctions are electrical synapses that do
not have chemical neurotransmitters and are known to affect overall brain dynamics, particularly
with respect to neuron coupling and rhythm synchronisation (Larsen
\cite{larsenBiologicalImplicationsGap1983}, Evans and Martin 
\cite{evansGapJunctionsStructure2002},
Traub et al.~\cite{traubHighfrequencyPopulationOscillations1999}, Simon et 
al.~\cite{simonGapJunctionNetworks2014}). Further explorations of the dynamics and 
solutions to these
systems forms an entire discipline of mathematical neuroscience in its own right. Many of these
concepts are discussed and expanded upon throughout stochastic neural field theory, which we will
briefly discuss in Section \ref{S:stochastic-nfms}.

\subsection{Reduction to neural mass models} 
\label{S:neural-mass-models}

It is not always desirable to consider significant spatial extent within a neural field
model. For smaller spatial scales that are relatively isotropic and homogeneous, having spatial
dependence and the incorporation of long-distance spatial connectivity is unnecessary, as this
heavily complicates the analysis of neural population models. The earlier work of Wilson and 
Cowan 
\cite{wilsonExcitatoryInhibitoryInteractions1972} initially maintained this smaller scope in a
spatially-invariant regime, before extending it to an early neural field model in 1973 (considered
in Section \ref{S:wc-derivation}). This earlier 1972 model would come to be known as a ``neural
mass model'' or NMM, which can be considered a special case of a neural field model at a single
point in space. This is mathematically equivalent to using delta distribution connectivity kernels
$w_{a,b}(x) = c_{a,b}\delta(x)$ for each population. Around the same time, NMMs were further
developed and formalised by Lopes da Silva et 
al.~\cite{lopesdasilvaModelBrainRhythmic1974} and
Freeman  \cite{freemanMassActionNervous1975}. NMMs share a concurrent history with 
neural
field models, with many authors publishing significant contributions in both model types. NMMs omit
spatial dependence and spatiotemporal time delays but share the same principles of aggregate neural
action arising from statistical mechanics. While NMMs still maintain a high degree of mathematical
complexity, they provide a simpler realisation of statistical mechanics in neuroscience, and are
crucial to understanding the dynamics of neural populations.

Inspired by the work of Katchalsky  \cite{katchalskyCarriersSpecificityMembranes1971}, 
Katchalscky et al.~\cite{katchalskyDynamicPatternsBrain1974}, Prigogine and Nicolis 
\cite{prigogineFluctuationsMechanismInstabilities1973} and others, it was Walter Freeman Jr who
pioneered what we might term today a ``systems approach'' to the study of neural masses 
(Freeman \cite{freemanMassActionNervous1975}) and from where the term ``neural mass 
model'' derives.
Freeman’s motivation for this approach was essentially to strike a balance between physiological
plausibility whilst remaining computationally tractable, such that the output of the model might be
directly linked to electrophysiological recordings (LFP data in Freeman’s case). Taking
Katchalsky’s extension of Liesegang rings from diffusion-coupled chemical reactions as a basis,
Freeman  \cite{freemanMassActionNervous1975} introduced a topological hierarchy of 
models
styled the $Ki$ formalism. The full detail of Freeman's work is beyond the scope of this section,
but further developments by Freeman led to analysis of neural mass models of the olfactory bulb in
the rabbit primary cortex.

Around the same time as   Freeman's formal establishment of neural mass action, an aligned approach
to studying the macroscale brain activity was proposed by Lopes da Silva et 
al.~\cite{lopesdasilvaModelBrainRhythmic1974} which they termed the ``lumped parameter 
approach''. To
briefly recap this development, Lopes da Silva put forward a computational model of a
two-population neural mass model to investigate spontaneous EEG signal generation and
alpha-frequency rhythms. To analyse this model within a mathematical framework, they reduced the
complexity of the model by deriving a dynamical systems model of aggregate activity within a single
cortical column. Utilising the simplest approach, Lopes da Silva and colleagues investigates
excitatory thalamocortical relay cells (TRCs) that receive external excitatory pulse density (from
neighbouring columns, not explicitly modelled), and inhibitory feedback from local interneurons.
External inputs to this system arise from pulse trains from peripheral excitatory neurons in
neighbouring columns, and are integrated as another input to the TRC population (typically modelled
as a Gaussian process).

\begin{figure}[htbp]
    \centering
    \includegraphics[width=0.98\textwidth]{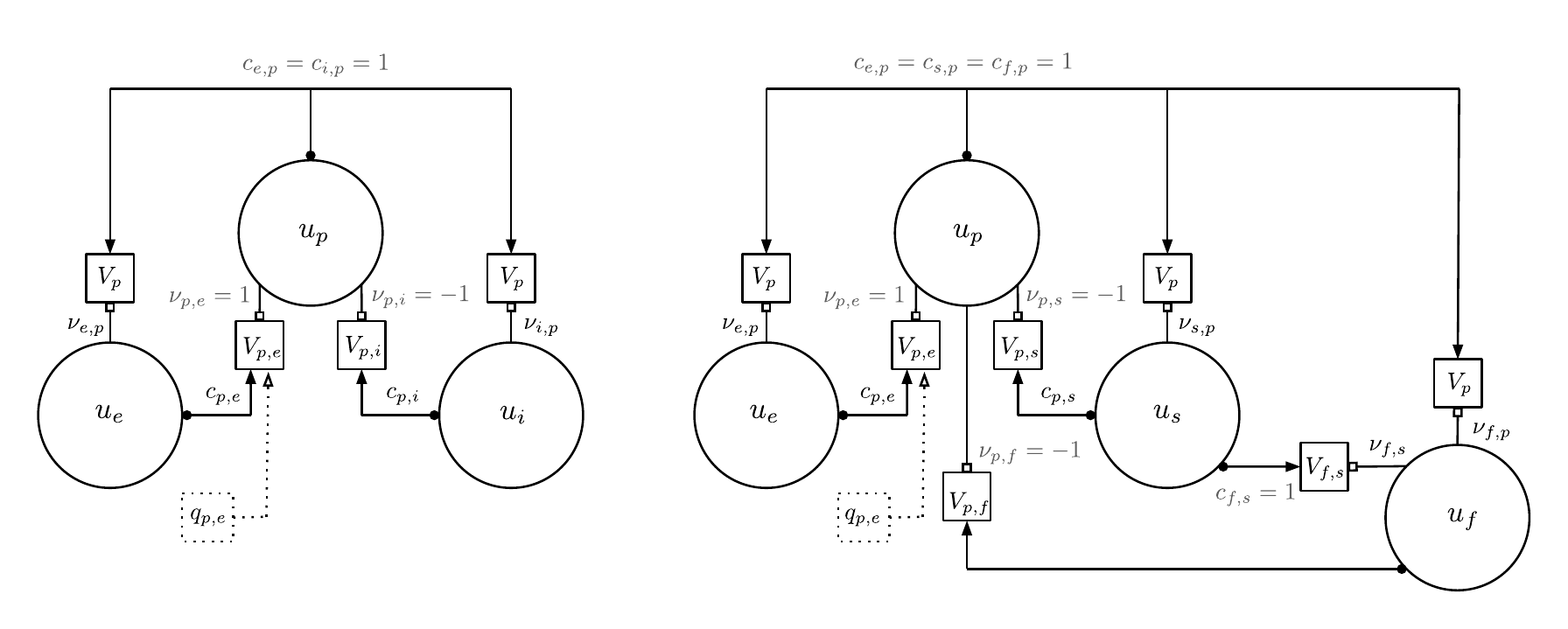}
    \caption{%
        \textbf{Diagram of Jansen-Rit (left) and Wendling (right) neural mass models.} The model by
        Jansen and Rit is an extension of the model by Lopes da Silva, and the model by Wendling is
        an extension of the model by Jansen and Rit. Each subsequent development adds more
        structure to the neural mass model and progressively more assumptions about the
        connectivity and temporal dynamics between neurons in a cortical column, which can be
        interpreted as a point in a neural field. These proposed models normalise certain
        connectivity strengths $\nu_{a,b}$ and $c_{a,b}$, and have been annotated in grey.
    }
    \label{F:jrw-nmm}
\end{figure}

The two-population model proposed by Lopes da Silva et 
al.~\cite{lopesdasilvaModelBrainRhythmic1974} was further developed by Zetterberg et 
al.~\cite{zetterbergPerformanceModelLocal1978} by extending it to a three-population 
model, which
during the resurgence of neural field models in the 1990's, was used by Jansen et 
al.~\cite{jansenNeurophysiologicallybasedMathematicalModel1993}. In their 
1993 
publication, Jansen and colleagues proposed that the mechanisms guiding the evolution of 
evoked potentials and spontaneous 
EEG
signal generation are the same. To show this, they constructed a model of pyramidal neurons in
cortical columns in the same fashion as done by Lopes da Silva and colleagues. The post-synaptic
potentials of the pyramidal population is identified as the EEG signal and also provides input to
surrounding interneurons. The model by Lopes da Silva utilises a second population of inhibitory
interneurons that provide feedback to the pyramidal neuron population. Jansen and Rit extend this
and include a third population of local excitatory interneurons that also provide feedback to the
pyramidal neurons. A diagram of the populations in this neural mass model is given in Figure
\ref{F:jrw-nmm}. The addition of this third population is motivated by incorporating additional
physiological realism of (long-range) excitatory connections. To support the hypothesis of the same
governing mechanisms, Jansen and Rit replaced the random input term with transient pulses and
reviewed the dynamical model response and found concordance with experimental observations.

Neural mass models can also be well understood within the mathematical framework we
derived in Section \ref{S:unifying-nfm-framwork}. Here we will show how the Jansen and Rit
equations can be derived using a general neural field model equation. Since this is a neural mass
model, we will immediately make the substitution $w_{a,b}(x) = c_{a,b} \delta(x)$, where $c_{a,b}$
is a connectivity constant and $\delta$ is the Dirac delta distribution. This removes spatial
dependence entirely, so we may omit the $x-$ordinate and express our neural mass model as
\begin{equation}
  V_{a,b}(t) =
    \int_{-\infty}^t
      \psi_{a,b}(t-T)
      \left[
        c_{a,b} \, f
        \left(
          \sum_c \nu_{b,c} V_{b,c}(T)
        \right)
        + q_{a,b}(T)
      \right]
    \di T,
\end{equation}
where it is also assumed that each population uses the same potential to rate transfer function
$f_b = f$ given by
\begin{equation}
  f(u) = \frac{2 e_0}{1+\exp[-r \, (u-v_0)]},
\end{equation}
where $e_0$ determines the maximal firing rate, $v_0$ is the post-synaptic potential for which 50\%
of maximal firing rate is achieved, and $r$ is the steepness parameter of the transfer function. A
minor deviation is made in this model from the framework we established when the external input is
concerned, though this is primarily a notational issue rather than conceptual. Jansen and Rit do
not use the mapping $u_a^0 = \sum_b \nu_{a,b} V_{a,b}^0$, instead setting $u_a^0 = 0$ for all $a$
and  considering only external inputs at the post-synaptic potential level. They consider three
populations: a pyramidal population $(p)$, an excitatory population $(e)$, and an inhibitory
population $(i)$. In the same vein as the model by Lopes da Silva et al., Jansen and Rit impose
that the only non-zero connections are from $p$ to $e$ and $i$, from $e$ to $p$, and from $i$ to
$p$. This can be seen in Figure \ref{F:jrw-nmm}. The lack of self-connections in this branch of
models can be attributed to Lopes da Silva intending to derive the simplest possible model for
spontaneous alpha EEG rhythms. Additionally, the temporal filtering of signals  is assumed to be
specific to the afferent population,  giving the notational simplification $\psi_{a,b}(t) =
\psi_b(t)$. Furthermore, it is also assumed that the temporal kernel is of the form of an alpha
function,
\begin{equation}
  \psi_{b}(t) = \frac{\Gamma_{b}}{\mu_{b}} \, t \exp(-t/\mu_b) \, \Theta(t),
\end{equation}
where $\Gamma_{b}$ is a gain parameter, $\mu_{b}$ denotes the time scale, and $\Theta$ is the usual
Heaviside function. Different time-scales are implemented for excitatory and inhibitory signals, as
there is physiological evidence for different synaptic timescales and gain parameters for common
excitatory and inhibitory neurotransmitters. That is, temporal kernels for post-synaptic potentials
are set to be
\begin{equation}
  \psi_{e}(t) =
    \psi_{p}(t) =
    \frac{\Gamma_e}{\mu_e}  \, t \exp(- t/\mu_e) \, \Theta(t),
    \qquad
  \psi_{i}(t) =
    \frac{\Gamma_i}{\mu_i} \, t \exp(- t/\mu_i) \, \Theta(t),
\end{equation}
where $(\Gamma_e, \mu_e)$ and $(\Gamma_i,\mu_i)$ are the synaptic gain and time constants of the
excitatory and inhibitory synapses, respectively. The physiological manifestation for the
difference in excitatory and inhibitory post-synaptic potential amplitudes and timescales arise
from the position of afferent synapses relative to the neuron's cell body. For instance, inhibitory
neurons are more likely to synapse near the cell body, producing larger potentials (Martin
\cite{martinCorticalNeuronsEEG1985}). This is accounted for by taking $\Gamma_e<\Gamma_i$. We can
equivalently express this system in terms of temporal differential operators. Using Fourier
transforms, it can be shown that the temporal differential operators corresponding to the
excitatory and inhibitory synaptic kernels are
\begin{equation}
  \mathbf{D}_t^{(e)} =
    \mathbf{D}_t^{(p)} =
    \frac{\mu_e}{\Gamma_e}
    \left[
      \ddx[2]{}{t} + \frac{2}{\mu_e} \ddx{}{t} + \frac{1}{\mu_e^2}
    \right],
    \qquad
  \mathbf{D}_t^{(i)} =
    \frac{\mu_i}{\Gamma_i}
    \left[
      \ddx[2]{}{t} + \frac{2}{\mu_i} \ddx{}{t} + \frac{1}{\mu_i^2}
    \right].
\end{equation}
Furthermore, we only consider external input to this system via the excitatory to pyramidal
synaptic connections, so we will set $q_{a,b}(t) = 0$ for all populations with exception to
$q_{p,e}$. Inverting the temporal kernel for each population and applying the connectivity between
these populations as proposed by Jansen and Rit gives rise to the following equations,
\begin{equation}
  \begin{split}
    \mathbf{D}_t^{(p)} \lbrace V_{e,p}(t) \rbrace &=
      c_{e,p} f
      \left( u_p \right), \\
    \mathbf{D}_t^{(p)} \lbrace V_{i,p}(t) \rbrace &=
      c_{i,p} f
      \left( u_p \right), \\
    \mathbf{D}_t^{(e)} \lbrace V_{p,e}(t) \rbrace &=
      c_{p,e} f
      \left( u_e \right)
      + q_{p,e}(t), \\
    \mathbf{D}_t^{(i)} \lbrace V_{p,i}(t) \rbrace &=
      c_{p,i} f
      \left( u_i
      \right),
  \end{split}
  \qquad
  \begin{split}
    u_p &=
      \nu_{p,e} V_{p,e}(t)
      + \nu_{p,i} V_{p,i}(t), \\
    \qquad&\\
    u_e &=
      \nu_{e,p} V_{e,p}(t), \\
    u_i &=
      \nu_{i,p} V_{i,p}(t).
  \end{split}
\end{equation}

\begin{framed}
    \textbf{Anatomical assumption:} It is assumed that the magnitude of inhibitory impulses is
    larger than  the magnitude of the excitatory impulses. This assumption and subsequent
    temporal kernels are inherited from van Rotterdam et 
    al.~\cite{vanrotterdamModelSpatialtemporalCharacteristics1982}. The difference in 
    amplitude was
    justified by considering that inhibitory neurons typically have axonal terminals that lie
    closer to the cell body, which results in larger perturbations upon inhibitory synaptic
    transmission (Kandel and Schwartz \cite{kandelPrinciplesNeuralScience1985}). 
    Additionally, it is also
    assumed that excitatory impulses occur on a faster timescale than inhibitory impulses.
\end{framed}
Jansen and Rit additionally assume that pyramidal neurons synapse onto both the excitatory and
inhibitory populations equally, giving us $c_{e,p} = c_{i,p}$. These constants are set to 1 for
simplicity. We can see that this assumption makes the second equation redundant, as we now have
$V_{e,p} = V_{i,p} = V_{p}$. Furthermore, the synaptic coefficients at the pyramidal dendrites are
set to $\nu_{p,e} = 1$ and $\nu_{p,i} = -1$. Upon making these substitutions and expanding the
operator on the left hand side gives rise to the familiar Jansen and Rit neural mass model seen
throughout the literature,
\begin{equation}
\begin{split}
  \ddot{V}_{p}(t) &=
    A a \, f
    \left(
      V_{p,e}(t) - V_{p,i}(t)
    \right)
    - 2a \, \dot{V}_{p}(t) - a^2 \, V_{p}(t)
    , \\
  \ddot{V}_{p,e}(t) &=
    A a \,
    \left[
      c_{p,e} f
      \left(
        \nu_{e,p} \, V_{p}(t)
      \right)
      + q_{p,e}(t)
    \right]
    - 2a \, \dot{V}_{p,e}(t) - a^2 \, V_{p,e}(t)
    , \\
  \ddot{V}_{p,i}(t) &=
    B b \, c_{p,i} f
    \left(
      \nu_{i,p}  V_{p}(t)
    \right)
    - 2b \, \dot{V}_{p,i}(t) - b^2 \, V_{p,i}(t)
    ,
\end{split}
\end{equation}
where $(A, B) = (\Gamma_e, \Gamma_i)$ and $(a,b) = (\mu_e^{-1}, \mu_i^{-1})$. A follow-up
publication by Jansen and Rit in 1995 also investigates a double cortical column model, which
effectively couples two copies of the above dynamical system together. However, no substantial
differences in model output was found, aside from an extension of time delays that give rise to a
slow-wave component in EEG simulation. An extended discussion on the connectivity constants
$c_{a,b}$ is given by Jansen and Rit  \cite{jansenElectroencephalogramVisualEvoked1995}, 
providing
more insight on the physiological and anatomical assumptions used to set the values of these
parameters. The typical values taken here replace each connection with a scaled global connectivity
parameter $C$, so the final version of this model sets $c_{p,e} = C, \nu_{e,p} = 0.8C, \nu_{i,p} =
0.25C, c_{i,p} = 0.25C$. The output of this neural mass model is given by the membrane potential of
the pyramidal population, $u_p(t) = V_{p,e} - V_{p,i}$, which is considered as  the primary
component underlying EEG signals.

\begin{framed}
    \textbf{Physiological assumption:} The output of this and the Wendling model assumes that the
    primary generator of the EEG signal is post-synaptic input to the pyramidal population, given
    by the sum of the excitatory and inhibitory post-synaptic potentials. This is a common
    convention when linking neural field models to EEG data, but it is putative and not used across
    all neural field models. For instance, the models by Liley and Steyn-Ross associate the
    excitatory post-synaptic potential amplitudes with an EEG signal whereas the models by Robinson
    et al. use the excitatory population activity. It is typically assumed that an EEG electrode
    measurement corresponds to the average membrane potential of the excitatory pyramidal
    population, though it is also argued that inhibitory populations also play a role (Nunez and 
    Srinivasan
    \cite{nunezElectricFieldsBrain2006}). From an electromagnetic perspective, the primary
    generators of the EEG are not explicitly known, and the largest microscopic measurements we
    have are of local field potentials, which are still several orders of magnitude away from what
    is being measured by the EEG. Hence, from a modelling perspective we cannot currently
    experimentally validate any assumptions about the primary generators of the EEG. Given, that we
    have no way of testing any of these assumptions, or if any of the models themselves correctly
    describe the large scale behaviour of populations of neurons, the choice of model output that
    corresponds to the EEG is somewhat arbitrary.
\end{framed}

The model proposed by Jansen and Rit successfully built on the work  Lopes da Silva and colleagues
and gave support for leading hypotheses on generation of spontaneous EEG signals and 
response to
evoked potentials. However, this model does not appear to replicate high frequency phenomena, which
is problematic for investigations in neuropathology. The temporal kernel can behave similarly as a
low-pass filter, which can inadvertently remove high-frequency phenomena. To challenge this, an
extension by Wendling et al.~\cite{wendlingEpilepticFastActivity2002},
physiologically motivated the addition of another inhibitory population into the framework provided
by Jansen and Rit in order to replicate the high-gamma activity seen in EEGs in subjects with
epilepsy. The Wendling model utilises the same pyramidal and excitatory populations as Jansen and
Rit, but asserts the incorporation of a slow inhibitory population and a fast inhibitory
population. The pyramidal population projects to each of the three peripheral populations, which
then each feed back into the pyramidal population. In addition to this, the slow inhibitory
population is hypothesised to down regulate the fast inhibitory population.

This extension done by Wendling et al. is motivated by links between
different inhibitory currents and the emergence of delta and theta rhythms in EEG (White et 
al.
\cite{whiteNetworksInterneuronsFast2000}) as well as the link between gamma oscillations in EEG
arising from the behaviour of inhibitory interneurons (Whittington et 
al.~\cite{whittingtonInhibitionbasedRhythmsExperimental2000}, Jefferys et 
al.~\cite{jefferysNeuronalNetworksInduced1996}). Wendling et al. also refer to the study by 
Miles et al.~\cite{milesDifferencesSomaticDendritic1996}, which highlight distinct fast and 
slow
responses to the inhibitory neurotransmitter GABA-A in CA1 pyramidal neurons. Specifically, the
neural mass model incorporates an additional level of detail on axonal projections of inhibitory
interneurons in cortical columns through the choice of different temporal kernels. The
inhibitory synapses that lie at the end of inhibitory interneuron axons can project onto the
dendrites of their target neurons, or onto the cell body (soma). Where these synapses project to
can greatly influence the way these inhibitory signals are integrated. For instance, a
somatic-projecting inhibitory signal will have a larger apparent magnitude and faster timescale
relative to a dendritic-projecting inhibitory signal when integrated at the axon hillock of the
target neuron due to the spatial origin of these signals. Experimental studies by Dichter 
\cite{dichterBasicMechanismsEpilepsy1997} and Cossart et 
al.~\cite{cossartDendriticNotSomatic2001}
highlight the importance of this distinction when investigating epilepsy. These aspects can
be included by expanding the number of neural populations and corresponding temporal kernels with
different parameters.

The Wendling neural mass model can be derived  using the modern neural field framework in a similar
fashion to the Jansen and Rit neural mass model, using the same delta distribution spatial kernels
$w_{a,b}(x) = c_{a,b} \delta(x)$, and alpha function temporal kernels
\begin{equation}
\begin{split}
  V_{a,b}(t) =
    \int_{-\infty}^t
      &\psi_{b}(t-T)
      \left[
        c_{a,b} \, f
        \left(
          \sum_c \nu_{b,c} \, V_{b,c}(T)
        \right)
        + q_{a,b}(T)
      \right]
    \di T, \\
  \psi_{b}(t) &= \frac{\Gamma_{b}}{\mu_{b}} \, t \exp(-t/\mu_b) \, \Theta(t).
\end{split}
\end{equation}
Instead, we consider 4 populations: a pyramidal population ($p$), an excitatory population ($e$), a
slow inhibitory population ($s$), and a fast inhibitory population ($f$), with the temporal kernels
\begin{equation}
\begin{split}
  \psi_{e}(t) &=
    \psi_{p}(t) =
        \frac{\Gamma_e}{\mu_e} \, t \exp(-t/\mu_e) \, \Theta(t),
    \\
  \psi_{s}(t) &=
        \frac{\Gamma_s}{\mu_s} \, t \exp(-t/\mu_s) \, \Theta(t),
    \\
  \psi_{f}(t) &=
        \frac{\Gamma_f}{\mu_f} \, t \exp(-t/\mu_f) \, \Theta(t).
\end{split}
\end{equation}
The connectivity between these populations follows the same arrangement as laid out by the Jansen
and Rit neural mass model: pyramidal neurons $p$ project onto the peripheral interneuron
populations $e,s,$ and $f$ and each of these populations feedback onto the pyramidal population
$p$. Crucially, Wendling et al. also refer to the study by Banks et 
al.~\cite{banksInteractionsDistinctGABAA2000}, which suggests that the slow inhibitory 
population $s$
also projects onto the fast inhibitory population $f$. The population connectivity is
diagrammatically represented in Figure \ref{F:jrw-nmm}. The Wendling model then `inherits' the
simplifications made by Jansen and Rit:
\begin{itemize}
    \item External input is only considered in the excitatory $e$ to pyramidal $p$ connections 
    as
        before, so we set $q_{a,b}(t) = 0$ for all $(a,b) \neq (p,e)$,
    \item The connectivity constants projected from the pyramidal population $p$ are set $c_{.,p} =
        1$ for simplicity, making separate equations for different pyramidal post-synaptic
        potentials redundant $V_{.,p}(t) = V_p(t)$,
    \item The pyramidal synaptic constants are set to $\nu_{p,e} = 1, \nu_{p,s} = -1,$ and
        $\nu_{p,f} = -1$,
    \item The connectivity constant between the inhibitory populations is set to $c_{f,s} = 1$.
\end{itemize}
From here, the procedure to derive the Wendling model ODEs is identical to the procedure we used
for the Jansen and Rit model: we invert the temporal integral operators as linear differential
operators and rearrange the equation to obtain
\begin{equation}
\begin{split}
  \ddot{V}_{p}(t) &=
    A a \, f
    \left[
      V_{p,e}(t)
      - V_{p,s}(t)
      - V_{p,f}(t)
    \right], \\
  \ddot{V}_{p,e}(t) &=
    A a \,
    \lbrace
      c_{p,e} f
      \left[
        \nu_{e,p} V_{e,p}(t)
      \right]
      + q_{p,e}(t)
    \rbrace
    - 2a \, \dot{V}_{p,e}(t) - a^2 \, V_{p,e}(t)
    , \\
  \ddot{V}_{p,s}(t) &=
    B b \, c_{p,s} f
    \left[
      \nu_{s,p} V_{s,p}(t)
    \right]
    - 2b \, \dot{V}_{p,s}(t) - b^2 \, V_{p,s}(t)
    , \\
  \ddot{V}_{p,f}(t) &=
    G g \, c_{p,f} f
    \left[
      \nu_{f,p} V_{f,p}(t) - \nu_{f,s} V_{f,s}(t)
    \right]
    - 2g \, \dot{V}_{p,f}(t) - g^2 \, V_{p,f}(t)
    , \\
  \ddot{V}_{f,s}(t) &=
    B b \, f
    \left[
      \nu_{s,p} V_{s,p}(t)
    \right]
    - 2b \, \dot{V}_{f,s}(t) - b^2 \, V_{f,s}(t)
    ,
\end{split}
\end{equation}
where $(A,B,G) = (\Gamma_e, \Gamma_s, \Gamma_f)$ and $(a,b,g) = (\mu_e^{-1}, \mu_s^{-1},
\mu_f^{-1})$. As with the preceding Jansen and Rit model, the mean membrane potential of the
pyramidal population $u_p(t) = V_{p,e}(t) - V_{p,s}(t) - V_{p,f}(t)$ is interpreted as the model
output, corresponding to the EEG signal. The connectivity parameters are handled in a similar
fashion to the Jansen and Rit model, where the connectivity is set to be proportional to a global
coupling parameter $C$ $\nu_{e,p} = C, c_{p,e} = 0.8C, \nu_{s,p} = 0.25C, c_{p,s} = 0.25C, $
$\nu_{f,p} = 0.3C, \nu_{f,s} = -0.1C, c_{p,f} = 0.8C$.

From perturbing the synaptic gain parameters $A,B,$ and $G$, six primary rhythms are identified by
Wendling et al.~\cite{wendlingEpilepticFastActivity2002}. A key conclusion from this model 
was the impact of
dendritic inhibition in epilepsy. That is, the model proposed that a key explanation for the
transition from `non-seizure' to `seizure' activity can be explained by the impairment of dendritic
inhibition. This neural mass model has thus become the focus of a substantial number of
publications in mathematical neuroscience and EEG modelling regarding epilepsy, which we will
review in Section \ref{S:applications} of this paper.

The neural field models that transpired from the second wave of neural field modelling in the
1990's continued to refine concepts and systems proposed in canonical publications from Wilson and
Cowan, Amari, Freeman, Lopes da Silva, Van Rotterdam, and Nunez. The work of Robinson, Liley, and
Steyn-Ross incorporated contemporary findings from neuroscience and proposed new explanations for
aggregate neural phenomena across brain regions by using mathematical concepts such as time delays.
The models proposed by these authors expanded on the framework used by Amari, and led to new
theories on the generation of rhythms in EEG signals and proposed and supported hypothesised
mechanisms for seizure generation and spread throughout the cortex. In tandem to this branch of
models, spatially invariant neural mass models also saw significant development throughout this
period with the work of Jansen, Rit, and Wendling. While these models focused on modelling smaller
regions of cortical tissue, they avoid difficulties in justifying spatially homogeneous kernels
that have been in use since the canonical field model by Wilson and Cowan. Modelling a single point
of cortical tissue as opposed to a continuum allowed for more efficient computation of results and
a more concentrated investigation on the aggregate circuitry required in cortical columns required
to sustain critical EEG rhythms in response to a random or deterministic input. These theoretical
developments were essential steps in realising neural field models as we recognise them today, and
provided the roots for contemporary approaches in modelling neural tissue and making clinical
inference from mathematical and computational methods.

\subsection{Contemporary clinical applications of neural fields} 
\label{S:applications}

Electroencephalography (EEG) is an essential tool in neuroscience used to monitor and assess
aggregate brain activity, and frequently used in clinical neurology for a variety of pathologies.
This technique involves using an array of electrodes  to detect and record neural activity via
fluctuations in the electrical potential on the order of micro-volts (Smith
\cite{smithEEGDiagnosisClassification2005}). It is one of the oldest modalities in neuroscience
with high millisecond temporal resolution  at the cost of low spatial resolution. Specific
questions where EEG is used include  investigations of the origin of common activity patterns in
neural tissue, in addition to a variety of clinical applications. EEG plays an important clinical
role  in the diagnosis, localisation, and management of epilepsy syndromes. Crucially, it can be
used to assess  emergence of slower rhythms and unprovoked discharges, appearing as ``spikes'' and
``waves'' in the signal, which can be indicative of epilepsy (Smith
\cite{smithEEGDiagnosisClassification2005}, St. Louis et 
al.~\cite{st.louisElectroencephalographyEEGIntroductory2016}). It is also an essential tool in
examining brain activity during sleep  and transitions in activity during the progression of the
sleep cycle to diagnose and treat sleep disorders. Additionally, it is an important tool in
monitoring the progression of dementia,  assessing severity of tumours and strokes,  as well as
examining the depth of  anaesthesia and coma St. Louis et 
al.~\cite{st.louisElectroencephalographyEEGIntroductory2016}.

Quasiperiodic waves in the EEG are fundamental  constituents of the recorded signal  and are used
to infer neurological conditions. Historically, this has been done using visual inspection  of the
EEG record,  but software tools, such as Persyst (Guillen et 
al.~\cite{guillenPerSystMonitoringTool2014}), can reveal rhythms through routine Fourier 
analysis.
These brain rhythms are partitioned by frequency ranges and are \textit{not} grouped by function.
Certain activities and levels of conscious activity are correlated with different frequency bands,
but it should be emphasised that our physiological understanding of the origin of these rhythms is
an ongoing topic of research. Slow rhythms include $\delta$ waves (1 -- 3 Hz) and $\theta$ (4 -- 7
Hz) waves, which tend to be prevalent in drowsy subjects. Higher frequency ranges are partitioned
into  $\alpha$ waves (8 -- 12 Hz) and $\beta$ waves (13 -- 25 Hz), and $\gamma$ waves (26 -- 100
Hz). The $\alpha$ rhythm is essential feature in every EEG report,  as healthy adults at rest with
eyes closed tend to display a low-amplitude mixed-frequency background  with a dominant $\alpha$
rhythm. Excessive $\beta$ rhythms are usually caused by drowsiness or sedative drugs. The presence
of intermittent slower rhythms in $\delta$ and $\theta$ bandwidths  can also arise from low states
of arousal (also common in children and elderly),  however they can also be indicative of an
epilepsy syndrome or progression of dementia (Medithe and Nelakuditi 
\cite{meditheStudyNormalAbnormal2016},
Tatum \cite{tatumHandbookEEGInterpretation2021}, Emmady and Anilkumar
\cite{emmadyEEGAbnormalWaveforms2021}). Rhythms of higher frequency such as $\gamma$ waves (26 --
100Hz) and ripples (over 100 Hz) are not common in scalp EEG,  but thought to reflect epileptiform
discharges. The compositions of rhythms observed on an EEG are unique to each individual, and vary
based on a variety of factors such as drowsiness, pharmacology, and neuropathologies.

The activity and rhythms recorded on an EEG are believed to arise from variations in
population-level activity in neural tissue, therefore neural field models provide a mathematical
framework to identify mechanisms for the emergence of rhythms. Investigations of rhythmogenesis
with neural field models appear at least as early as 1974 with the work of
Lopes da Silva et al.~\cite{lopesdasilvaModelBrainRhythmic1974}, which proposed a two 
population neural mass model
alongside a typical neural network simulation. The output of the neural field model used resembles
the $\alpha$-rhythm on an EEG, providing a relatively simple physiological hypothesis for the
origin of neurological rhythms as a consequence of neural population arrangements and activity in
the cortex. As models of neural fields matured, the questions and investigations of rhythmogenesis
became more prevalent. The work by Liley et al.~\cite{lileyContinuumModelMammalian1997} 
shows how neural field
approaches were refined over the decades and provided progressively more detailed hypotheses
between the simulations and the underlying neurophysiology. Rhythms in neural fields are determined
by the differences in timescales in difference postsynaptic temporal kernels, but the external
input term driving many of these models is also responsible for key aspects of rhythmic output of
these models. External inputs to neural field models tend to be simple white noise inputs to an
excitatory population, which represents neural activity not explicitly modelled. Although a white
noise process is the simplest process to use for this purpose, a study by
Freyer et al.~\cite{freyerBiophysicalMechanismsMultistability2011} identifies that using a 
more elaborate
stochastic process with state-dependence and feedback produces a more physiologically accurate
power spectrum. Pharmokinetic influence, among other physiological states, are interpreted as
different sets of parameter values for neural field models throughout these studies, which provides
methods of generating theoretically motivated hypotheses for the prevalence, emergence, and
persistence of rhythms. This is an essential step in understanding the dynamics of these types of
models, and leads onto crucial questions of stability of rhythms and state transitions, such as
regular dynamical changes observed during sleep as well as pathological rhythms of epilepsy.
Rhythmogenesis, both endogenous such as sleep or pathological such as seizures is a
fundamental part of neuroscience that neural field models can be used to investigate.

Neural field models can also be used to investigate mechanisms of regular activity state
transitions in rhythms that occur during sleep.
Despite being widely studied, there is relatively little known about the neurophysiological
processes involved during the sleep cycle. Sleep can broadly be partitioned into three stages:
Waking stage,  the Non Rapid Eye-Movement (NREM) stage, and the Rapid Eye-Movement (REM) stage. The
NREM stage comprises approximately 80\% of the duration of sleep, and can be distinguished on an
EEG by the presence of  slow oscillations, increased power in $\delta$ frequencies, and distinct
bursts of activity such as ``spindles'' and ``K-complexes''
(Purves et al.~\cite{purvesStagesSleep2001}, Nayak and Anilkumar 
\cite{nayakEEGNormalSleep2021}). The REM stage,  sometimes referred to as
paradoxical sleep,  is the deepest stage of sleep with little or no arousal in subjects  and
commonly associated with dreaming. EEG activity recorded during this stage resembles the higher
frequency $\beta$ and $\gamma$ patterns  seen during the waking phase
(Purves et al.~\cite{purvesStagesSleep2001}, Al Sawaf et 
al.~\cite{alsawafEEGBasalCortical2021}). During sleep,  brain activity alternates
between NREM and REM sleep in several cycles before waking,  exhibiting discernible state
transitions on an EEG (Nayak and Anilkumar \cite{nayakEEGNormalSleep2021}) . Neural field 
models provide a mathematical
route in which  these transitions can be modelled and studied in terms of  the potential migration
of parameters governing mean quantities of neural tissue.

Events that occur during the sleep cycle, such as sleep spindles and K-complexes, provide useful
targets for modelling non-linearities and state transitions on EEG. Subjects are typically
motionless during sleep, which results in EEG recordings with less extraneous noise compared to an
EEG recording during wakefulness. The thalmocortical neural field model summarised in Section
\ref{S:modern-nfms-derivation} has also been used to investigate rhythms in sleep. A publication by
Abeysuriya et al.~\cite{abeysuriyaPredictionVerificationNonlinear2014} finds a region of 
parameter
space from which the model output resembles sleep spindles. They suggest that these
electrophysiological events arise from the thalamic relay nuclei, instead of the reticular nucleus
and predict that the power in the harmonic of the spindle scales with the oscillation of the
spindle. Alongside this, other investigations by Weigenand et 
al.~\cite{weigenandCharacterizationKComplexesSlow2014} and Costa et 
al.~\cite{costaModelingEffectSleep2016} derive a two-population cortical neural mass model 
with
conductance-based synapses to investigate transitions between sleep stages. They find a
mathematical formulation that replicates patterns and transitions seen on sleep EEGs which is
modulated by the neurotransmitter input to the excitatory pyramidal population. Their model was
spatially extended to a four-population cortico-thalamic neural mass model by Costa et 
al.~\cite{costaThalamocorticalNeuralMass2016}, which was verified by comparing 
perturbations in the
model to auditory stimulation of subjects during sleep.

Rhythms observed during an EEG recording can be used to inform the depth of sleep and level of
consciousness of a subject during a sleep study. This is also a critical component of clinical
evaluation when assessing and diagnosing brain death or assessing a comatose state. Additionally,
rhythms on EEG can also be used to assess the depth of anaesthesia, which must be carefully
monitored before and during a surgery by highly trained professionals. The development of neural
field models with non-linear conductance-based temporal dynamics by Liley et al. and Steyn-Ross et
al. were highly motivated by this particular application and used to investigate the phase
transitions that occur in cortical activity. General anaesthetics, such as isoflurane and
benzodiazipines, generally increase the level of inhibition in neural systems. The effects of which
can be investigated through bifurcation analysis of a neural field model.
Steyn-Ross et al.~\cite{steyn-rossTheoreticalElectroencephalogramStationary1999} 
investigate the action of
modulation of inhibition in a conductance based neural field model with spatial homogeneity and
find three distinct states that correspond to different EEG morphologies than can be related 
to the
level of anaesthetic action. As inhibition decreases, the models transitions from a ``comatose''
state, through an ``active'' state, and eventually reaching a hyperactive ``seizure'' state,
relating to several studies of inhibitory-dominated regimes in microscale neural network models
(Brunel \cite{brunelDynamicsSparselyConnected2000}). Their results predict a rapid transit 
between comatose
states and active states at a critical value of inhibition, and they propose a multistability
hypothesis for the switching between these two states. They find concordance between the behaviour
of clinical EEGs and model output, where the frequency spectrum peaks before collapsing during the
transition from active to comatose. A later publication by
Bojak and Liley  \cite{bojakModelingEffectsAnesthesia2005} alternatively proposes a 
bifurcation mechanisms, instead
of multistability, for the transitions observed during anaesthetic intervention. The competing
hypotheses of multistability vs. bifurcation is common to investigations of epilepsy as well, and
remains highly contested. In this study, spatial homogeneity is not assumed and they investigate
the output of over 70,000 parameter configurations in order to try map the model parameter space to
physiologically plausible regions. Investigations of the connection between clinical data and the
parameter space are essential to one day utilising these models for directing 
anaesthetic dosages in
clinical practice and further understanding these neurological mechanisms.

Another neurological phenomenon involving state transitions is found in epilepsy, conceptualised as
changes between a ``seizure'' state and a ``non-seizure'' state. Epilepsy is another extensive
sub-class of intermittent brain pathologies which encapsulates a variety of neurological syndromes
characterised by recurrent seizures, which can manifest as uncontrolled disturbances to movement,
consciousness, perception and/or behaviour. The type of epilepsy is often classified according to
the seizures that occur,  which in turn indicate which parts of the brain and neural pathways are
affected. Epilepsy encompasses a wide variety of different brain disorders that result in recurrent
and/or unprovoked seizures. However, like many clinical sciences,  there is often much dispute over
the terminology and classification of these syndromes that spans  decades of discussions across a
wide variety of international neurology groups. The pathologies underlying epilepsy are not always
clear, but the syndrome appears to be related to abnormalities in brain connectivity. Different
people with epilepsy often present different symptoms, levels of severity, and responses to
anti-epileptic drugs. Hence, this condition must be treated on an individual case approach instead
of a one-size-fits-all strategy,  which requires regular consultation with neurologists. Diagnosis
of epilepsy from EEG is often  a precarious clinical decision making process. A false positive can
result in mis-prescription of  anti-epilepsy drugs with side effects, in addition to impacts on
anxiety and overall mental well-being. On the other hand, a false negative may lead to  an
untreated epilepsy patient having a seizure without warning  during daily activities, which can
lead to serious injury or death. Severe cases of epilepsy that cannot be controlled with medication
often require highly invasive surgical intervention, which involves mapping of epileptogenic
sections of cortex using intracranial EEG and stereotactic EEG. These discussions within the
neurological sciences encourages the development of novel approaches to data analysis and
mathematical modelling  in order to achieve subject-precision treatments and prognosis and to
further develop our understanding of state transitions in brain dynamics.

Phase transitions in neural field models have been applied to epilepsy since the 1970s with the
works of Kaczmarek  \cite{kaczmarekModelCellFiring1976} and
Kaczmarek and Babloyantz  \cite{kaczmarekSpatiotemporalPatternsEpileptic1977}, which 
identified bifurcation points that give
rise to a transition to oscillatory activity in their model. Bifurcation theory continues to be
used to explore potential physiological causes behind epileptic state 
transitions. An illustration
of how these concepts are broadly applied is seen in the work of
Lopes da Silva  \cite{lopesdasilvaEpilepsiesDynamicalDiseases2003}, which describes the 
transition between
non-seizure (interictal, steady-state with intrisic fluctuations) dynamics and seizure (ictal,
synchronous oscillations) dynamics via 3 mechanisms acting on a bistability in a neural field model
with comparison to data from subjects with photosensitive epilepsies. A study by
Breakspear et al.~\cite{breakspearUnifyingExplanationPrimary2006} does a thorough 
bifurcation analysis on the
cortico-thalamic neural field model by Robinson et al. and identify regions of the parameter space
where patterns resembling generalised seizures arise. The choice of bifurcation parameter was
motivated by physiological hypotheses, which asserted that the excitatory influence of cortical
pyramidal neurons on the specific thalamic nuclei was responsible for seizure activity
(McCormick and Contreras \cite{mccormickCellularNetworkBases2001}). Much of the 
pathological activity seen in the
cortico-thalamic neural field model by Robinson et al. arises from the time delay between the
cortical populations and the thalamic populations. Marten et 
al.~\cite{martenOnsetPolyspikeComplexes2009}
investigates physiological rationales for the addition of a time delay and identified the
difference in timescales of GABA-A and GABA-B neurotransmitter receptors as the likely explanation.
A variety of dynamical characteristics, such as Hopf bifurcations and false bifurcations, were
identified in physiologically plausible regions of the model and thus proposes more potential
mechanisms for polyspike complexes observed on an epileptiform EEG recording. In a following
publication, Marten et al.~\cite{martenDerivationAnalysisOrdinary2009} convert this system 
of delayed
differential equations to a system of ordinary differential equations by using the `linear chain
trick' (Smith \cite{smithDistributedDelayEquations2011}), which produces equivalent 
dyanmics to the
original cortico-thalamic neural field model.

The neural mass model by Wendling has also been used to study the generation of polyspike events in
EEG recordings. A bifurcation analysis by Blenkinsop et al.  
\cite{blenkinsopDynamicEvolutionFocalonset2012}
connected specific clinical features in EEG recordings to dynamical features present in the
Wendling neural mass model. Contemporary results on generalised epilepies suggest that temporal
evolution of seizures may arise from gradual changes in underlying physiological mechanisms, which
bifurcation theory is well suited for investigating. A number of Hopf bifurcations, false
bifurcations, SNICs and eigenstructure changes are identified in the neural mass model of Wendling
et al., which result in the onset of polyspike complexes as observed in EEGs of subecjects with
epilepsy. These bifurcations are related to specific features of clinical EEG data by comparison
with locally stationary segments of patient EEGs. It is imperative to note that a number of
different pathways produce the same transitions to epileptic activity, which is critical to be
mindful of when interpreting results from neural field modelling for directing clinical strategies
for epilepsy and other neurological conditions.

An ongoing research effort in neural field modelling attempts to improve our understanding of
macroscale dynamics by investigating connections with data and models on the microscale. For
example, a study by Eissa et al.~\cite{eissaCrossscaleEffectsNeural2017} investigates the
connections between different spatial scales upon the onset of seizure activity. As simultaneous
recording techniques improve in accuracy and reliability, the mathematical connection between
different scale of modelling can be elucidated. The interplay between global and local inhibition
is examined as another potential physiological mechanism for seizure onset. Bifurcation analysis of
a modified Wilson and Cowan neural field model is used to describe macroscale dynamics, which can
then be identified in the data and related to physiological aspects through relevant perturbations
in parameter values. Another study by Martinet et 
al.~\cite{martinetHumanSeizuresCouple2017} uses
simltaneous microscale and macroscale recordings in tandem with a Steyn-Ross neural field model to
examine mid-seizure dynamics. Guided by data across multiple scales and neural field modelling,
they propose candidates for physiological mechanisms that lead to the increased coherence during
seizures.

Neural field modelling has played a crucial role in demonstrating consequences from physiological
hypotheses on transitions of neural activity states, and continues to further our understanding of
brain state dynamics. However, many of the conclusions and results of neural field modelling result
in hypothetical and theoretical conclusions that are difficult, if not impossible, to verify due to
limitations on experimental neuroscience. Neural field modelling at this stage is showing more
promising outcomes by directing our focus to particular dynamical structures, such as certain
bifurcations, that emerge from these mathematical frameworks of mean macroscale neural activity.
Results from Kramer et al.~\cite{kramerHumanSeizuresSelfterminate2012} illustrates this by
identifying a Hopf bifurcation in a neural field model that presents similar patterns of
``slowing'' in the mean activity directly before a seizure as observed in the data. Dynamical
pathways for seizure onset and offset and biophysically realistic explanations are identified from
the neural field model used, and shows concordance between modalities on different spatial scales.
This phenomena of ``critical slowing'' is another ongoing topic of discussion within epilepsy
modelling in contemporary literature. There are a variety of publications that develop this
hypothesis further, and may provide a method of forecasting critical transitions from recordings of
EEG (Maturana et al.~\cite{maturanaCriticalSlowingBiomarker2019}). Overall, identifying 
bifurcations
from data is a promising direction for neural field modelling. Results from Jirsa et 
al.~\cite{jirsaNatureSeizureDynamics2014} utilises an approach of classifying bifurcations by 
patterns
observed in dynamical transitions and uses this as a framework for constructing an oscillatory
dynamical system that exhibits the same transitions. Higher order bifurcations are also being
analysed and connected to data (Saggio et al.~\cite{saggioFastSlowBursters2017}). 
Throughout the
next Section, we discuss the limitations of neural field modelling and the prospective
phenomenological approaches used in the contemporary literature.

\section{Discussion and future directions} 
\label{S:discussion}

This review paper on mathematical models of brain dynamics has focused on \linebreak population-scale models.
Throughout we have highlighted a number of critical theoretical developments and the underpinning
assumptions necessary to construct neural field models. Compared to traditional bottom-up
approaches, neural field modelling uses average quantities of populations of neurons  instead of
properties of individual neurons. This approach provides an essential balance between biophysical
realism and mathematical tractability, however there are some crucial shortcomings and consequences
that warranted a more thorough examination. This was addressed through  considering the historical
context, motivations and derivations of neural field models and explicitly detailing their
underpinning mathematical, physiological and anatomical assumptions. Then a general unifying
framework was derived from which, in principle, all neural field models (and neural mass models)
can be constructed and related back to. Following this, some of the key applications of NFM's were
also explored such as rhythmogenesis, sleep, and epilepsy. We conclude our review with a brief
discussion on the limitations of NFM's and by highlighting some recent research directions that
provide more biophysically realistic details whilst maintaining mathematical tractability.

\subsection{Limitations of neural field modelling} 
\label{S:limitations}

In order to maintain a balance between biological plausibility and mathematical tractability when
constructing models, compromises are inevitably made and it is essential these are carefully
considered. Whilst many of these assumptions are straightforward and justifiable, others are more
subtle and their validity less clear, especially when they are not currently experimentally
verifiable. This ambiguity leads to issues with interpreting results generated by NFM's. For
example, when brain state transitions are linked to connectivity parameters.

A theory should be constructed with `a clear physical picture of the process' being modelled and/or
`a precise and self-consistent mathematical formalism' (Dyson
\cite{dysonMeetingEnricoFermi2004}).

Unfortunately, the current status of NFM's permits neither of these. Unlike statistical physics,
there is not yet a precise and self-consistent mathematical framework that links the microscopic
with the macroscopic. In fact, much of the mathematical toolbox from theoretical physics that is
used to deal with complex interacting systems are primarily for closed equilibrium systems. In
contrast, like many biological systems, the brain is an open non-equilibrium system with scales
that are interdependent and not easily separable.

A primary challenge is the many degrees of freedom that NFM's have, even with the use of averaged
and lumped parameters. This is often due to lack of data and an incomplete understanding of the
macroscopic structures and processes. The resulting parameter space is often so large and
unconstrained that almost any dynamical behaviour can be constructed.

As famously quoted by John Von Neumann:
\begin{quote}
  ``With four parameters I can fit an elephant,  and with five I can make him wiggle his trunk!"
  --- Dyson  \cite[p.~297]{dysonMeetingEnricoFermi2004}.
\end{quote}

This paints an ambiguous physical picture of what is being modelled, as there is often no direct
correspondence between model and experimental parameters. 
 For example, it is not currently possible to measure the average synaptic connectivity
between  two neural populations in the cortex. At best the values used in the literature are order
of magnitude estimates that are currently unverifiable.

Given the lack of a precise mathematical formalism and/or a clear physiological picture at the
macroscopic scale,  neural field models could effectively be considered as biologically inspired
phenomenological models. As observed throughout the applications outlined in the previous section,
the majority of experimental verification is based on a visual comparison to EEG data. This is
partly due to the limitations of our understanding of the EEG signal, which is also impacted by the
disconnect between activities on different spatial scales. In order for neural field theory to
progress from descriptive to predictive, their hypotheses need to be experimentally validated so
that the theoretical framework can be successively iterated with experimental results. A
fundamental goal for the field then is the generation of testable predictions at the macroscale,
similar to those achieved at the neuronal level with the Hodgkin-Huxley equations. Although these
equations are phenomenologically derived, they are both descriptive and predictive. This is because
the parameters and state variables of the Hodgkin-Huxley model have a one-one correspondence with
those measurable in experiments, providing a predictive framework for single neuron dynamics. In
order for NFM's to generate novel results that are reproducible and predictive, a concordance
between theory and experiment is required.

\subsection{Future directions} 

Neural field theory is a very active area of research with many directions being explored and
developed. In this section we present four relatively recent examples: stochastic approaches which
extend NFM's to have state-dependent stochastic inputs, random neural networks which studies
networked dynamical systems using random matrix theory, abstracted network models which use more
abstract models to understand functional neuroimaging data and next-generation field models which
develop ensemble solutions that connect the micro and macro-scales.

\subsubsection{Stochastic neural field modelling} 
\label{S:stochastic-nfms}

Stochastic approaches are an active sub-discipline of research within modern neural field
modelling. Complex and chaotic variations in cellular and sub-cellular processes and interactions
across substantially large populations of neurons are more naturally modelled by a stochastic
process, instead of a detailed deterministic description, which is a theme prevalent across
mathematical biology as a whole. The incorporation of stochastic descriptions can capture
additional system interactions due to finite size effects and variations in parameters, which often
leads to models with more elaborate dynamics at the cost of greater mathematical complexity. In
neural field modelling, stochastic extensions significantly change the previously observed
behaviour of solutions in neural fields, often changing their stability and the propagation of
travelling waves (Bressloff 
\cite{bressloffStochasticNeuralField2009,bressloffInvasionExtinctionHeterogeneous2012}, 
Kilpatrick and Ermentrout
\cite{kilpatrickWanderingBumpsStochastic2013}).
More details on the principles behind stochastic modelling is found in the works by Bressloff
 \cite{bressloffStochasticNeuralField2014} and Faugeras and Inglis 
\cite{faugerasStochasticNeuralField2015}.

Stochastic neural fields arise from two approaches: stochastic extensions of existing neural field
frameworks, or population density approaches based on interacting microscale neuron models
(Bressloff \cite{bressloffStochasticNeuralField2014}). Extensions typically begin with a 
simpler neural field
framework, such as the Amari field equation \eqref{eq:amari-field} with one neural population and a
first order temporal kernel. In using one population, the deterministic dynamics are constrained to
exhibiting only the essential dynamics typical of a neural field while enabling more focus on the
impacts of the inclusion of a stochastic term. This approach results in Langevin-type equations of
the form
\begin{equation}
\di U(x,t) =
    \left(
        -U(x,t) +
        \int_\Omega
        w(x-X) \, f[U(X, t)]
        \di X
    \right) \,
    \di t +
    \sigma
    \left(
        U(x,t)
    \right) \,
    \di W(x,t),
\end{equation}
where $U(x,t)$ represents the mean membrane potential at position $x\in\Omega$ and time $t$, $f$ is
the wave-to-pulse transfer function, $w$ is the homogeneous spatial kernel, $\sigma$ is a
state-dependent diffusion coefficient, and $W(x,t)$ is a canonical Wiener process.
Multi-population extensions are also possible (Lima and Buckwar 
\cite{limaNumericalSolutionNeural2015}, Bressloff 
\cite{bressloffStochasticNeuralField2019}), as well as stochastic delay
differential equation perspectives (Buckwar
\cite{buckwarIntroductionNumericalAnalysis2000}, Riedler and Buckwar
\cite{riedlerLawsLargeNumbers2013}, Ableidinger et 
al.~\cite{ableidingerStochasticVersionJansen2017}).

Population density approaches to stochastic neural field modelling give a greater focus on the
randomness that arises from variation amongst neurons (Buice and Cowan
\cite{buiceFieldtheoreticApproachFluctuation2007}, Bruice et 
al.~\cite{buiceSystematicFluctuationExpansion2010}, Bressloff
\cite{bressloffStochasticNeuralField2009}, Bressloff and Lai
\cite{bressloffStochasticSynchronizationNeuronal2011}). There are a variety of formulations of this
fashion that incorporate randomness arising from synaptic activity in the form of a point process.
With careful analysis, the limiting behaviour can be derived using various methods, which recovers
classical neural field equations with additional stochastic terms. The continued development of
mathematical methods that prioritise the link between microscale and macroscale activity  is
essential for resolving issues in neural field modelling. A detailed discussion of stochastic
models in neuroscience is beyond the scope of this review given the overall richness and complexity
underpinning the mathematical formulation An advantage of models of this nature comes from a
greater reliance on more phenomenological descriptions and thus finding alternate methods of
quantifying and modelling neural activity (Sompolinsky et 
al.~\cite{sompolinskyChaosRandomNeural1988}, Ostojic 
\cite{ostojicTwoTypesAsynchronous2014},
Mastrogiuseppe and Ostojic 
\cite{mastrogiuseppeIntrinsicallygeneratedFluctuatingActivity2017}, Ipsen and Peterson 
\cite{ipsenConsequencesDaleLaw2020}). In addition, these models can be used to explain
phenomena arising from finite-size population effects and the occurrence of large deviations and
rare events (Faugeras and Maclaurin \cite{faugerasAsymptoticDescriptionStochastic2014}, 
Kuehn and Riedler
\cite{kuehnLargeDeviationsNonlocal2014}, Bressloff and Newby 
\cite{bressloffPathIntegralsLarge2014},
Lang and Stannat \cite{langFiniteSizeEffectsTraveling2017}). Stochastic inputs can be useful 
for
understanding spontaneous state transitions and characteristic changes in time series statistics in
systems close to criticality. For example, a study by Freyer et 
al.~\cite{freyerCanonicalModelMultistability2012} showed that using a multiplicative 
state-dependent
stochastic process results in a more realistic model output closer to the structure of experimental
EEG data. Abstractions and stochastic extensions of neural field models can further enhance the
focus of these discussions by giving a closer correspondence between experimental data and
mathematical structure, allowing for more data-driven approaches (Kalitzin et 
al.~\cite{kalitzinStimulationbasedAnticipationControl2010}, Benjamin et 
al.~\cite{benjaminPhenomenologicalModelSeizure2012}, Schmidt et 
al.~\cite{schmidtComputationalBiomarkerIdiopathic2016}).

\subsubsection{Random neural network models}
\label{S:random-neural-networks}

In describing the dynamics of large nonlinear networks of neurons, another important mathematical
approach is the study of random neural networks (Stern et 
al.~\cite{sternDynamicsRandomNeural2014}).
Instead of using a continuous spatial kernel to describe the connectivity between neurons, a matrix
operator with a discrete topology can be used. This weight matrix connects neural mass models and
is mathematically equivalent to a neural field model with a discretised field. This approach treats
the dynamics of neural networks as a directed graph dynamical system composed of $N$ first-order
neural mass models with instantaneous synapses that are nonlinearly coupled via a random
connectivity matrix $W$. The connectivity weights, which are the elements of $W$ are chosen from
identically and independently distributed (IID) random variables such as a normal distribution with
mean $\mu$ and variance $\sigma$. The dynamics of the $i$-th node is given by
\begin{equation}
    \dot{x}_i = -\frac{x_i}{\tau} + \sum_{j=1}^N w_{i,j} \, \phi(x_j), \label{eq:ind_node}
\end{equation}
where the state variable $x_i$ represents the average membrane potential of the $i$-th neural
population, $\tau$ is the membrane time-constant, $w_{i,j}$ is the synaptic weight from population
$j$ to population $i$ and constitutes a single entry in the $N \times N$ connectivity matrix $W$,
and $\phi(x_j)$ is a nonlinear coupling function, similar to a wave-to-pulse transfer function in
typical neural field models.

As $N$ becomes large, it is not mathematically tractable (due to nested nonlinear coupling
functions) to compute the fixed points and linearise the system around them to quantify the local
stability, as is typically performed with dynamical systems. To avoid these complications, random
matrix theory (RMT) is used to compute the statistical properties of the eigenspectrum from the
statistics of the connectivity matrix via the circular law (Tao et 
al.~\cite{taoRandomMatricesUniversality2010}). This analyses establishes an examinable 
mathematical
relationship between the network connectivity structure and phase transitions of the system. This
work was originally pioneered in neural networks by Sompolinsky et 
al.~\cite{sompolinskyChaosRandomNeural1988} using dynamical mean-field techniques from 
spin glass
models. However, computations of phase transitions in large nonlinear networked systems can be
traced back to the work of May  \cite{mayWillLargeComplex1972} from theoretical ecology 
who
demonstrated that the larger a system becomes, the more unstable it also becomes, which is known as
a May-Wigner transition (Ipsen \cite{ipsenMayWignerTransition2017}).

In such a networked system of the type shown in Equation \eqref{eq:ind_node}, it was found that the
system undergoes a phase transition into a `chaotic' like state with highly fluctuating and complex
dynamics (Stern et al.~\cite{sternDynamicsRandomNeural2014}), where the order parameter 
is in fact
the variance of the connectivity matrix. The microscopic mechanisms underlying this phase
transition have also been mathematically examined (Wainrib and Touboul
\cite{wainribTopologicalDynamicalComplexity2013}) and it was found that there is an explosion from
a single or simple set of fixed points to an exponential increase in the number of fixed points.
The phase transition can be computed approximately using a Kac-Rice formalism that counts the
number of fixed points. However, these results are based on networks with a single randomly
connected population drawn from an IID which is not anatomically realistic.

\begin{figure}[htbp]
    \centering
    \includegraphics[width=0.98\textwidth]{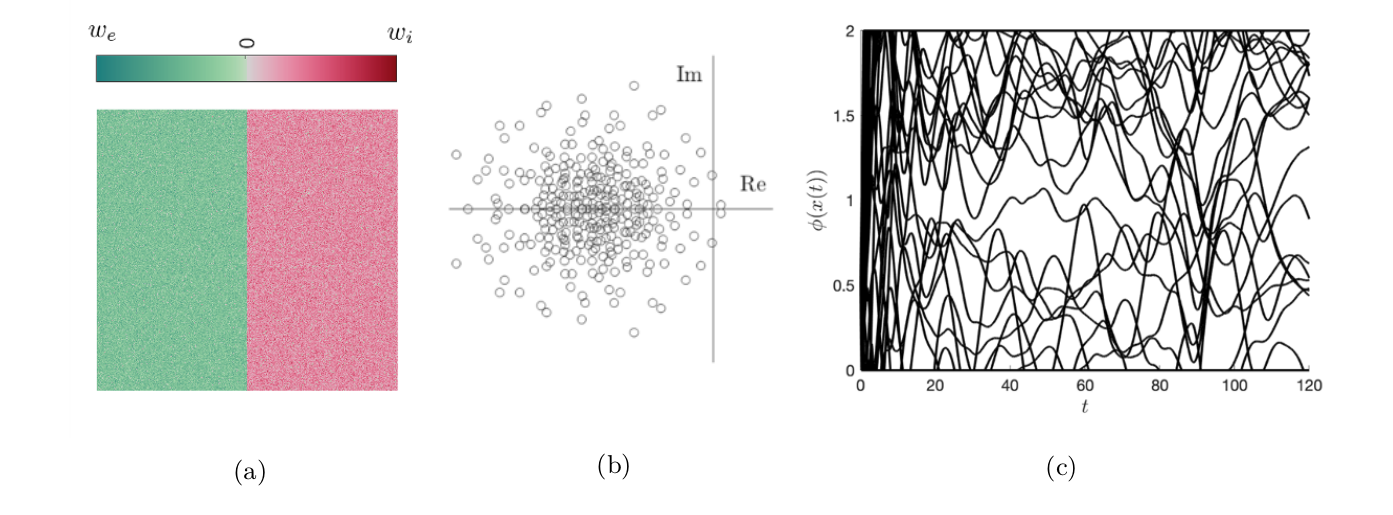}
    \caption{%
        \textbf{Instance of a random two-population neural network model.} (a) Heat map of random
        connectivity matrix. The block structure of the connectivity matrix incorporating Dale's
        law with positive excitatory weights $w_e$ and negative inhibitory weights $w_i$ can be
        seen. (b) Eigenspectrum of connectivity matrix, showing a compact circular support that is
        non-uniform in density due to Dale's law (Rajan and Abbott
        \cite{rajanEigenvalueSpectraRandom2006}) and a pair of unstable eigenvalues. (c) Simulation
        of networked system. The unstable eigenvalues cause a phase transition as can be seen in
        simulations, which shows the firing rate activity of individual nodes over time.
    }
    \label{F:NFM-RNN-pic}
\end{figure}

Rajan and Abbott  \cite{rajanEigenvalueSpectraRandom2006} demonstrated that when a 
more anatomically
realistic connectivity in the form of Dale's law is added there is a significant change in the
nature of the phase transition. Dale's law states that excitatory (inhibitory) neurons can only
give excitatory (inhibitory) signals (Eccles et 
al.~\cite{ecclesElectricalChemicalTransmission1976}). This introduces non-random structure 
that makes
the connectivity matrix entries no longer identically distributed and only partially random, as
matrix columns corresponding to the excitatory (inhibitory) elements are positive (negative)
entries. The matrix entries for different populations are drawn from different distributions with
their own means, variances and proportions with respect to each other e.g. no. of excitatory vs no.
of inhibitory. This consequently alters where and how rapidly the phase transition takes place in
parameter space. An example ``two-population'' connectivity matrix is shown in Figure
\ref{F:NFM-RNN-pic} (a). Further, because of the different variances of the populations, the
eigenspectrum computed from the circular law is no longer of uniform density (Rajan and 
Abbott
\cite{rajanEigenvalueSpectraRandom2006}), as seen in Figure \ref{F:NFM-RNN-pic} (b). The non-random
connectivity of Dale's law was also recently incorporated into a Kac-Rice formalism by Ipsen 
and Peterson 
\cite{ipsenConsequencesDaleLaw2020} to study the effects it has on the microscopic mechanism of the
phase transition. The location of the phase transition critically depended on the different
connectivity parameters of the different neural populations. Such a technique enables mathematical
investigations to be performed between the network connectivity and network dynamics that would not
be possible using standard techniques from classical dynamical systems.

Another technique that can be employed in networked dynamical systems that are coupled via
heterogeneous matrices is Dynamical Mean-field theory (DMF). In this approach, a many-body
deterministic problem is approximated and reduced to a one-body stochastic system. For example, the
coupling term in Equation \eqref{eq:ind_node} is replaced by a stationary Gaussian stochastic
process $\eta$,
\begin{equation}
    \dot{x}_i = -\frac{x_i}{\tau} +\eta(t).
\end{equation}
The individual trajectories $x_i$ are now stochastic processes, which in the limit of a large
network become uncorrelated and statistically equivalent (Ostojic
\cite{ostojicTwoTypesAsynchronous2014}), so that the network dynamics can be effectively described
by a single process. The first and second order moments of the network activity can be
non-trivially computed for certain cases and are strongly dependent on the network parameters. This
enables us to investigate the relationship between the statistics (first two moments) of the
activity and the network structure.

\subsubsection{Abstracted networked models} 
\label{S:abstract-networks}

Recordings from the brain (e.g. EEG) are effectively a spatially discrete representation of 
its
activity. Rather than consider spatial kernels that are continuous, more recent approaches have
utilised coupled neural mass models to create so-called network dynamic models. Effectively, this
is conceptually equivalent to a neural field model with a discretised neural field. Each channel of
an EEG measures bulk neuronal activity across centimetres, and neural mass models target the same
level of description. Recent developments investigate the coupling of EEG channels by using this
approach, providing an illustration of new methods of investigating spatial dependence of neural
activity and seizure dynamics (Carvallo et al.~\cite{carvalloBiophysicalModelingBrain2019}, 
Ers\"oz et al.~\cite{ersozNeuralMassModeling2020}). An example of this in the work of 
Goodfellow et al.~\cite{goodfellowEstimationBrainNetwork2016}, where intracranial EEG is 
modelled using a neural
field model comprised of coupled neural mass models (each corresponding to an EEG channel). This
framework allowed for the inference of a discrete spatial kernel based on measured EEG channel
correlations during seizure events, and was used to direct surgical strategies in severe cases of
epilepsy. A further development on this work by Lopes et 
al.~\cite{lopesOptimalStrategyEpilepsy2017} simplifies the neural mass model at
each channel to a canonical normal form model containing the SNIC bifurcation responsible for
epileptic transitions.

Networked dynamical systems have been prevalent in recent literature outside of neural field theory
as well, such is as the works of Kalitzin et 
al.~\cite{kalitzinStimulationbasedAnticipationControl2010}, Benjamin et 
al.~\cite{benjaminPhenomenologicalModelSeizure2012}, Ashwin et 
al.~\cite{ashwinFastSlowDomino2017} and
Woldman et al.~\cite{woldmanEvolvingDynamicNetworks2019}.
These more abstracted approaches employ a more direct top-down approach and aim to model the
properties or features of the data rather than the underlying physical processes, which remains
mostly inaccessible given the current state of technology. An example of an abstracted networked
dynamical system for neurosurgical planning in cases of severe epilepsy is given in Jirsa et 
al.~\cite{jirsaNatureSeizureDynamics2014}, who constructed a coupled oscillator (called the
``Epileptor'') based on a set of bifurcations that can give rise to the onset and off-set of
seizures as recorded in EEG. A saddle-node/homoclinic bifurcation pair is used to describe the
majority of seizure-like events based on frequency and amplitude scaling and direct current (DC)
shifts upon a state transition. The fast oscillator is based on a system proposed by 
Hindmarsh et al.~\cite{hindmarshModelNeuronalBursting1984} whereas the slow oscillator is 
based on a 
 system
proposed by Roy et al.~\cite{royPhaseDescriptionSpiking2011}. The coupling between these
oscillators is mediated by an ultraslow ``permittivity'' variable, linear inhibitory coupling from
the slow to fast subsystem, and low-pass filtered excitatory coupling from the fast to slow
subsystem. The equations describing this model are shown below:
\begin{equation}
    \begin{split}
        \dot{x} &= y - f_1(x,u,z) - z + I_1, \\
        \dot{y} &= 1 - 5x^2 - y, \\
        \dot{z} &= \frac{1}{\tau_0} (4(x-x_0) - z),
    \end{split}
    \qquad
    \begin{split}
        \dot{u} &= -v + u - u^3 + I_2 + 0.002g(x) - 0.3(z - 3.5), \\
        \dot{v} &= \frac{1}{\tau_2}(-v + f_2(x,u)), \\
    \end{split}
\end{equation}
where $x$ and $u$ are driven by independent white noise processes with standard deviation 0.025,
$x_0 = -1.6$, $\tau_0 = 2857$, $\tau_2 = 10$, $I_1 = 3.1$, $I_2 = 0.45$, $\gamma = 0.01$, and
\begin{equation}
\begin{split}
  &f_1(x,u,z) =
    \begin{cases}
      x^3 - 3x^2,  &  x<0, \\
      x(u - 0.6(z-4)^2),  &  x\geq 0,
    \end{cases}
  \quad \\
  &f_2(x,u) =
    \begin{cases}
      0,  &  u < 0.25, \\
      6x(u+\frac{1}{4}),  &  u \geq 0.25,
    \end{cases}
  \quad
  g(x) = \int_{t_0}^{t}e^{\gamma(t - s)} x(s) ds.
\end{split}
\end{equation}
The parameters and construction of the Epileptor is phenomenologically motivated, however the
dynamical landscape captures a great deal of detail of state transitions that are observed on EEG
(Saggio et al.~\cite{saggioTaxonomySeizureDynamotypes2020}). It should be noted that 
there are
ongoing efforts to identify relations between the model parameters and underlying neurophysiology
(Chizhov et al.~\cite{chizhovMinimalModelInterictal2018}). The Epileptor is used as a 
constituent of
the networked dynamical system of brain dynamics, called The ``Virtual Epilepsy Patient'', by 
Jirsa et al.~\cite{jirsaVirtualEpilepticPatient2017}. Each Epileptor node represents a different
brain region, and connectivity between nodes in the model is inferred from stereo-EEG data.
Despite the lack of identifiable parameters from neuronal activities, the framework presented
maintains a close relationship with higher levels of neurological data recording, such as EEG and
MEG techniques. A top-down focus may lead to new discoveries in the dynamics of brain state
transitions that were previously unidentifiable with classical approaches to mean field modelling.

Another application that utilised abstracted networked dynamical systems is in epilepsy diagnosis
and the work of Schmidt et al.~\cite{schmidtComputationalBiomarkerIdiopathic2016} and 
Woldman et al.~\cite{woldmanDynamicNetworkProperties2020}, which investigates 
correlations between 
 EEG
channels in resting state as a potential marker for epilepsy using an abstracted multi-level
Kuramoto model  \cite{kuramotoInternationalSymposiumMathematical1975,
 kuramotoChemicalOscillationsWaves1984}. These authors use a multi-level Kuramoto
approach in modelling emergent synchronisation from resting state EEG. Consider $P$ populations of
oscillators, where each population $p \in \br{1,\cdots, P}$ contains $N$ Kuramoto 
oscillators, each
with a natural frequency drawn from a population-specific distribution $g_p(\omega)$. We introduce
a $P \times P$ coupling matrix $\rho$, with entries $\rho_{p,q}$ describing the interactions
between populations. These entries may be binary or represent a weighted network. Let $C$ and $K_p$
be the global coupling parameter and the local coupling parameters. The evolution of oscillator $j
\in \br{1, \cdots, N}$ in population $p \in \br{1, \cdots, P}$ is given by,
\begin{equation}
  \ddx{\theta_j^p}{t} =
    \omega_j^p +
    \frac{K_p}{N} \sum_{k=1}^{N}
      \sin(\theta_k^p - \theta_j^p) +
    C \sum_{q=1}^P
      \frac{\rho_{p,q}}{N}
      \sum_{k=1}^N
        \sin(\theta_k^q - \theta_j^p).
\end{equation}
In this modular network, synchronisation can emerge locally within a single population as well as
across the global network. This was proposed as a model of EEG activity generated by the cortical
columns in the study by Schmidt et al.~\cite{schmidtDynamicsNetworksRole2014}. The 
degree of
synchrony in the individual nodes is measured using the same order parameters as for the simple
Kuramoto model, $r_p e^{i \psi_p} = \frac{1}{N} \sum_{j=1}^N e^{i \theta_j^p}$, and global coupling
can be quantified by the absolute mean of these constituent order parameters, $r_g =
\abs{\frac{1}{P} \sum_{p=1}^{P} r_p e^{i \psi_p}}$. In this context, the non-synchronised
incoherent state in the Kuramoto formulation corresponds to healthy background activity observed in
EEG, whereas emergent synchronous activity across the network is taken as a proxy for epileptiform
activity. As with the Epileptor, this phenomenological framework is based on the oscillatory
dynamics observed in high-level descriptions of neural activity rather than emergent properties of
large neural networks, unlike the next generation of neural field models. The abstracted nature of
these models may provide more accessible options for identifying models to clinical data, rather
than explaining physiological mechanisms of neural tissue. For further details on the principles of
oscillator models in neuroscience, we refer the reader to the review by Ashwin et 
al.~\cite{ashwinMathematicalFrameworksOscillatory2016}.

\subsubsection{Next-generation neural field modelling} 
\label{S:next-gen-nfms}

Recently, a ``next-generation'' approach to neural field modelling has emerged, which is derived
from microscopic principles (Coombes and Byrne 
\cite{coombesNextGenerationNeural2019}). This recent
sub-discipline of mathematical neuroscience re-imagines the principles of neural population models,
deriving macroscale states and parameters from the thermodynamic limit of the well-established
quadratic integrate-and-fire model. Using the work of Byrne et 
al.~\cite{byrneNextgenerationNeuralField2019}, we can give a brief overview of the modelling 
principle
used and the derivation of the model, which is based on the work of Luke et 
al.~\cite{lukeCompleteClassificationMacroscopic2013} and Montbri\'o et 
al.~\cite{montbrioMacroscopicDescriptionNetworks2015}. Consider $N$ neurons equally 
spaced across a
one-dimensional domain $\Omega$. The quadratic integrate-and-fire model is given by
\begin{equation}
    \ddx{u_i}{t}
    = u_i^2
    + \eta_i
    + \sum_{m} g_m^i(t) \, \left[
        u_m^{\tx{rev}} - u_i
    \right],
\end{equation}
where $u_i$ represents the membrane potential of neuron $i$, $\eta_i$ is a background input to
neuron $i$, $u_m^{\tx{rev}}$ is the reversal potential of synapse group $m$, and $g_m$ is the
conductance of synapse group $m$. A Lorentz distribution is chosen for $\eta_i$, since this allows
exact evaluation of the integrals that follow in this derivation. Note that so far in this model
the connectivity between neurons is all to all and the population is purely excitatory. A reset
condition is also imposed on each equation, so that the membrane potential $u_i$ is set to a resent
potential $u_r$ whenever the $u_i$ reaches a threshold potential $u_{\tx{th}}$. Using results from
Ermentrout and Kopell  \cite{ermentroutParabolicBurstingExcitable1986}, the membrane 
potential state
variables of quadratic integrate-and-fire model can be transformed to phase state variables of the
theta model (using $u_i = \tan(\theta_i/2)$), given by
\begin{equation}
    \ddx{\theta_i}{t}
    = 1 - \cos(\theta_i)
    + (1+\cos(\theta_i))\eta_i
    + \sum_{m} g_m^i(t) \, \left[ (1+\cos(\theta_i))u_m^{\tx{rev}} - \sin(\theta_i) \right],
\end{equation}
Each conductance is influenced by inputs from other neurons, as we have seen in classical neural
field modelling,
\begin{equation}
    \left( 1 + \tau_m \ppx{}{t} \right)^2 g_m^i(t)
    = \frac{2\kappa_m}{N} \sum_{j=1}^N w_m(x_i - x_j)\, \delta(\theta_j - \pi),
\end{equation}
where $\tau_m$ is the synaptic time constant, $\kappa_m$ is the global coupling strength of synapse
$m$, $w_m(x_i - x_j)$ is the homogeneous spatial kernel that depends on the distance between neuron
$i$ and neuron $j$. Taking the thermodynamic limit of the system of neurons and letting
$N\to\infty$, it can be established that the state of the mean field of a system of coupled
oscillators can be described by a probability density function $\rho(x,\eta,\theta,t)$ that obeys
the continuity equation (Landau and Lifshitz \cite{landauFluidMechanicsLandau1959}, 
Kuramoto
\cite{kuramotoInternationalSymposiumMathematical1975, 
kuramotoChemicalOscillationsWaves1984}, Lancellotti
\cite{lancellottiVlasovLimitSystems2005}), which can be used to derive the partial differential
equation,
\begin{equation}
    \left( 1 + \tau_m \ppx{}{t} \right)^2 g_m(t)
        = \frac{\kappa_m}{\pi} \sum_l
        \int_{\Omega}
            \int_{0}^{2\pi}
                w_m(x-y) \, e^{i l \theta}
                \left\lbrace
                    \int_{-\infty}^\infty
                        \rho(y, \eta, \theta, t)
                    \di \eta
                \right\rbrace
            \di \theta
        \di y
    .\label{eq:nexgennfm1}
\end{equation}
This system can be simplified by using the Ott-Antonsen ansatz 
\cite{ottLowDimensionalBehavior2008} by restricting the form of $\rho$ to a product of 
factors
$L(\eta)$ and $\sum_n \alpha(x,\eta,t)^n e^{i n \theta}$. This can be used to derive a new
expression of $\rho$ and another partial differential equation for $\alpha$. In addition to this, a
Kuramoto order parameter is defined for the mean field,
\begin{equation}
    z(x,t) =
    \int_0^{2\pi}
        \int_{-\infty}^{\infty}
            \rho(x,\eta, \theta, t) \,
            e^{i\theta}
        \di \eta
    \di \theta.
\end{equation}
These results are used to transform Equation \eqref{eq:nexgennfm1} as
\begin{equation}
    \left( 1 + \tau_m \ppx{}{t} \right)^2 g_m(t)
        = \kappa_m
        \int_{\Omega}
            w_m(x-y) \, f[z(y,t)]
        \di y,
        \quad\quad
        f(z) = \frac{1}{\pi} \frac{1 - \abs{z}^2}{\abs{1+z}^2},
\end{equation}
giving us the form of a classical neural field model, where $f$ is a real-valued function that
depends on the complex-valued population synchrony measure $z$, which if analogous to the
wave-to-pulse transfer function of classical neural field models. With further computation, a
differential equation for $z$ can also be derived, which completes the derivation of the 
next-generation
neural field model.

Here we only give a surface-level overview of the principles and equations involved, an in-depth
review can be found in the works of Coombes and Byrne  
\cite{coombesNextGenerationNeural2019} and Bick et 
al.~\cite{bickUnderstandingDynamicsBiological2020}. Although this approach is relatively 
new, it
has already been applied to problems in rhythmogenesis (Byrne et 
al.~\cite{byrneMeanFieldModel2017},
Keeley et al.~\cite{keeleyFiringRateModels2019}, Taher et 
al.~\cite{taherExactNeuralMass2020},
Segneri et al.~\cite{segneriThetaNestedGammaOscillations2020}), synchronisation of neural 
activity
(Luke et al.~\cite{lukeCompleteClassificationMacroscopic2013}), and propagation of epileptic
seizures (Gerster et al.~\cite{gersterPatientspecificNetworkConnectivity2021}). Furthermore, 
this
framework can also be extended readily to include neural connectivity via gap junctions 
(Laing \cite{laingExactNeuralFields2015}) and multiple neural populations (Segneri et 
al.~\cite{segneriThetaNestedGammaOscillations2020}), which alters the dynamical landscape
significantly.

This new branch of neural field modelling is readily adaptable to many methods of modern
statistical mechanics and brings a significant step forward on connecting microscopic dynamics to
macroscopic dynamics.

\subsection{Concluding remarks} 
\label{S:conclusion}

Despite their limitations, neural field models remain the most descriptively accurate models of
both normal and abnormal brain dynamics such as those observed in EEG data. They are relatively low
dimensional, computationally inexpensive to simulate and can be constructed within an analytical
framework. This framework facilitates rigorous mathematical analysis that enables us to carefully
map and explain the models behaviour. The parameters of the model, even in the absence of a one to
one correspondence with experimental measures, can then be interpreted in a physiologically and
anatomically meaningful way. We again reiterate the importance of transparency of the underpinning
assumptions needed to construct neural field models, in order to enable careful and meaningful
interpretation of the results. This is essential for future developments that aim to address the
limitations of neural field theory. Most importantly, the evolution of this type of modelling in
parallel with neuroscience experiments and technology means that they are capable of generating
testable hypotheses that can be experimentally verified. This gives them the potential to be not
only predictive, but to also have significant explanatory power that elucidates some of the unknown
mechanisms behind brain function and dysfunction.






\providecommand{\bysame}{\leavevmode\hbox to3em{\hrulefill}\thinspace}
\providecommand{\MR}{\relax\ifhmode\unskip\space\fi MR }
\providecommand{\MRhref}[2]{%
  \href{http://www.ams.org/mathscinet-getitem?mr=#1}{#2}
}
\providecommand{\href}[2]{#2}



\ACKNO{%
    Andre D.H. Peterson acknowledges financial support in the form a fellowship from the Graeme Clark
    Institute. Blake J. Cook acknowledges the financial support of the EPSRC through a DTP
    studentship award to the University of Birmingham.  Wessel Woldman acknowledges the financial
    support of Epilepsy Research UK through an Emerging Leader Fellowship award F2002, and the NIHR
    through grant AI01644. John R. Terry acknowledges the financial support of the EPSRC through
    grants EP/N014391/2 and EP/T027703/1, and the NIHR through grant AI01646. Andre D. H. Peterson
    and John R. Terry acknowledge the financial support of the Royal Society through International
    Exchange grant  IE170112. All authors gratefully acknowledge the comments of Stephen Coombes on
    an early draft of the manuscript.
}

\nocite{*}
\end{document}